\documentclass[12pt,preprint]{aastex}

\newcommand{\kms}{km s$^{-1}$}
\newcommand{\etal}{{\it et al.} }

\slugcomment{To appear in Ap. J.}

\shorttitle{Multi-epoch polarimetry of SiO masers toward TX Cam}
\shortauthors{Kemball et al.}

\begin{document}

\title{Multi-epoch Imaging Polarimetry of the SiO Masers in the
Extended Atmosphere of the Mira Variable TX Cam}

\author{Athol J. Kemball}
\affil{Department of Astronomy, and \\
  Institute for Advanced Computing Applications and Technologies \\
  University of Illinois, 1002 W. Green Street, Urbana, IL 61801}
\email{akemball@illinois.edu}

\author{Philip J. Diamond}
\affil{Jodrell Bank Center for Astrophysics \\
  University of Manchester, Manchester, M13 9PL, United Kingdom}
\email{philip.diamond@manchester.ac.uk}

\author{Ioannis Gonidakis}
\affil{Department of Astrophysics, Astronomy, \& Mechanics, \\
University of Athens, GR-157 83, Athens, Greece} 
\email{gonidakis@gmail.com}

\and

\author{Modhurita Mitra, Kijeong Yim, Kuo-Chuan Pan, and Hsin-Fang Chiang}
\affil{Department of Astronomy \\
  University of Illinois, 1002 W. Green Street, Urbana, IL 61801}


\begin{abstract}

We present a time series of synoptic images of the linearly-polarized
$v=1,\ J=1-0$ SiO maser emission toward the Mira variable, TX
Cam. These data comprise 43 individual epochs at an approximate
biweekly sampling over an optical pulsation phase range of $\phi=0.68$
to $\phi=1.82$. The images have an angular resolution of $\sim 500
\mu$as and were obtained using the Very Long Baseline Array (VLBA),
operating in the 43 GHz band in spectral-line, polarization mode. We
have previously published the total intensity time series for this
pulsation phase range; this paper serves to present the
linearly-polarized image sequence and an associated animation
representing the evolution of the linear polarization morphology over
time. We find a predominantly tangential polarization morphology, a
high degree of persistence in linear polarization properties over
individual component lifetimes, and stronger linear polarization in
the inner projected shell than at larger projected shell radii. We
present an initial polarization proper motion analysis examining the
possible dynamical influence of magnetic fields in component motions
in the extended atmospheres of late-type, evolved stars.

\end{abstract}

\keywords{masers - polarization - stars: magnetic fields -
stars: individual (TX Cam)}

\section{Introduction}

The extended atmospheres of late-type, evolved stars on the asymptotic
giant branch (AGB) are host to complex underlying astrophysical
processes of vital importance. Mass-loss from these stars originates
in this region and is an important enrichment mechanism for the
interstellar medium. The large-amplitude, long-period variables
(LALPV) span a range of variability sub-classes, including the Mira
variables, with a median pulsation period of several hundred days
\citep{habing96}. Their near-circumstellar environments (NCSE) are
permeated by pulsation shocks from the central star \citep{hinkle82,
hinkle97, alvarez00} and complex convective motions in the outer
envelope \citep{porter97,freytag02}. These stars have significant
mass-loss rates and are often obscured in visible optical bands. SiO
masers at 43 GHz are however ubiquitous in the extended atmospheres of
these objects. As compact high-brightness components that are
significantly polarized, they act as important probes of the
astrophysics in the close circumstellar environment, including the
morphology and relative dynamical influence of magnetic
fields. Spectral-line polarization VLBI techniques allow direct
imaging of this region in Stokes $(I,Q,U,V)$ at a spatial resolution
of $\mu$as, unmatched at other wave-bands. The power of this technique
is significantly enhanced if synoptic monitoring is conducted over a
range of stellar pulsation phase, so that the key dynamical drivers
can be studied in this region.

There remain significant uncertainties in integrated astrophysical
models of the NCSE. Stellar pulsation hydrodynamical models are either
semi-analytic \citep{bertshinger85}, or confined to spherical or
axi-symmetric numerical studies \citep{bowen88, bessell96,
humphreys02}. Theoretical studies of magnetic fields in AGB stars have
produced a range of predicted magnetic field magnitudes and
morphologies. \citet{blackman01} have proposed a conventional $\alpha
\omega$ dynamo driven by an inner, differentially-rotating stellar
core. Photospheric fields of the order of $\sim 400$ G are predicted
in this model, but for the case of an isolated (non-binary) AGB star,
this model would need a dynamo resupply mechanism to sustain a
magnetic field for long enough to shape global mass loss
\citep{nordhaus07}. The role of magnetic fields in shaping the global
mass-loss process is considered by \citep{garcia-segura99}. In
contrast, a convective $\alpha^2 \omega$ dynamo has been proposed by
\citet{soker02}; in this model, turbulent dynamo fields may produce
magnetic cool spots ($\sim 10-100$ G) that regulate local dust
formation, but have no global shaping role in mass-loss
\citep{soker02a}. Broader arguments against the proposition that
global magnetic fields can shape planetary nebulae are summarized by
\citet{soker06}. In a numerical MHD study, \citet{dorch04} have
examined a convective dynamo in a supergiant atmosphere and report
predicted local fields of up to $\sim 500$ G. These fields would have
significant dynamical influence, albeit localized with low filling
factors.

Recent water maser polarimetry of late-type, evolved stars has found
evidence for $> 10^2$ mG magnetic fields at the shell radius of the
H$_2$O masers; this extrapolates to a surface field of $\sim 10^2$
G \citep{vlemmings05}. Recent optical spectroscopy of unobscured
central stars in planetary nebulae (PNe) has measured
photospheric magnetic fields of several kG in these
stars \citep{jordan05}. Magnetic fields of this magnitude would be
dynamically significant and of great astrophysical importance to
related studies of collimation mechanisms for PPN \citep{meixner04}. A
small subset of stellar water maser sources that show linear jets also
appear to be magnetically collimated \citep{vlemmings06}.

As part of a synoptic monitoring campaign of the $J=1-0$ SiO maser
emission toward the Mira variable TX Cam at fine time-sampling, this
paper presents linear polarization images for the first 43 epochs
observed, over a pulsation phase range $\phi=0.68$ to $\phi=1.82$. The
reduction and analysis of the corresponding total intensity data was
published earlier by \citet[hereafter Paper I]{diamond03}. This paper
serves to present the time-series of linear polarization images, and
provides an analysis of the linear polarization morphology and early
results on associated proper motions. The associated circular
polarization data will be presented in a future paper. TX Cam is a an
M8-M10 Mira variable with a pulsation period of 557.4 days
\citep{kholopov85}, and has a Mira period-luminosity distance estimate
of $\sim 390$ pc \citep{olivier01}. It was first observed in total
intensity VLBI observations by \citet{diamond94}. Recent simultaneous
total intensity VLBI imaging of the $v=\{1,2\},\ J=1-0$ transitions is
reported by \citet{yi05}.

The first polarimetric VLBI imaging of the $v=1,\ J=1-0$ SiO maser
emission toward TX Cam was conducted at a single epoch by
\citet{kemball97}, who found a predominant orientation of the electric
vector position angle (EVPA) tangent to the arc of the projected maser
shell. This will be referred to as tangential polarization in what
follows. Subsequent SiO VLBI imaging polarimetry of TX Cam and
IRC+10011 at a later single epoch was carried out by
\citet{desmurs00}, who found a similar tangential linear polarization
morphology. A combined optical interferometric and VLBI polarimetric
monitoring campaign for a sample of O-rich AGB stars, but not
including TX Cam, is currently underway and is described in a series
of papers by \citet{cotton04}, \citet{cotton06}, and
\citet{cotton08}. Recent single-dish polarimetry of the SiO $J=2-1$
transition at 86 GHz for a sample of late-type, evolved stars is
reported by \citet{herpin06}.

The structure of this paper is as follows. Section 2 describes the
observations and data reduction methods. The results and discussion
are presented in Section 3, and conclusions in Section 4.

\section{Observations and Data Reduction}

\begin{figure}
\epsscale{0.9}
\plotone{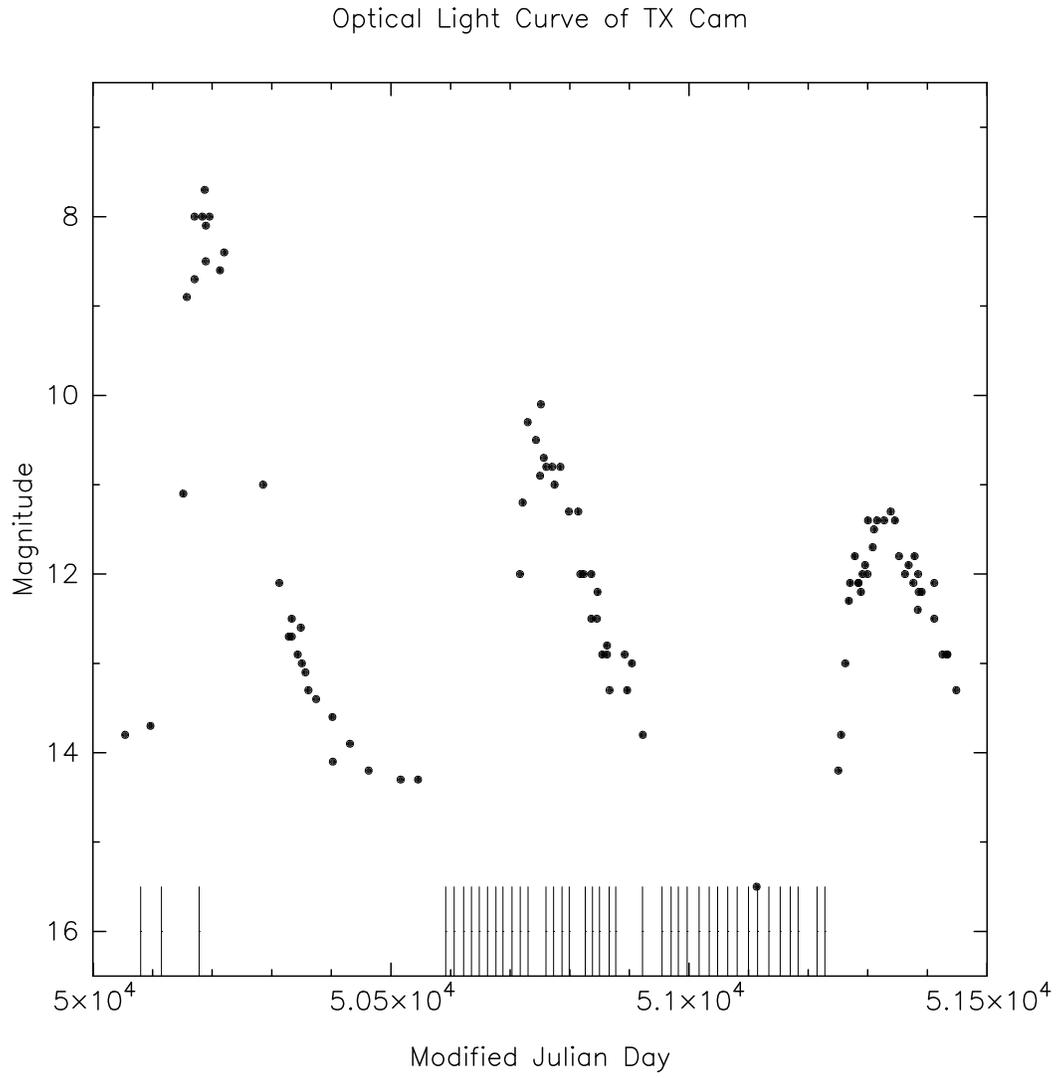}
\caption{The optical light curve of TX Cam as provided by the AAVSO.
The markers at the base of the figure indicate the dates of the VLBA 43 GHz 
observations. Reproduced from Figure 1 of Paper I.}
\label{fig-aavso}
\end{figure}

In this paper we present the polarization reduction of the time-series
of 43 synoptic observations of the $v=1,\ J=1-0$ SiO maser emission
toward TX Cam, previously analyzed and reported in total intensity in
Paper I. The list of epoch codes, observing dates, and associated TX
Cam optical pulsation phases are repeated here for reference as
Table~\ref{tbl-vlba-epochs}. The observations in this paper cover the
time interval from 24 May 1997 to 19 February 1999, corresponding to a
TX Cam pulsation phase range $\phi=0.68$ to $\phi=1.82$. The data were
observed using the Very Long Baseline Array
(VLBA\footnote{http://vlba.aoc.nrao.edu}) operating in the 43 GHz
band, and augmented at most epochs by a single antenna from the Very
Large Array\footnote{http://vla.aoc.nrao.edu}. Both telescopes are
operated by the National Radio Astronomy Observatory
(NRAO\footnote{The National Radio Astronomy Observatory is a facility
of the National Science Foundation, operated under cooperative
agreement by Associated Universities, Inc.}). Auxiliary VLA
interferometer observations were scheduled throughout the period of
the VLBA observations, to allow absolute calibration of the EVPA of
the linearly-polarized SiO emission. Both the VLBA and VLA
observations and data reduction are discussed in further detail
below. Optical variability data, showing the pulsation phases in
relation to the scheduling of these VLBA observations, is shown in
Figure~\ref{fig-aavso}, reproduced here from Paper I.

\subsection{VLBA Observations and Reduction}

The data were recorded at each antenna in dual-circular polarization
in two 4 MHz baseband spectral windows, each digitally sampled at the
full Nyquist rate of 8 Mbps in 1-bit quantization. The lower spectral
window was centered at a fixed topocentric frequency corresponding to
the $v=1,\ J=1-0$ SiO transition, at an assumed rest frequency
$\nu_0=43.122027$ GHz and a systemic velocity $V_{LSR}=+9$ \kms toward
TX Cam, Doppler-shifted to the center of the array and the midpoint of
the observations at each epoch. No further real-time Doppler
corrections were applied; these were performed in post-processing. For
the majority of the VLBA epochs in Table~\ref{tbl-vlba-epochs}, the
$v=2,\ J=1-0$ transition was simultaneously observed, and centered in
the second spectral window; the analysis of these data will be
presented in a future paper

Each VLBA observing epoch was scheduled in a comparable local sidereal
time (LST) range, as permitted by telescope operations. Three
fixed schedule formats were used over the course of these
observations. For epochs \{A-O\} a schedule of 6-hour duration was
used, balanced between the target source, TX Cam (3.5 h), the primary
continuum calibrator, J0359+509 (1.1 h), and secondary calibrators
3C454.3, J0609-157, and 3C286 (each 0.22 h). A single scan on 3C286
was included initially to explore an alternative absolute EVPA
calibration method, as discussed further below. This was not
successful however, and for epochs \{P-X\} the 3C386 scan was
re-allocated to TX Cam (3.6 h), and the schedule re-balanced. For
epochs \{Z-AQ\}, an 8-hour schedule format was used, but 2 hours of
each run were shared with a 22 GHz project, leaving the total 43 GHz
observing time, and its division between sources, unchanged.

The data were correlated at the VLBA correlator in Socorro, NM,
adopting a field center position for TX Cam of ($\alpha_{J2000}=
05^h00^m51^s.186,\ \delta_{J2000}=56^d10^m54^s.341)$. The correlator
accumulation interval $\Delta t_c$ was constrained by output data rate
limits in place at the correlator at that time. A value $\Delta
t_c=6.03$ s was used for epochs \{A-G,I-K\}, and reduced to $\Delta
t_c=4.98$ s for epochs \{H,M-AR\}. All polarization correlation
products (RR,RL,LR,LL) were formed in each correlation accumulation
interval from the dual-circular data acquired at each antenna. In this
spectral-line polarization mode the VLBA correlator has a channel
count limit of $N_{chan}=128$. This produced auto- and cross-power
spectra in each 4 MHz baseband with a nominal velocity spacing of
$\sim 0.2$ \kms.

The measured visibility data were reduced using spectral-line
polarization calibration methods described by \citet{kemball95} and
\citet{kemball97}, as implemented in a semi-automated pipeline using a
modified version of the AIPS\footnote{http://www.aips.nrao.edu}
package maintained by the NRAO. These analysis methods are optimized
for the reduction of spectral-line polarization VLBI observations of
the type described here; in particular they make no implicit
assumption that Stokes $V=0$. The SiO molecule is non-paramagnetic,
and the mean degree of circular polarization in the SiO maser
rotational transitions is small but not zero, $m_c \sim 1-3\%$
\citep{barvainis87,kemball97}. In the data reduction method employed
here, all antenna-based group delay, fringe-rate, and phase
calibration, including final phase self-calibration, was performed
relative to a reference antenna in a reference receptor
polarization. This calibration was transferred to the orthogonal
receptor polarization by solving for all relevant phase and group
delay offsets using continuum calibrator observations. This
generalized reduction method retains positional coincidence between
Stokes $I$ and $V$ images, while making no implicit assumption that
Stokes $V$ is identically zero \citep{kemball95}. Amplitude
calibration was performed with the same goal of preserving the low
Stokes $V$ signature. Each circular receptor polarization was
calibrated independently using the method of \citet{reid80} to fit the
autocorrelation spectra in each receptor polarization to a template
reference spectrum in that polarization. The template spectra in each
receptor polarization were calibrated on an absolute scale using
system temperature measurements reported during the observations and
known point-source sensitivities and gain curves published for the
VLBA antennas. A final single differential polarization gain
correction was determined by cross-fitting the spectra between
receptor polarization at the template scan to measure a relative scale
factor. For the low integrated $m_c$ known for SiO masers, this
approximation is warranted for the integrated autocorrelation
spectrum; it would not be appropriate for individual cross-power
spectra, for example, as some SiO maser components can be
substantially circularly polarized \citep{kemball97}. This final
differential polarization gain correction adjusts for errors in the
reported system temperatures and point source sensitivities, as used
in the a priori absolute calibration of the template spectrum in each
receptor polarization. We estimate these uncertainties very
approximately as 5-10\% for the VLBA at this frequency.  We do not
correct for atmospheric opacity at the pointing position and time of
observation of the template spectrum for the data reported
here. However, this has no polarization dependence, and accordingly
does not affect our measurement of linear or circular polarization
percentage or orientation; it only introduces a modest uncertainty in
our absolute flux density scale. This uncertainty is mitigated by the
fact that the template spectrum is chosen for its signal-to-noise
ratio (SNR) and peak magnitude; these factors ensure that it is almost
always at a sufficiently high elevation to minimize atmospheric
opacity contributions.

As phase self-calibration is used, our final VLBA images of the SiO
maser emission toward TX Cam retain no absolute astrometric
information or alignment. In Paper I, zeroth frequency-moment total
intensity images from successive epochs were initially aligned using
spatial cross-correlation.  This initial alignment was then refined
using feature-based registration in which the deviation in proper
motion paths predicted for individual maser components was
minimized. For the polarization and total intensity images produced
and reported in the current reduction, we adopt the alignment frame
developed in Paper I as an absolute reference. Each of the
zeroth-moment total intensity images produced in the current work was
registered relative to its corresponding epoch in the Paper I frame
using Fourier-based matched filtering. A robust, weighted-sum
estimator of the filter output peak was used to measure the
two-dimensional spatial offsets. This was found to out-perform an
elliptical Gaussian fit to the peak, likely because it is less
sensitive to small morphological differences in the total intensity
images being aligned. Based on manual cross-checks of the positions of
individual SiO maser components, we believe the registration of our
polarization images relative to the alignment frame determined in
Paper I has an accuracy of $100 \mu as$ or less. Note that the images
in Paper I used to produce the Stokes $I$ movie published there were
enhanced in contrast by applying the transformation $\sqrt{I}$ and a
subsequent threshold cutoff $\sqrt{I} < 0.7$. For consistency, the
same transformation was applied to the total intensity images reduced
here before the matched filter alignment.

The calibrators, J0359+509, 3C454.3, and J0609-157 were reduced
separately using continnuum polarization VLBI techniques for
millimeter wavelength observations described by \citet{kemball96}. As
the only calibrator source with sufficient parallactic angle coverage,
continuum instrumental polarization leakage terms were determined from
the observations of J0359+509 alone and applied to the other
calibrator sources. Continuum polarization images in Stokes $(I,Q,U)$
were then produced for each calibrator at each epoch. As noted above,
the calibrators 3C454.3 and J0609-157 were only observed for a single
13-min scan at each epoch, primarily as visibility calibrators, and
have limited image sensitivity and poor $uv$-coverage as a result. The
image quality for these sources was accordingly often poor, especially
for 3C454.3, which has more complex structure \citep{kemball96} than
the more compact J0609-157.

\subsection{VLA Observations and Reduction}

The absolute EVPA of any linearly-polarized emission measured in our
VLBA observations is unknown as no instrumental measurement of the
absolute R-L phase difference at the reference antenna is
possible. Although the VLA similarly lacks the capability to make
this measurement, it can provide absolute astronomical calibration of
EVPA relative to a small subset of primary polarization calibrators
given the lower spatial resolution of the array.

We have adopted the traditional approach of calibrating the EVPA of
our VLBA observations by using the VLA to transfer absolute
polarization calibration to these data via a sufficiently compact
secondary polarization calibrator suitable for imaging with both
arrays. We use the continuum source J0359+509 as this transfer
calibrator. Short (1-2 h) VLA observations were accordingly requested
across the time-span of our VLBA observations, and were scheduled when
logistically possible on the VLA. The dates of the auxiliary VLA
observation associated with our TX Cam monitoring campaign are listed
in Table~\ref{tbl-vla}.

In these VLA observations we measured the absolute EVPA of J0359+509
relative to the primary polarization calibrator 3C138, which is the
only primary polarization calibrator available in this right ascension
range. Our observations were primarily scheduled in continuum mode,
observing with two 50 MHz spectral windows and full
polarization. Instrumental polarization calibrators were chosen
separately for each VLA run, as it was not possible to schedule these
observations in identical local sidereal time ranges. Each observation
included J0359+509 and 3C138 however, in order to address the primary
scientific goal.

Two dominant sources of systematic error affected these
observations: i) their short duration, which at times limited the
fidelity of the instrumental polarization calibration due to the
reduced parallactic angle range; and ii) the extended structure of
3C138 at this angular resolution and observing frequency, making it a
sub-optimal polarization reference in the long-baseline VLA
configurations. The deficiencies in 3C138 notwithstanding, it remained
our only viable primary polarization calibrator in this LST range.

The data were reduced using standard heuristics for Q-band VLA
observations, as described in an appendix of the AIPS cookbook
\citep{greisen07}. During data loading, an opacity correction was
applied using a weighted estimate of the zenith opacity based on the
surface weather, with a relative weight of 0.75, and a seasonal model,
weighted by 0.25. A standard gain curve supplied for the VLA was
used. Post-facto antenna-position corrections were then applied as
needed. All subsequent reduction was performed for each 50 MHz
spectral window independently. The data were edited interactively and
the primary flux scale established for 3C138 using the model provided
within the AIPS package. An initial phase calibration over a solution
interval of 20 s was performed, using an a priori model for 3C138 (to
mitigate source resolution) and any uv-range limits recommended for
other calibrators in the schedule, but which were otherwise assumed to
be point sources. An amplitude gain correction was then determined for
each antenna, scan, and receptor polarization on each continuum
source. This allowed the flux density of the continuum calibrators
other than 3C138 to be determined relative to the prior flux density
scale established for 3C138. A joint solution for instrumental
polarization was performed across all continuum calibrators, and final
images in Stokes $(I,Q,U)$ produced for each continuum source in each
50 MHz spectral window separately. An image pixel spacing of
$\frac{1}{10}$th of the expected spatial resolution in each VLA
configuration was adopted, assuming that the VLA configurations have
the following spatial resolutions at 43 GHz: A) 0.05\arcsec; B)
0.15\arcsec; C) 0.47\arcsec; and, D) 1.5\arcsec
\citep{ulvestad07}. For hybrid configurations, such as AB, the
geometric mean of the nominal resolution in each configuration was
used. The images were cleaned to the $1\sigma$ thermal noise limit
predicted for each source based on the number of calibrated
visibilities, time and frequency averaging, and known system
equivalent flux density.

\begin{figure}
\epsscale{0.9}
\plotone{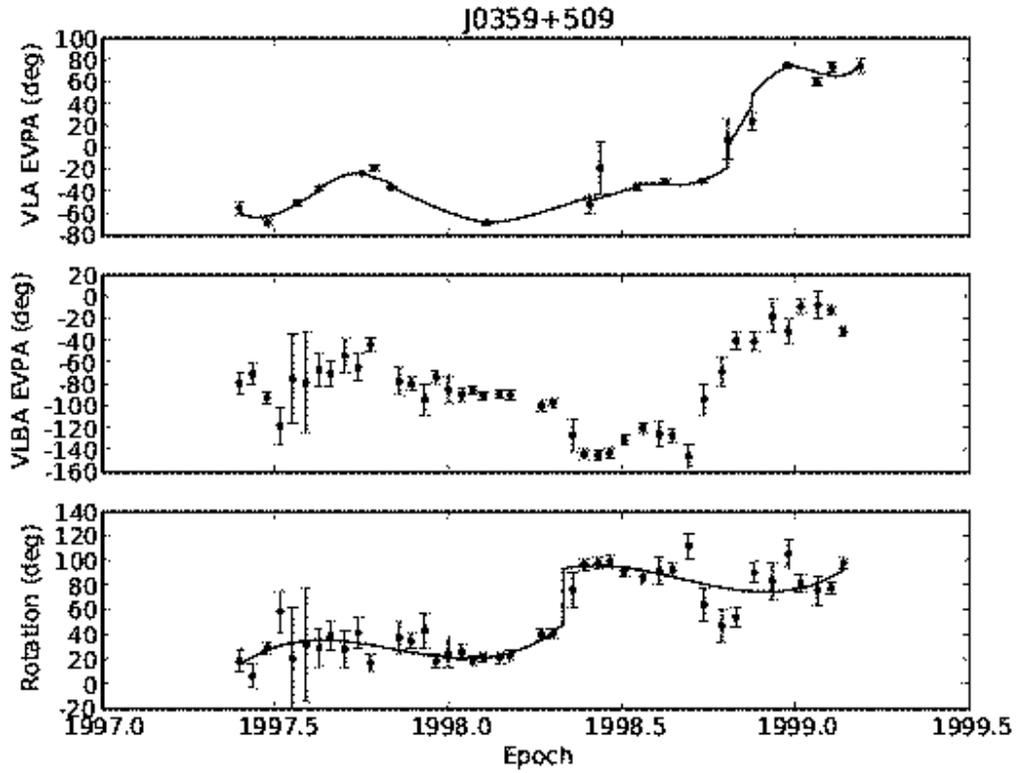}
\caption{The measured VLA absolute EVPA of J0359+509 ({\it upper
panel}), the VLBA EVPA ({\it middle panel}), and the difference ({\it
lower panel}), equivalent to the EVPA rotation needed to calibrate the
absolute EVPA of the VLBA polarization images.}
\label{fig-evpa}
\end{figure}

The absolute EVPA of J0359+509 was determined separately in each 50
MHz spectral window, relative to the value measured for 3C138 in that
window, here assumed to have an absolute EVPA $\chi_{EVPA} =
-14^\circ$ \citep{perley03}. The final absolute EVPA $\chi$ of
J0359+509 was computed as the mean of the EVPA measured independently
in each 50 MHz spectral window, here denoted by subscripts 1 and 2,
with an estimated standard error $\epsilon_\chi =
\frac{1}{\sqrt{2}}|\chi_1 - \chi_2|$. This error estimate is
approximate due to the small number of independent measurements. The
measured absolute EVPA for J0359+509 is listed in Table~\ref{tbl-vla}
and plotted in the top panel of Figure~\ref{fig-evpa}. A weighted
cubic polynomial fit over a scrolling window five samples wide is
drawn as a solid line in the upper panel. This fit was used to
interpolate the absolute EVPA measured with the VLA to the dates of
the VLBA observations, as listed in Table~\ref{tbl-vlba-epochs}. We
note however that the VLA data are sparsely-sampled in time,
especially during a sharp rise in EVPA in late 1998. Continuum
extra-galactic sources are significantly variable in linear
polarization, and this interpolation scheme does not capture all
source variability at the current sampling.

The EVPA measured for J0359+509 in the VLBA observations is plotted in
the middle panel in this figure. This measured EVPA is not absolutely
calibrated and is instead uncertain by an additive rotation equal to
half the RCP-LCP phase difference at the VLBA reference antenna, which
in the VLBA reduction was chosen to be Los Alamos. Error bars
$\sigma_\chi$were estimated approximately for the VLBA continuum
polarization images from the measured off-source rms, $\sigma_Q$ and
$\sigma_U$, in Stokes $Q$ and $U$ respectively as $\sigma_\chi =
\frac{1}{2(Q^2 + U^2} \sqrt{\sigma_u^2 Q^2 + \sigma_q^2 U^2}
\sqrt{\frac{1}{N}}$, where $N$ is the number of pixels over which
$(Q,U)$ were measured, and $\sigma_Q$ and $\sigma_U$ were corrected
for noise correlation resulting from convolution with the restoring
beam (of mean FWHM $\sim 300\mu$as) using the relation given by
\citet{condon91}.

The predicted EVPA rotation for each epoch, obtained as the difference
between the interpolated VLA and measured VLBA EVPA at each epoch, is
plotted as a series of points in the lower panel. We expect, from
instrumental considerations, that the RCP-LCP phase difference at a
VLBA antenna should be relatively stable over long periods, barring
any changes in receivers or electronics. We note the bimodal
distribution of points that has a vertical break coinciding with the
only receiver change at the reference antenna (Los Alamos) during the
period of these observations. This occurred on 30 April
1998. Accordingly, we fit a cubic polynomial to the data separately on
either side of this break to derive the final EVPA rotation per
epoch. This was the lowest-order polynomial to yield a reasonable fit
to the data. The significant deviations in late 1998 are likely due to
the systematic under-sampling of the VLA EVPA measurements during this
period, as noted above. We estimate the peak-to-peak error in absolute
EVPA alignment at $\sim 10^\circ - 20^\circ$ using this method.

\section{Results and Discussion}

\begin{figure}[h] 
\advance\leftskip-1cm
\advance\rightskip-1cm
$\begin{array}{cc} 
\includegraphics[width=70mm, height=70mm]{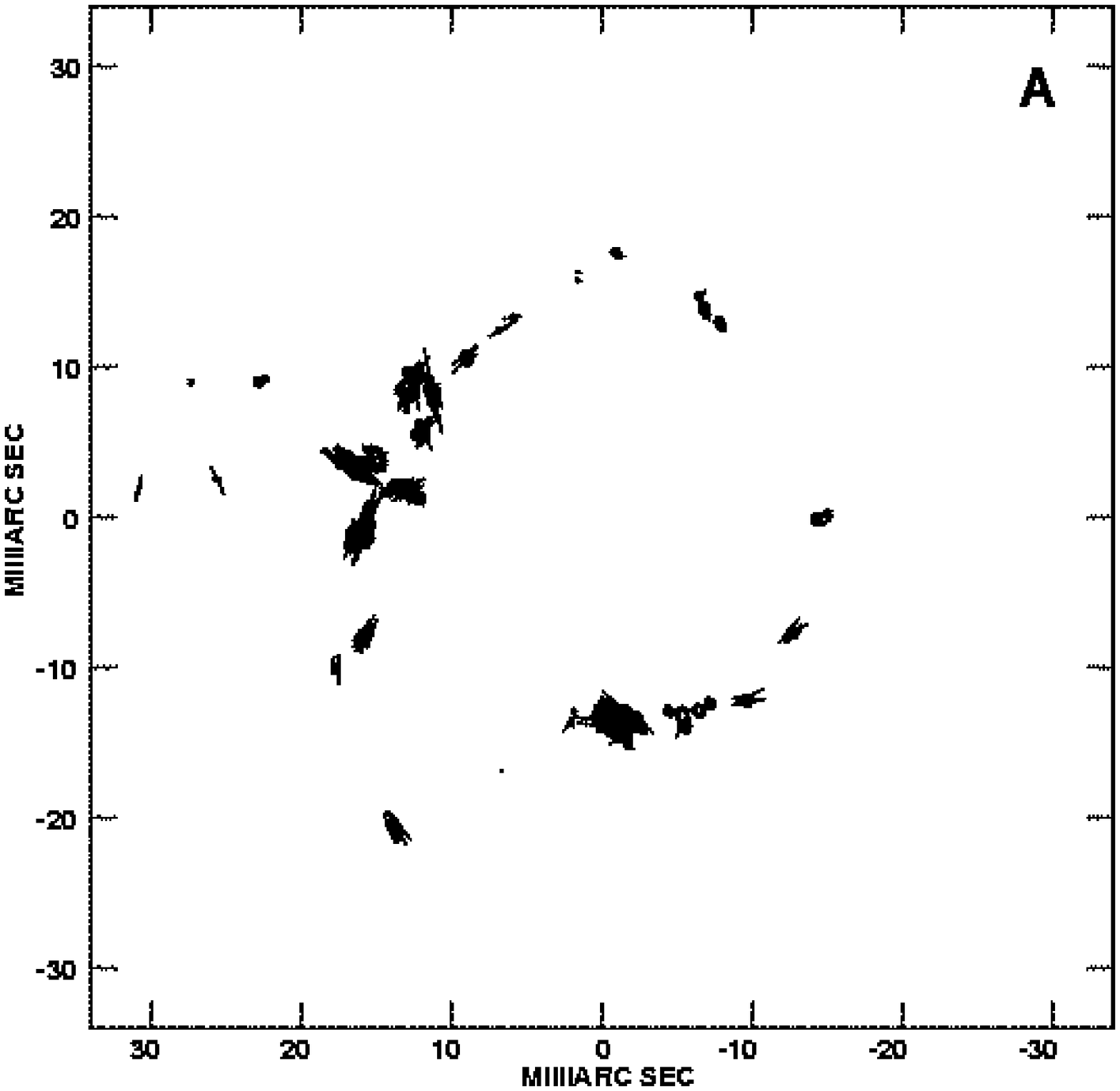} & 
\includegraphics[width=70mm, height=70mm]{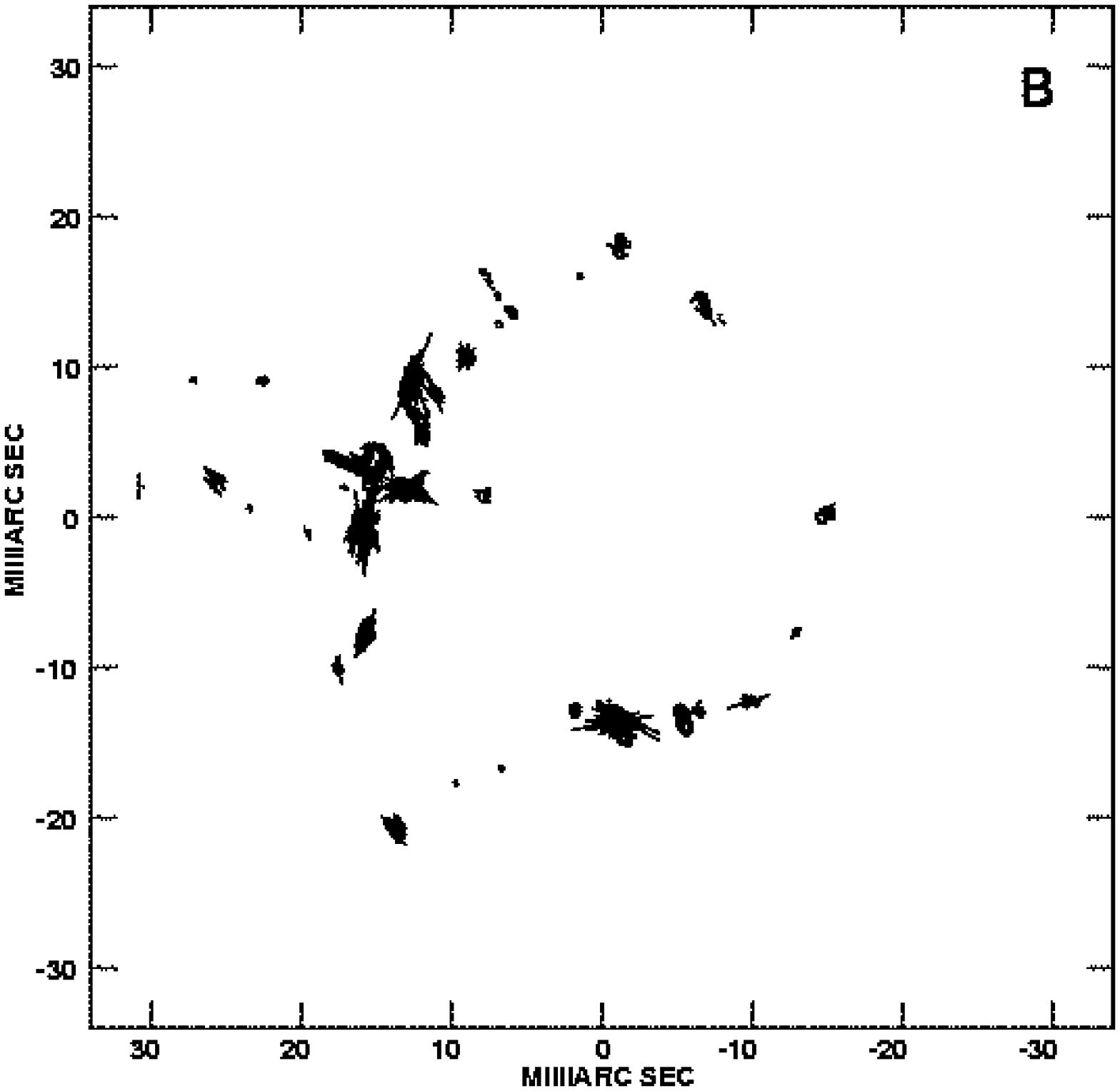} \\ 
\includegraphics[width=70mm, height=70mm]{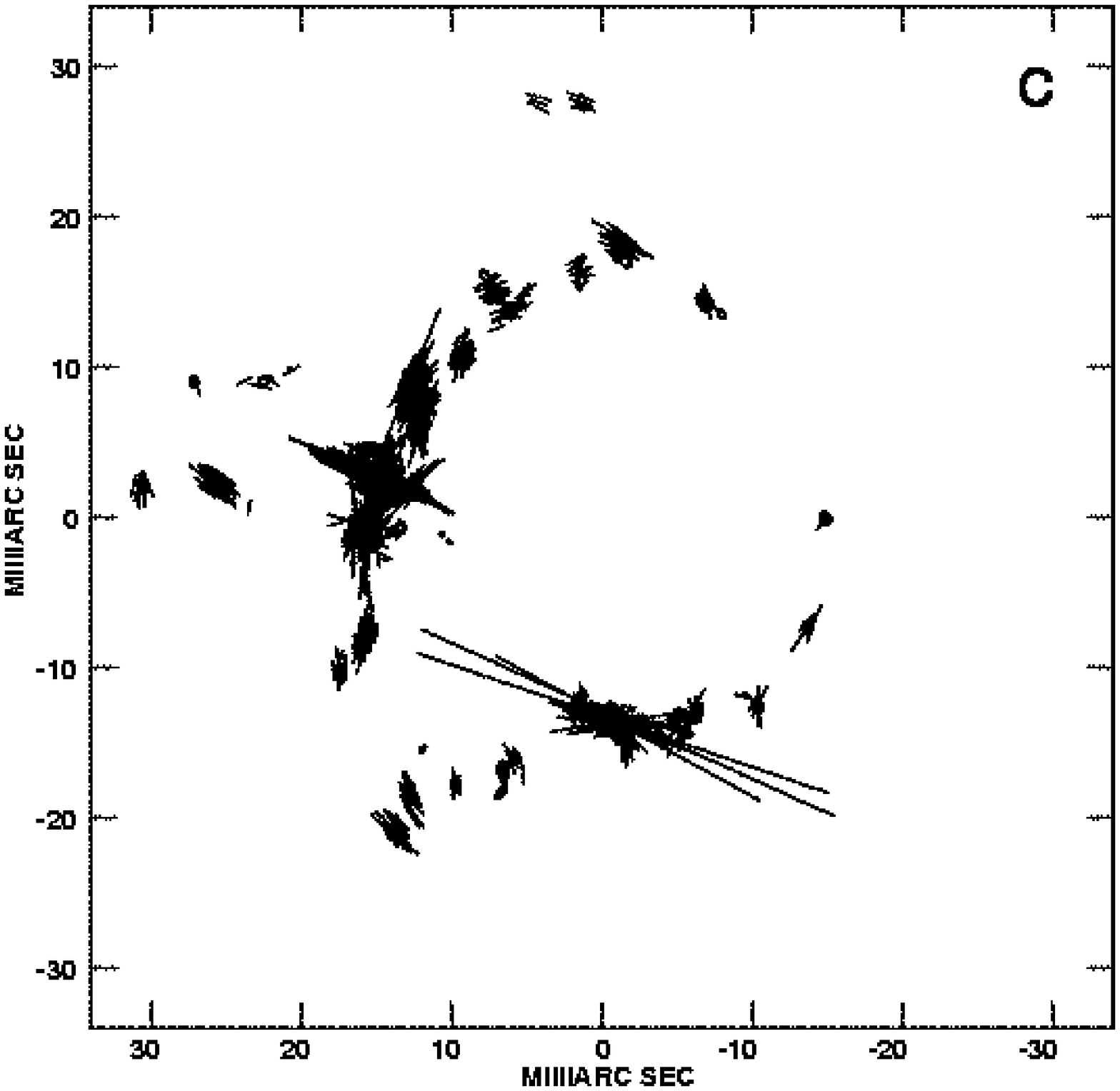} & 
\includegraphics[width=70mm, height=70mm]{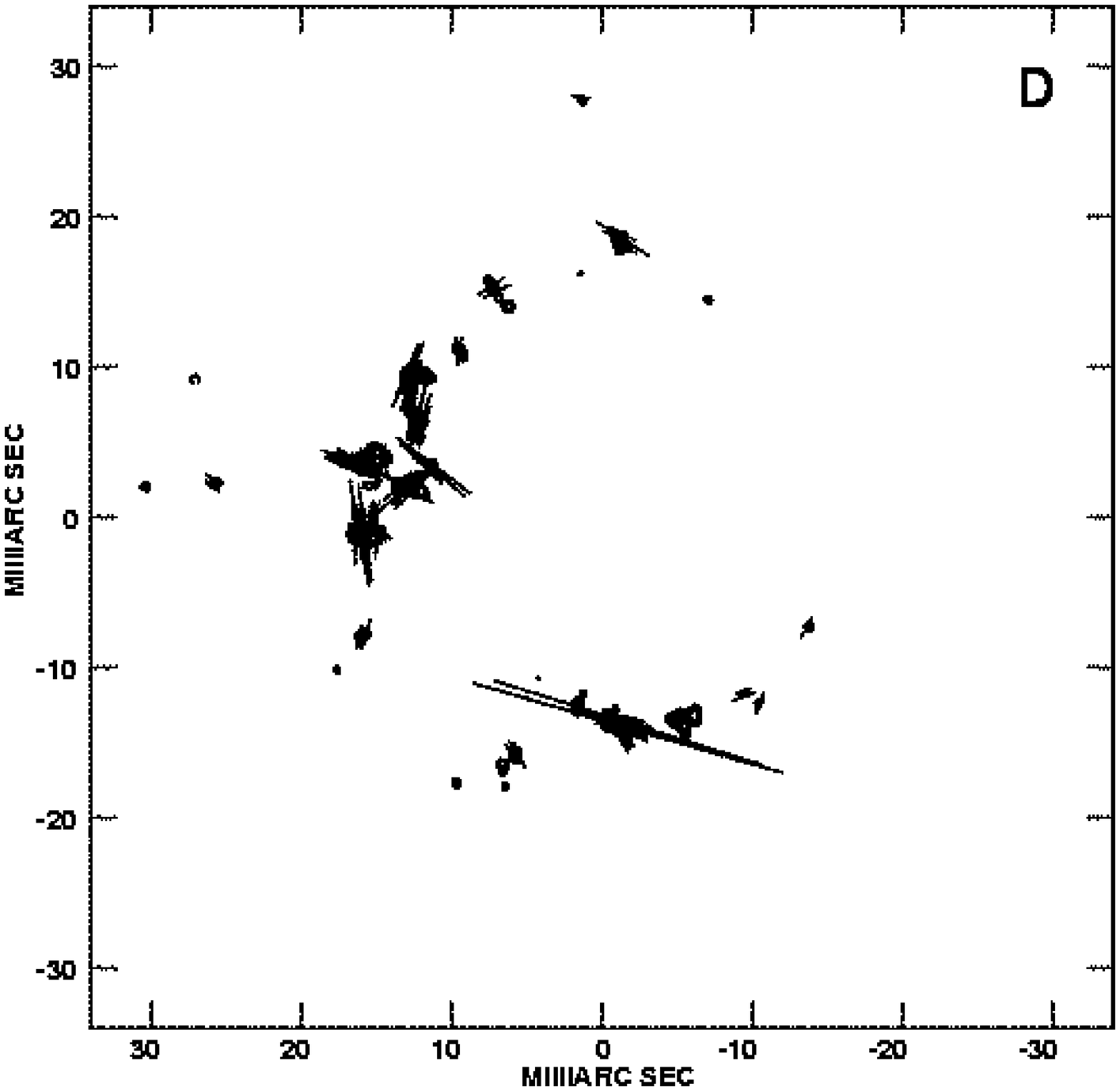} \\ 
\includegraphics[width=70mm, height=70mm]{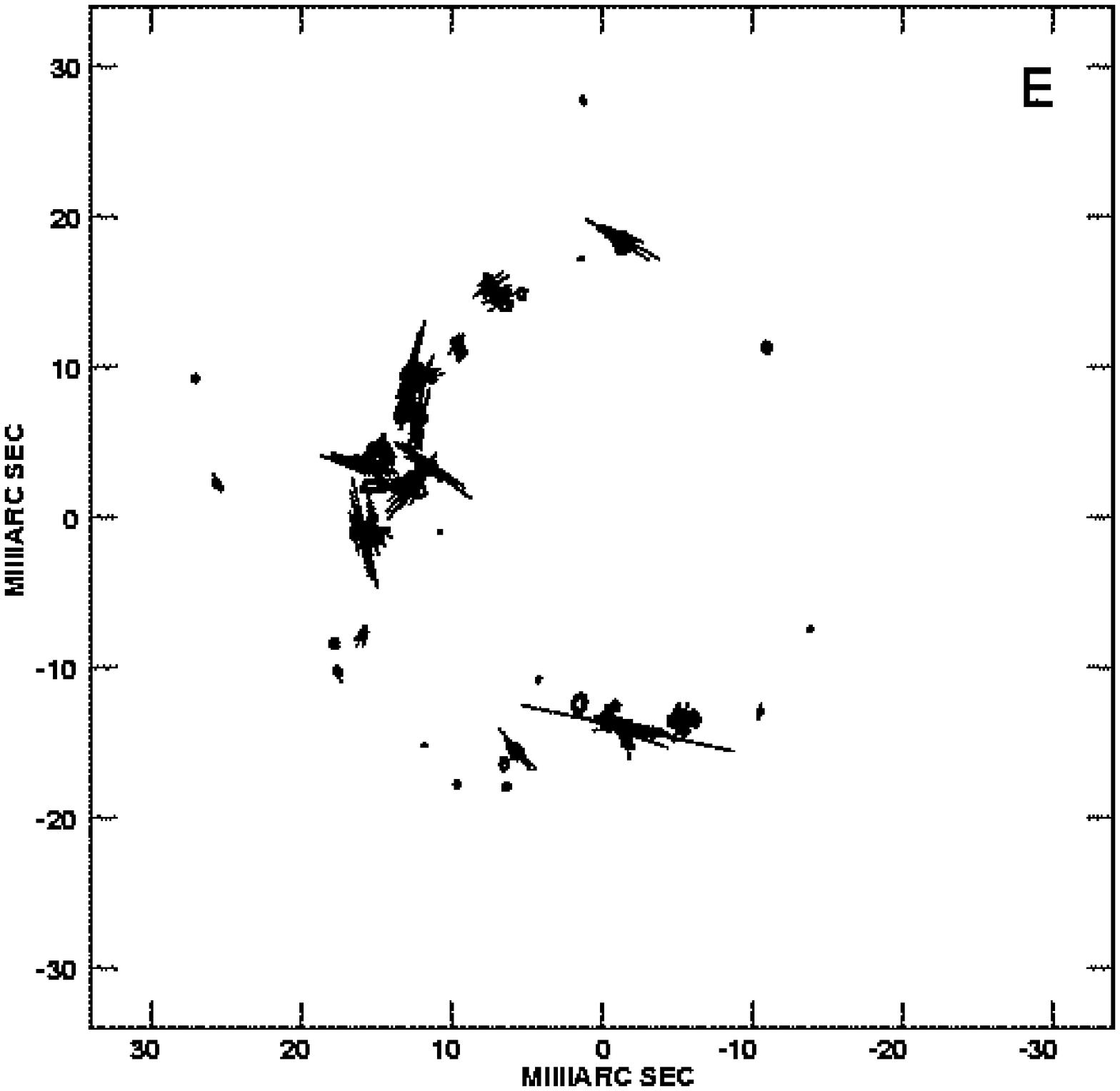} & 
\includegraphics[width=70mm, height=70mm]{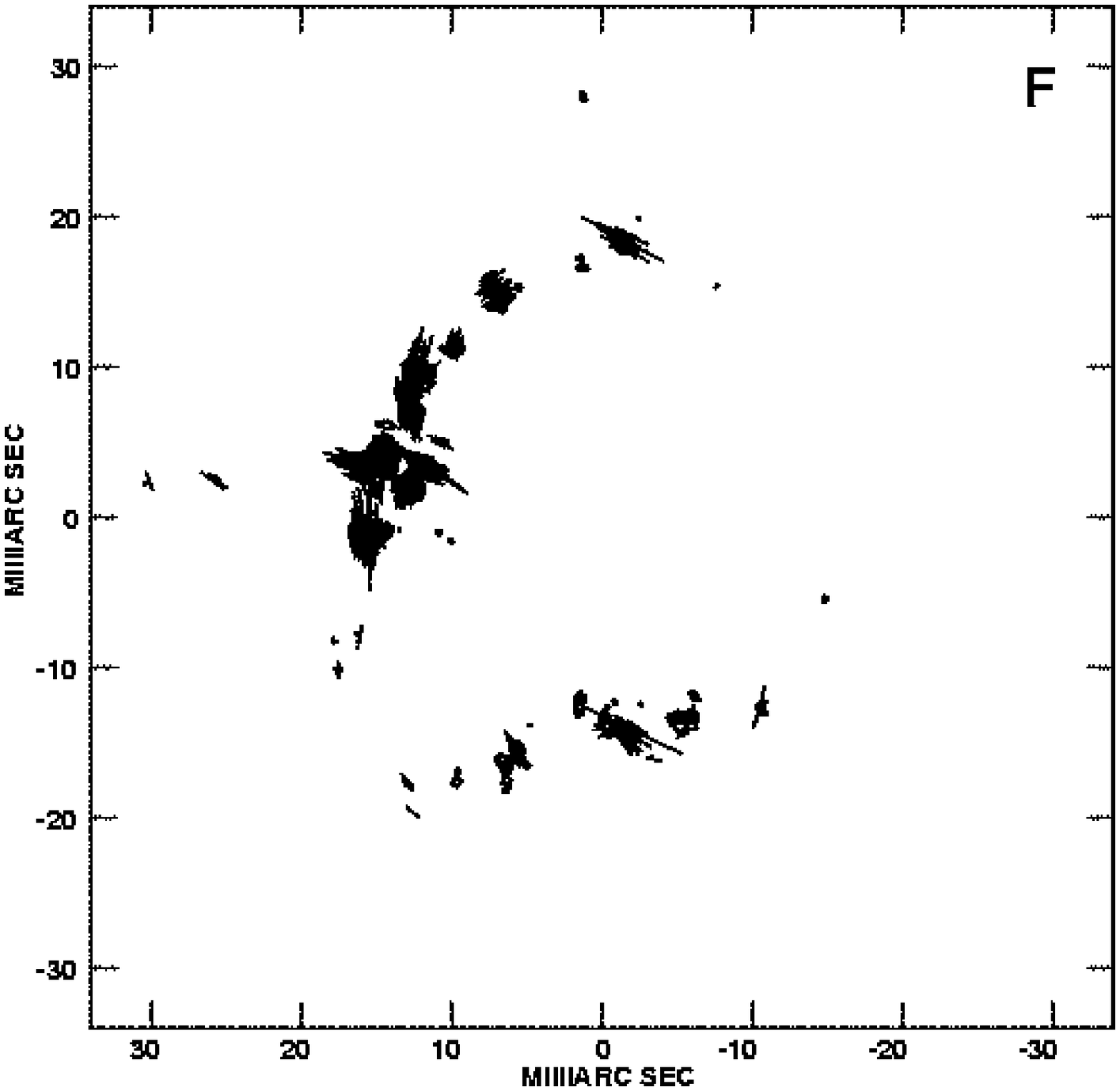} \\ 
\end{array}$ 

\caption{{\tiny For each epoch, a contour plot of the enhanced
contrast Stokes $I$ image at levels [1,2,5,10,20,40,80,100] \% of the
peak. Vectors are overlaid proportional to linearly-polarized
intensity (on a scale 10 mas = 18.08 Jy/beam) and drawn at a position
angle of the EVPA. All Stokes parameters $(I,Q,U)$ are summed over
velocity. The figures are $1360\times1360$ pixels in size with a pixel
spacing of 50$\mu$as. An fixed elliptical Gaussian restoring beam of $540
\times 420\mu$as at a position angle of 20 $\deg$ was used for all
images, representative of the point-spread function for the typical
$u-v$ coverage obtained during this observing campaign.}}

\label{fig-pcntr-3} 
\end{figure}

\begin{figure}[h] 
\advance\leftskip-1cm
\advance\rightskip-1cm
$\begin{array}{cc} 
\includegraphics[width=70mm, height=70mm]{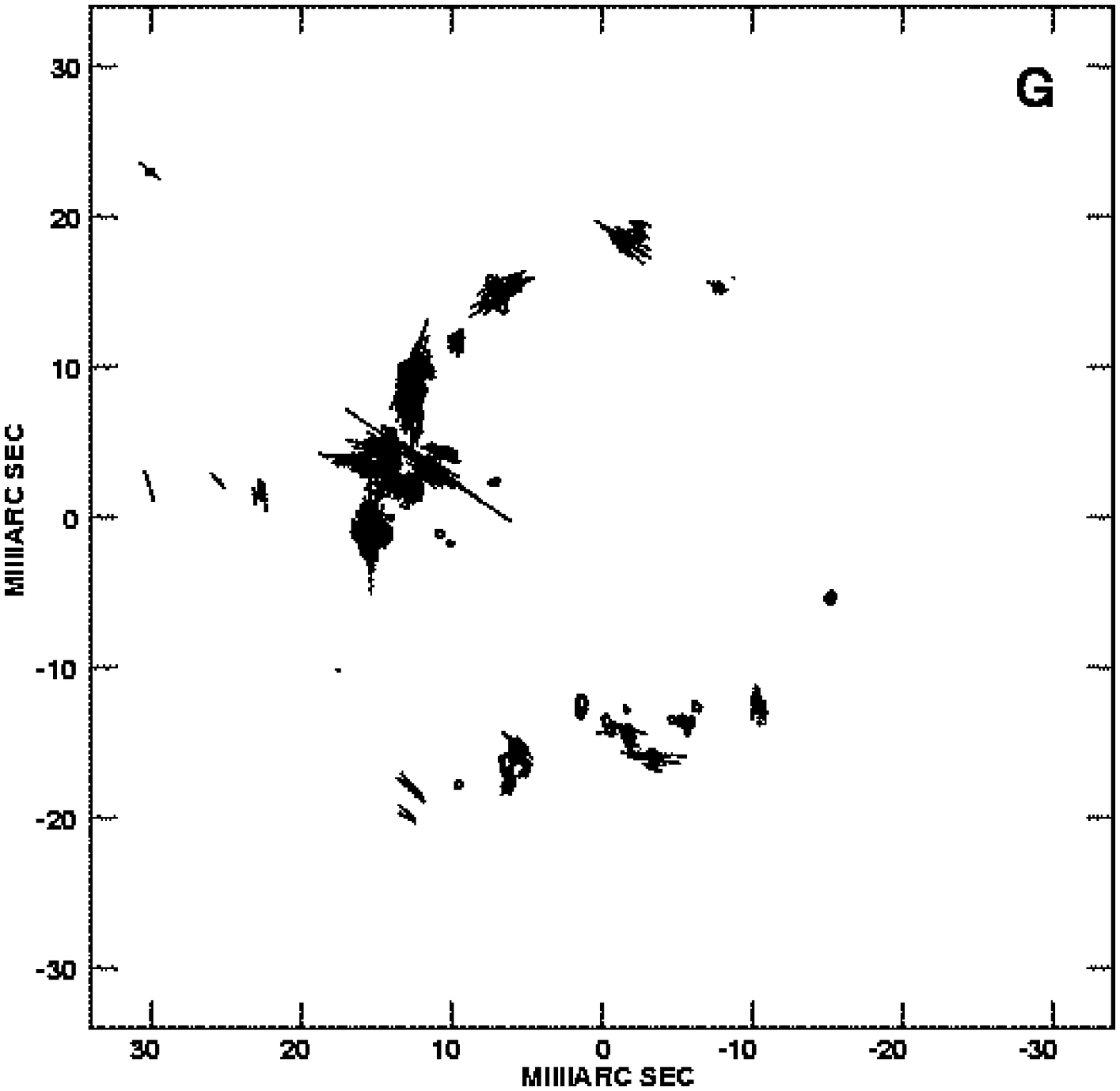} & 
\includegraphics[width=70mm, height=70mm]{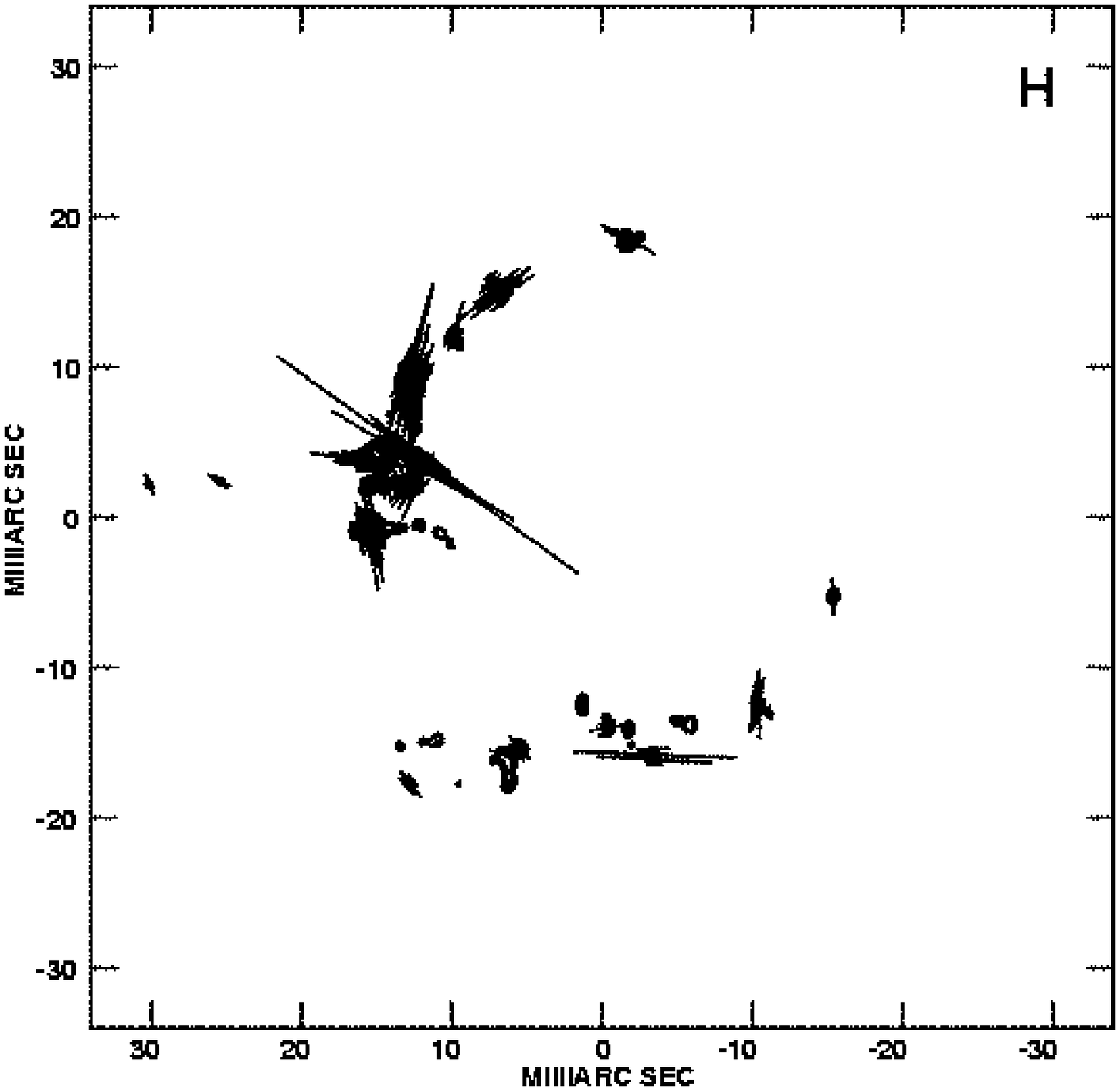} \\ 
\includegraphics[width=70mm, height=70mm]{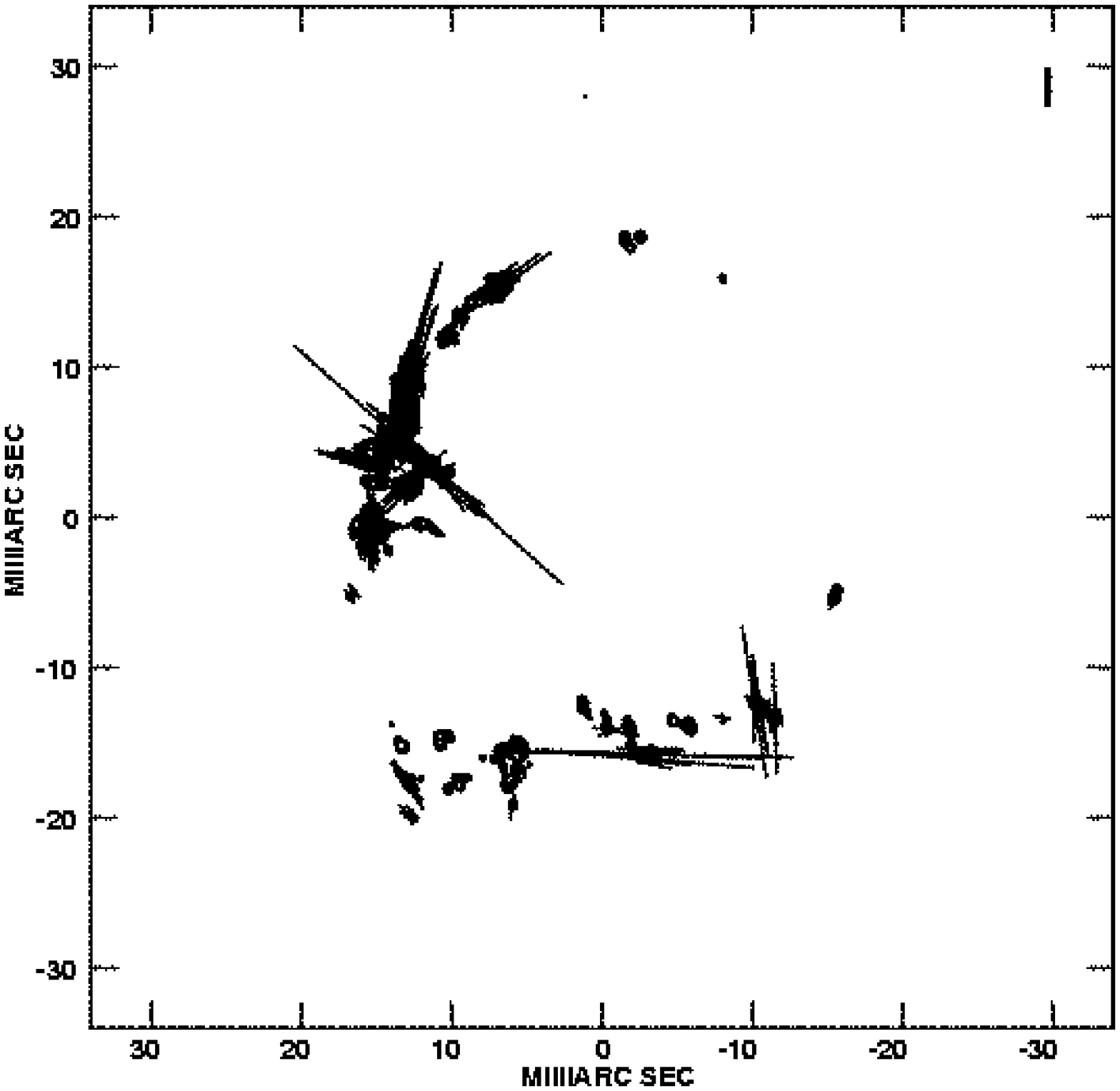} & 
\includegraphics[width=70mm, height=70mm]{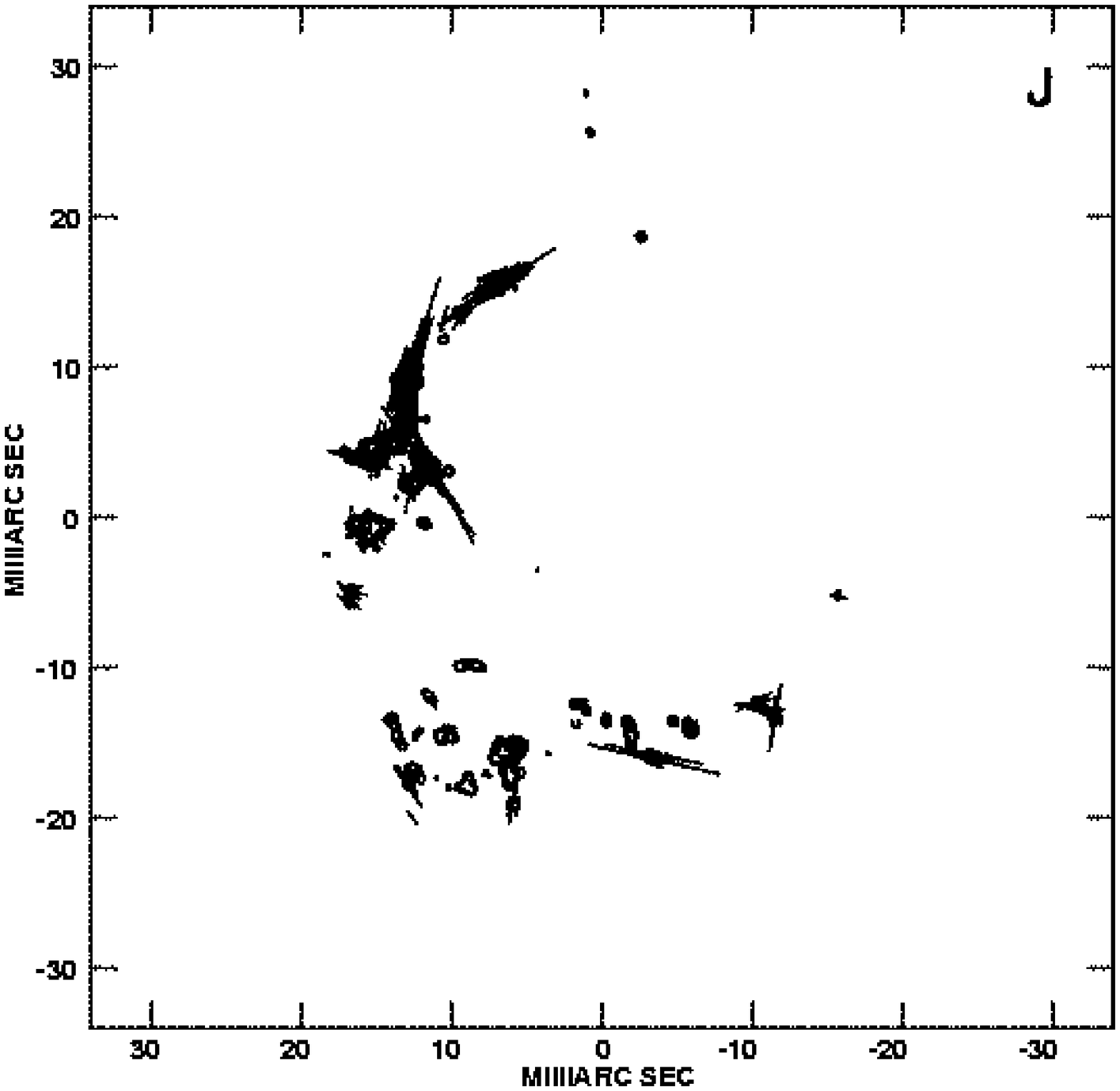} \\ 
\includegraphics[width=70mm, height=70mm]{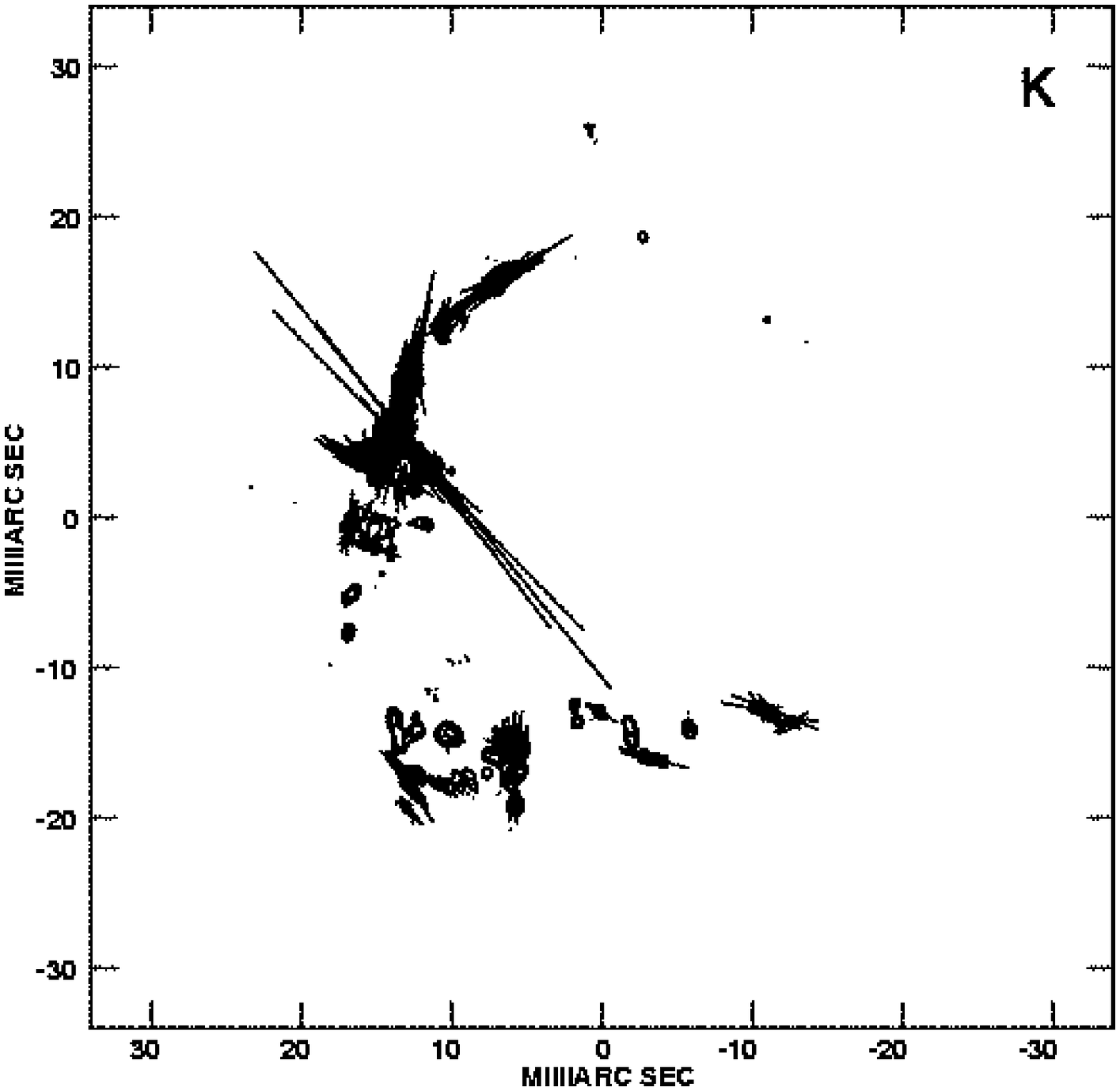} & 
\includegraphics[width=70mm, height=70mm]{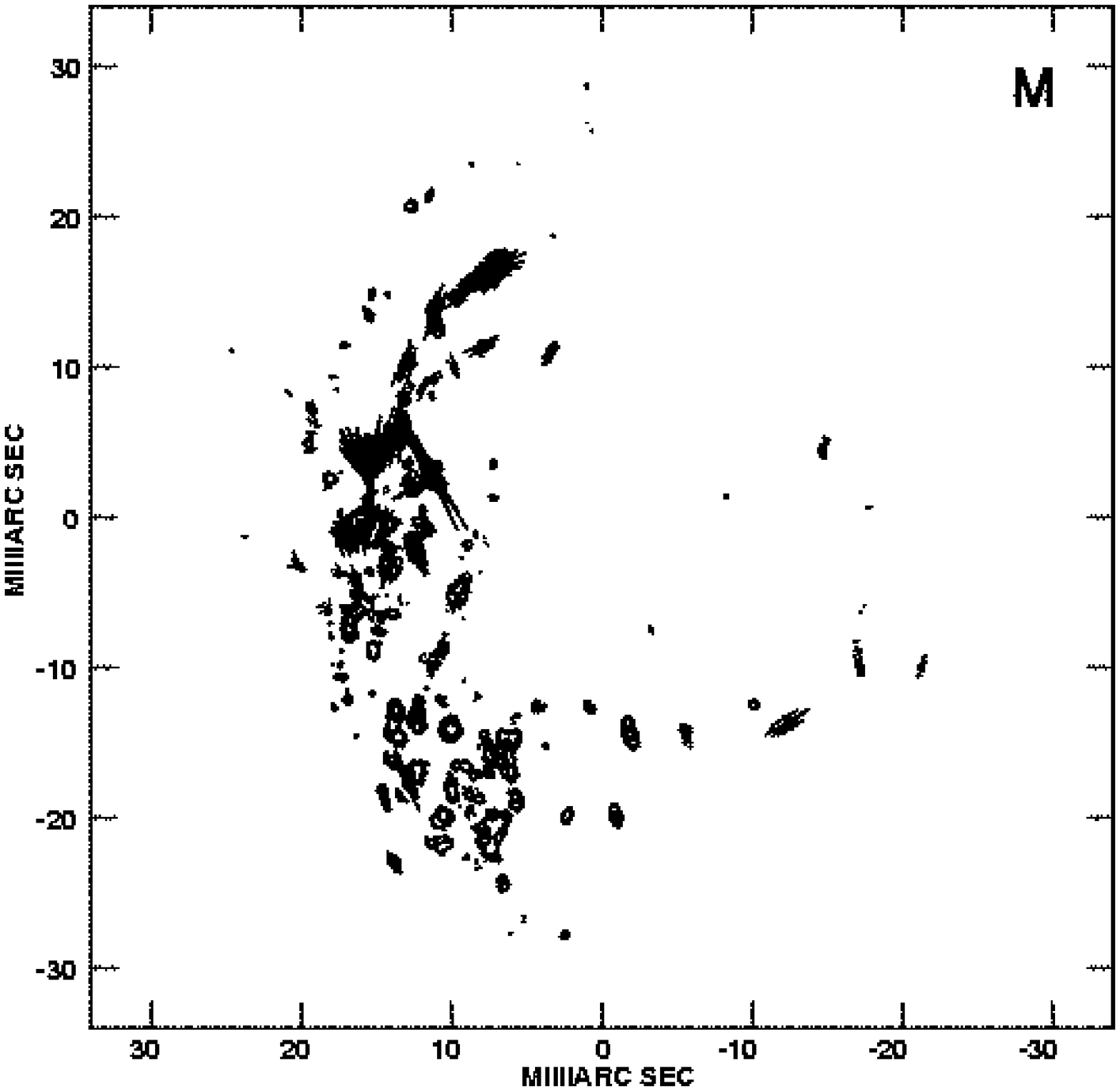} \\ 
\end{array}$ 

\caption{Same as Figure~\ref{fig-pcntr-3}.}

\label{fig-pcntr-4} 
\end{figure}

\begin{figure}[h] 
\advance\leftskip-1cm
\advance\rightskip-1cm
$\begin{array}{cc} 
\includegraphics[width=70mm, height=70mm]{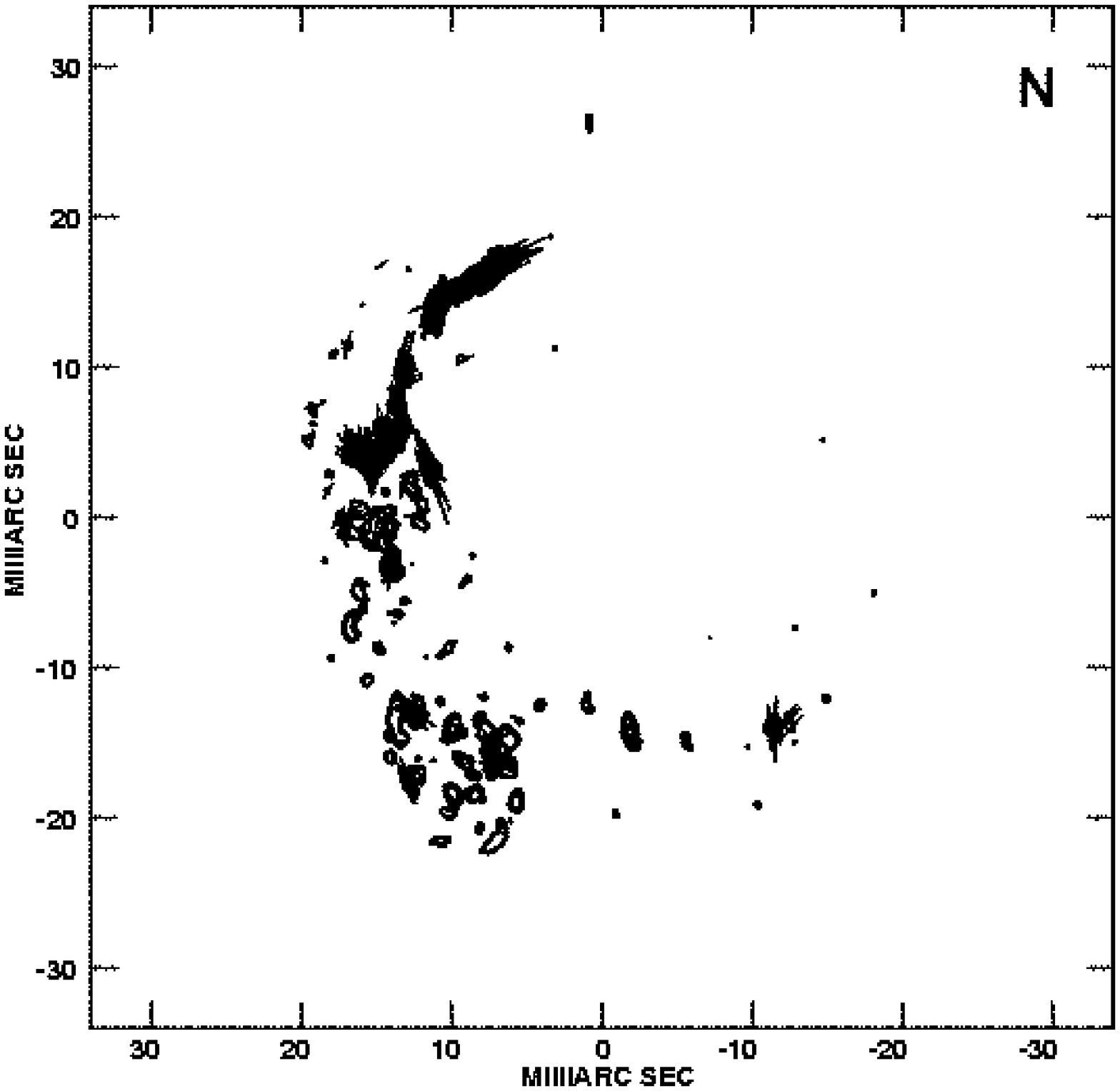} & 
\includegraphics[width=70mm, height=70mm]{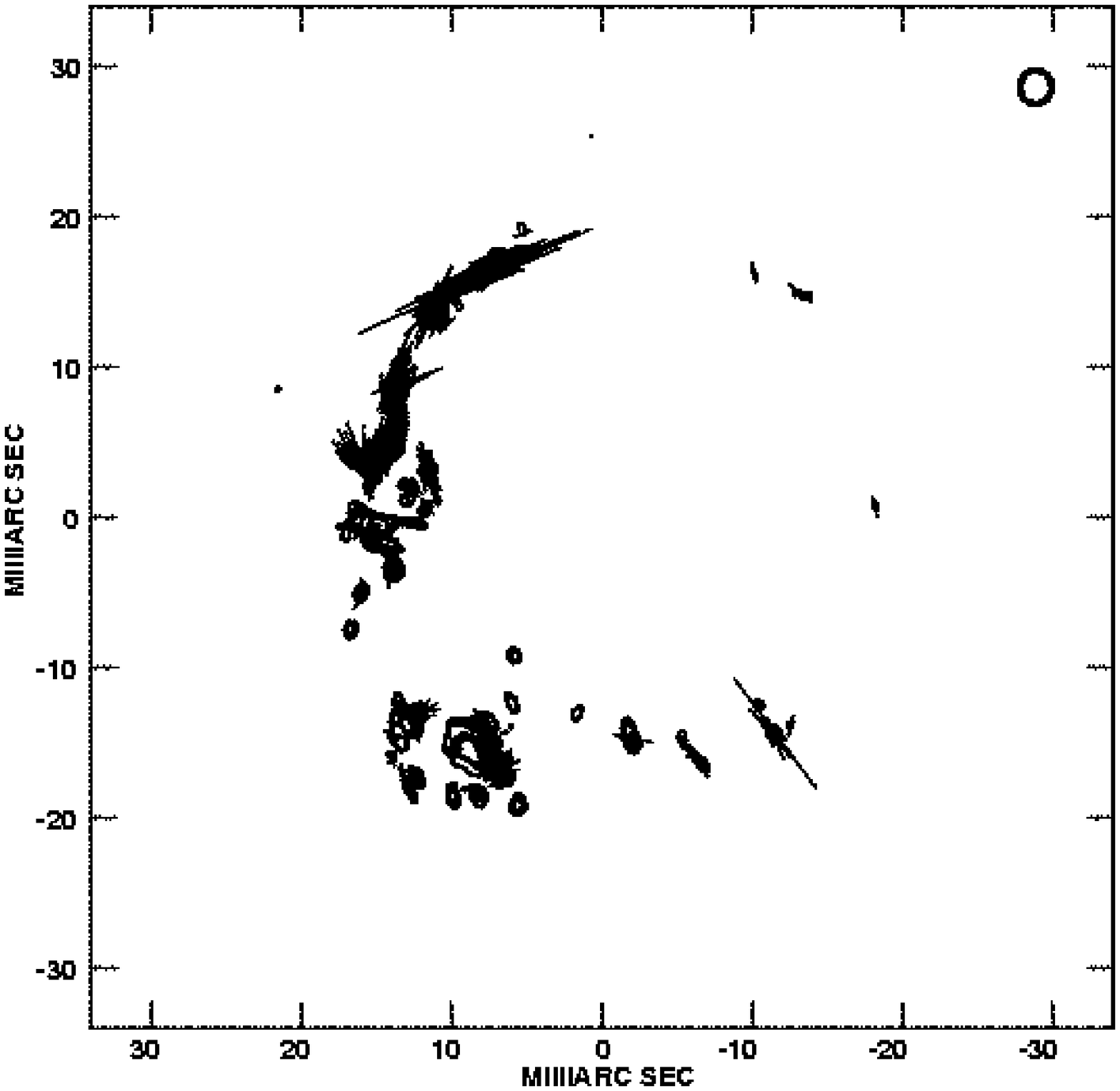} \\ 
\includegraphics[width=70mm, height=70mm]{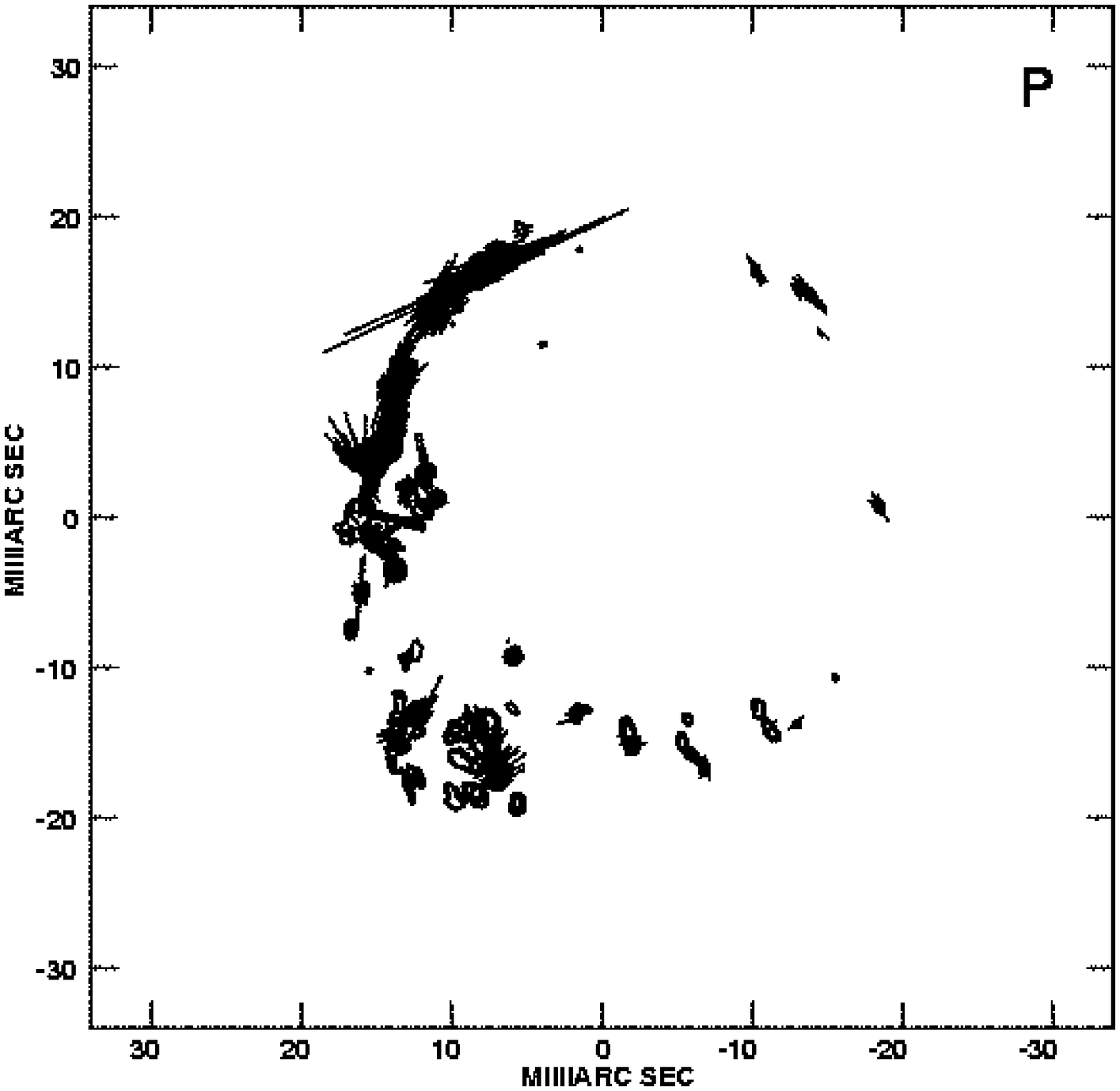} & 
\includegraphics[width=70mm, height=70mm]{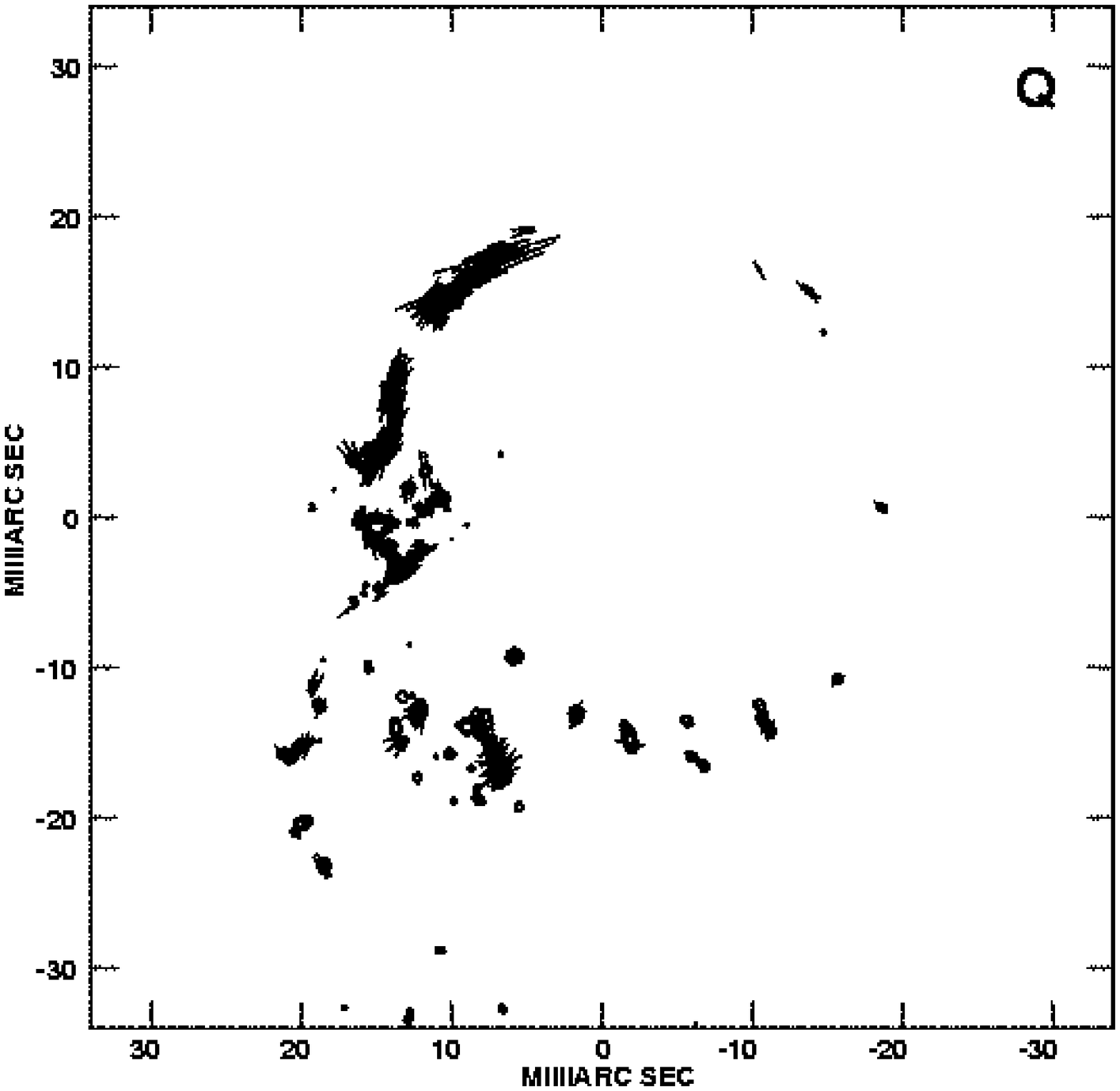} \\ 
\includegraphics[width=70mm, height=70mm]{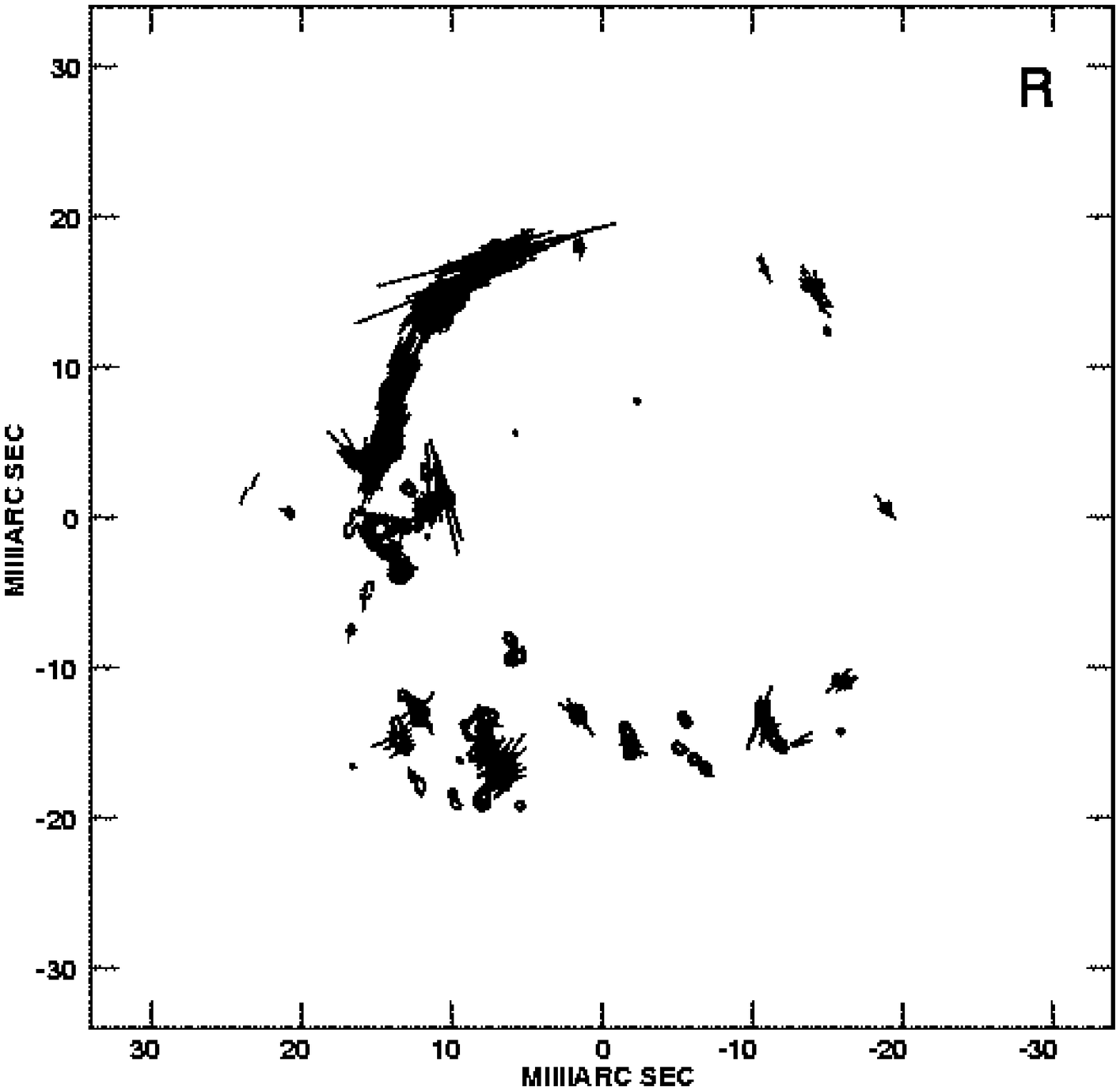} & 
\includegraphics[width=70mm, height=70mm]{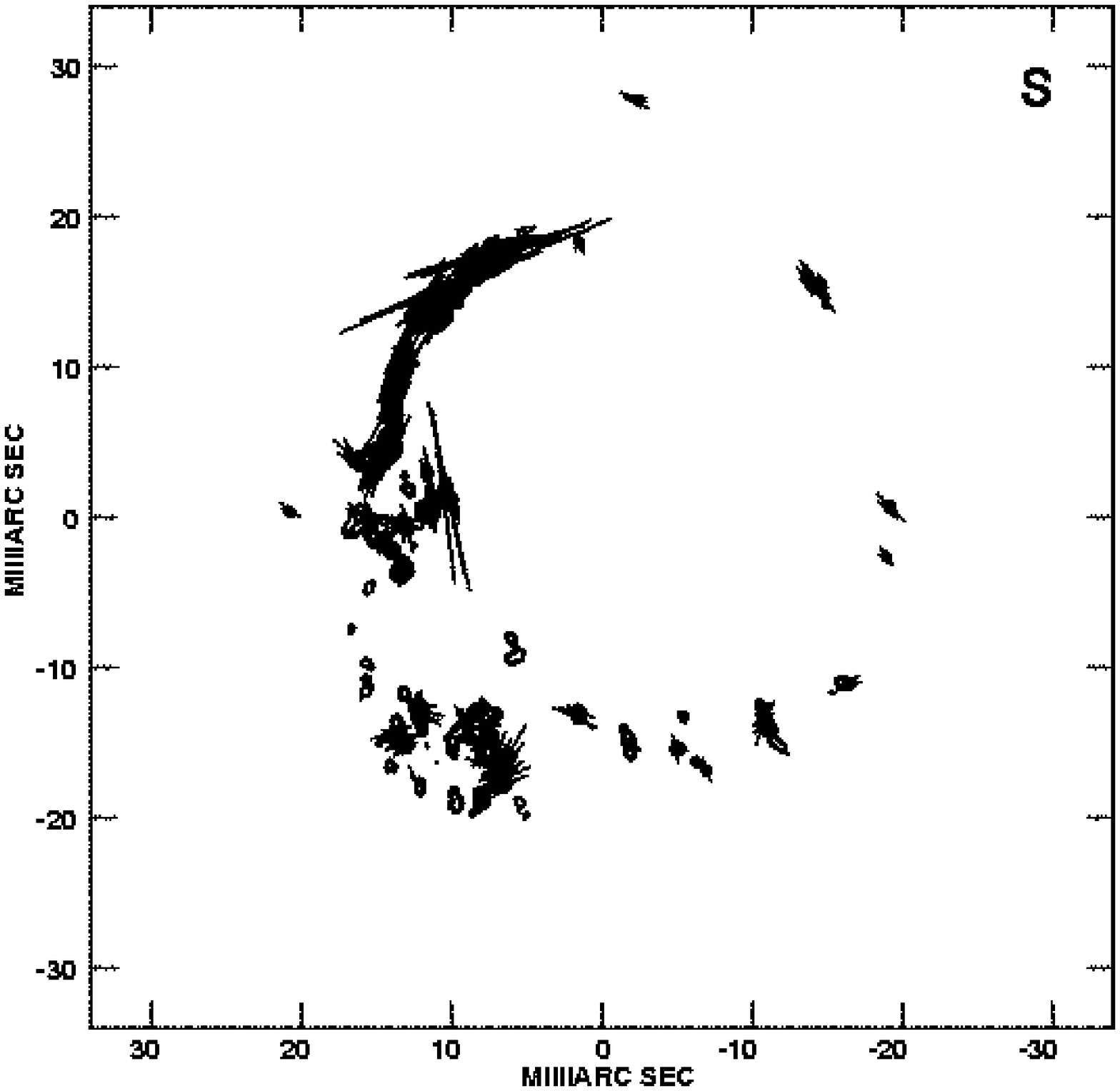} \\ 
\end{array}$ 

\caption{Same as Figure~\ref{fig-pcntr-3}.}

\label{fig-pcntr-5} 
\end{figure}

\begin{figure}[h] 
\advance\leftskip-1cm
\advance\rightskip-1cm
$\begin{array}{cc} 
\includegraphics[width=70mm, height=70mm]{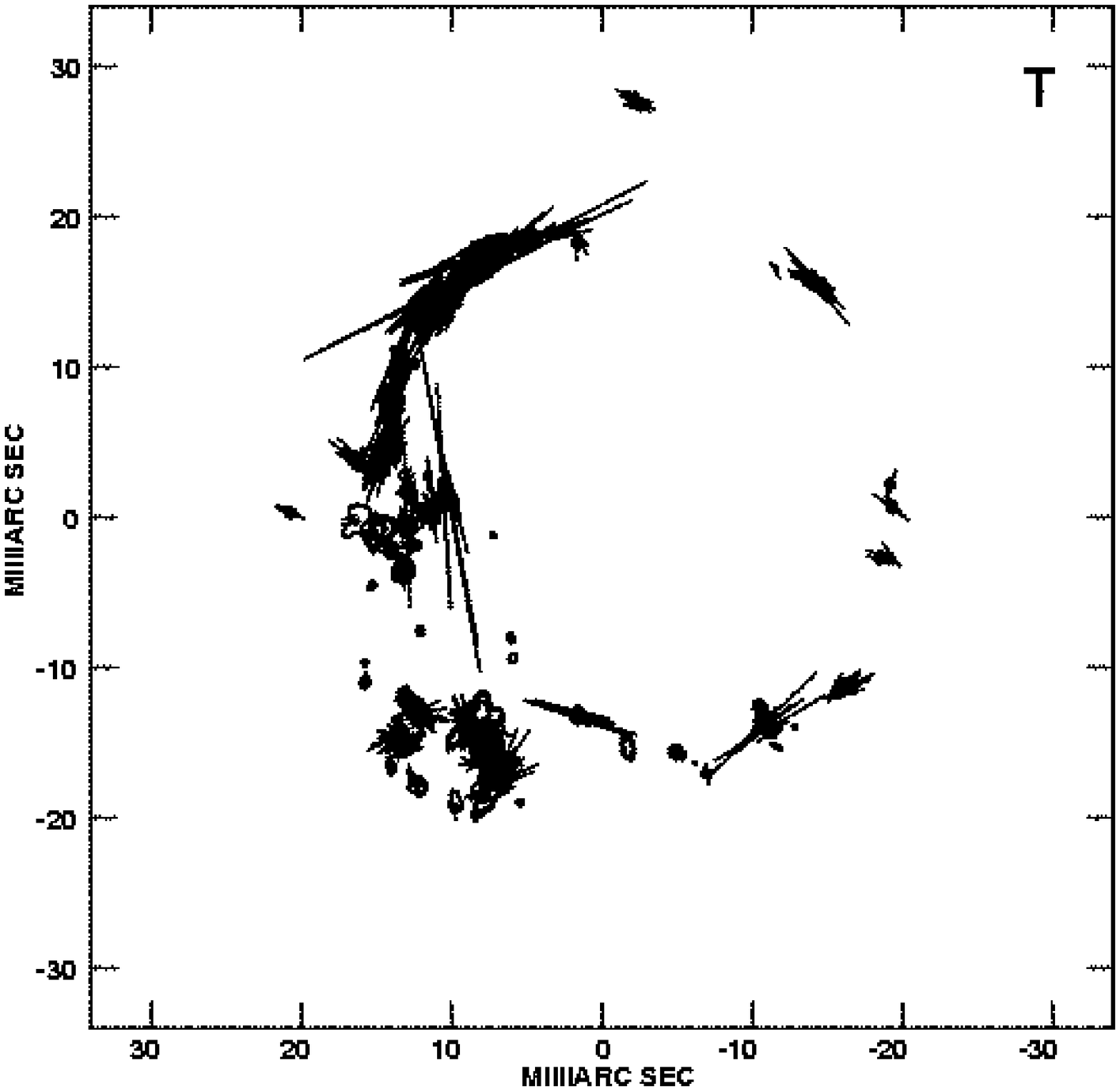} & 
\includegraphics[width=70mm, height=70mm]{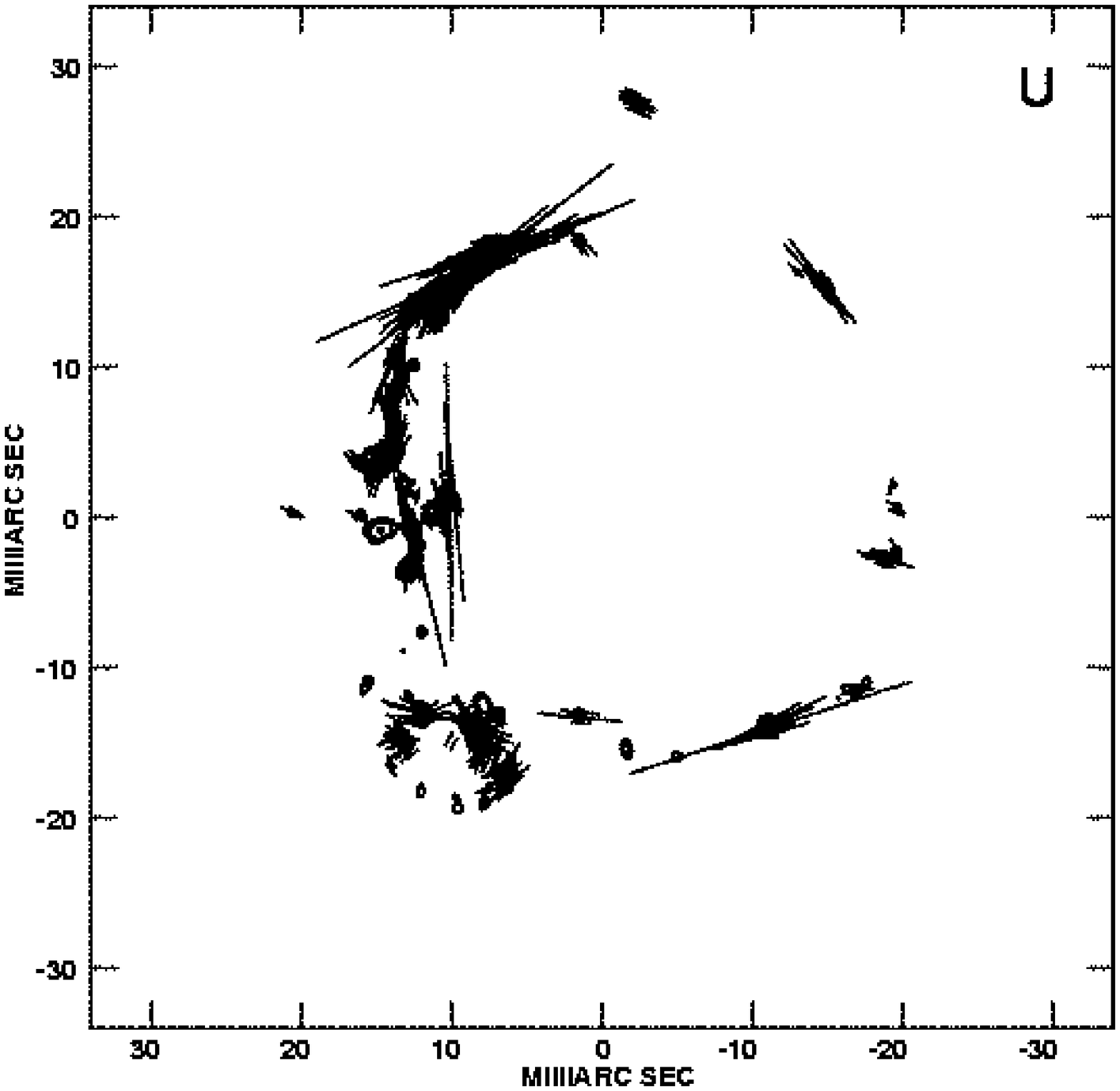} \\ 
\includegraphics[width=70mm, height=70mm]{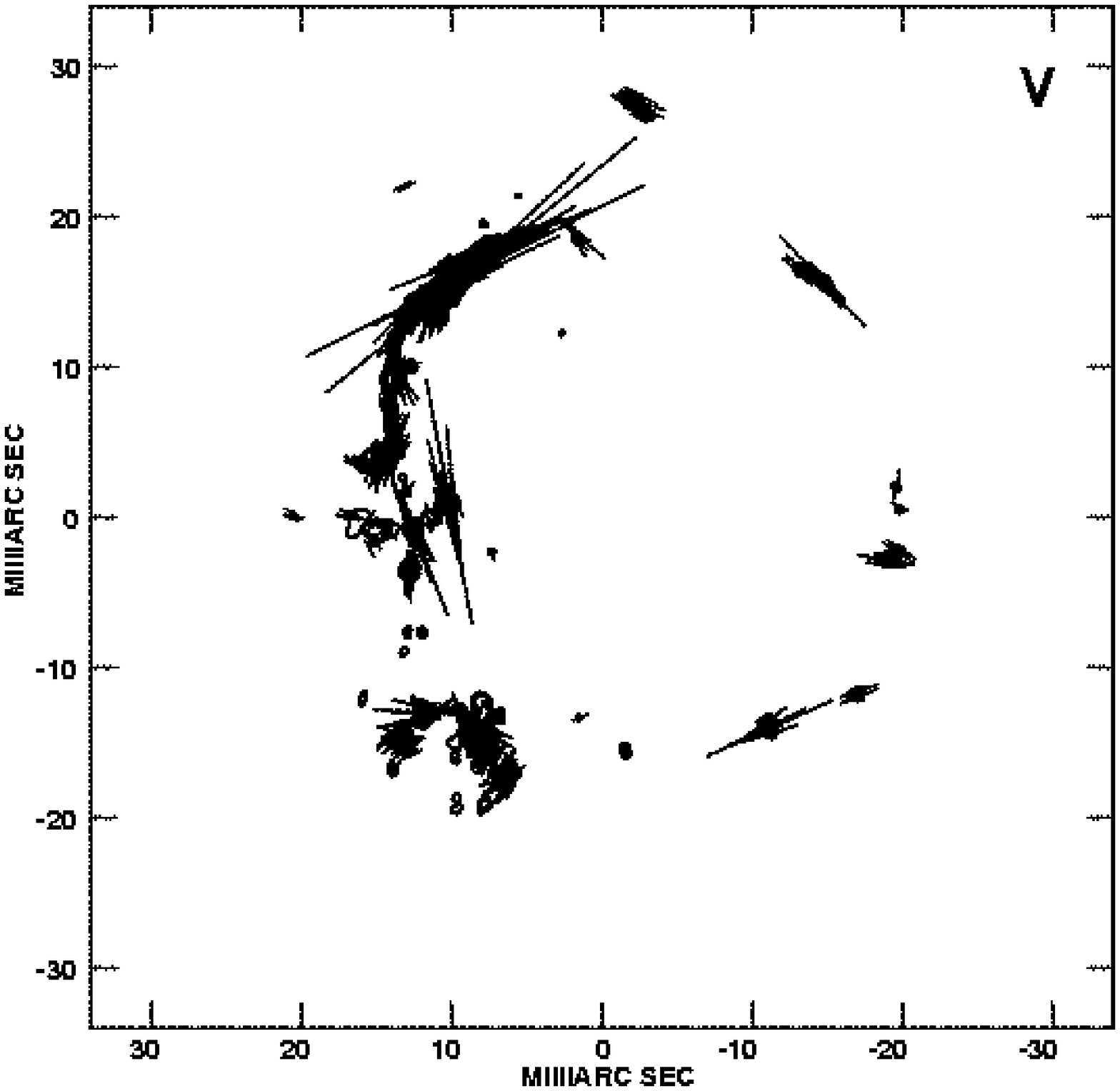} & 
\includegraphics[width=70mm, height=70mm]{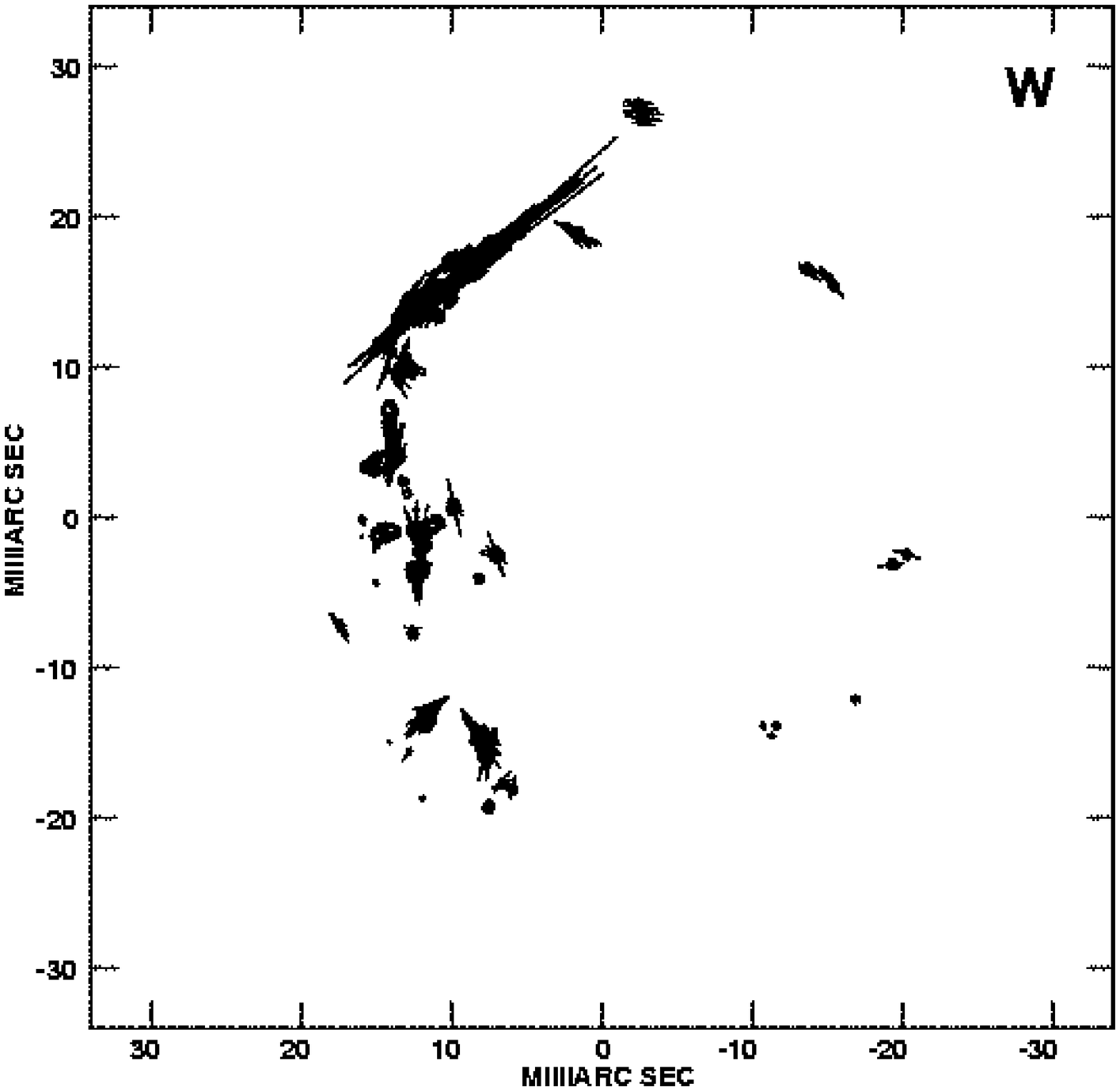} \\ 
\includegraphics[width=70mm, height=70mm]{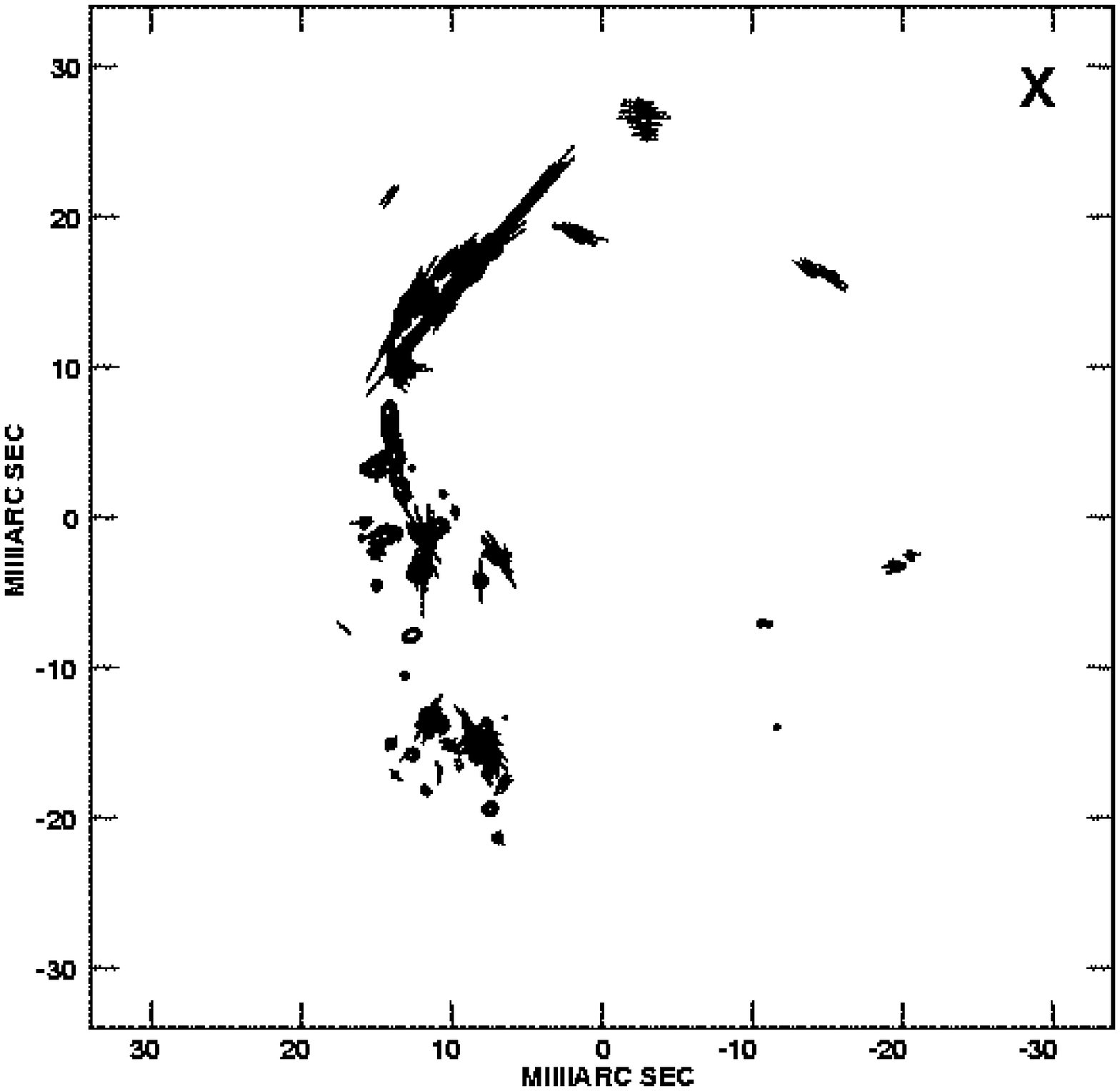} & 
\includegraphics[width=70mm, height=70mm]{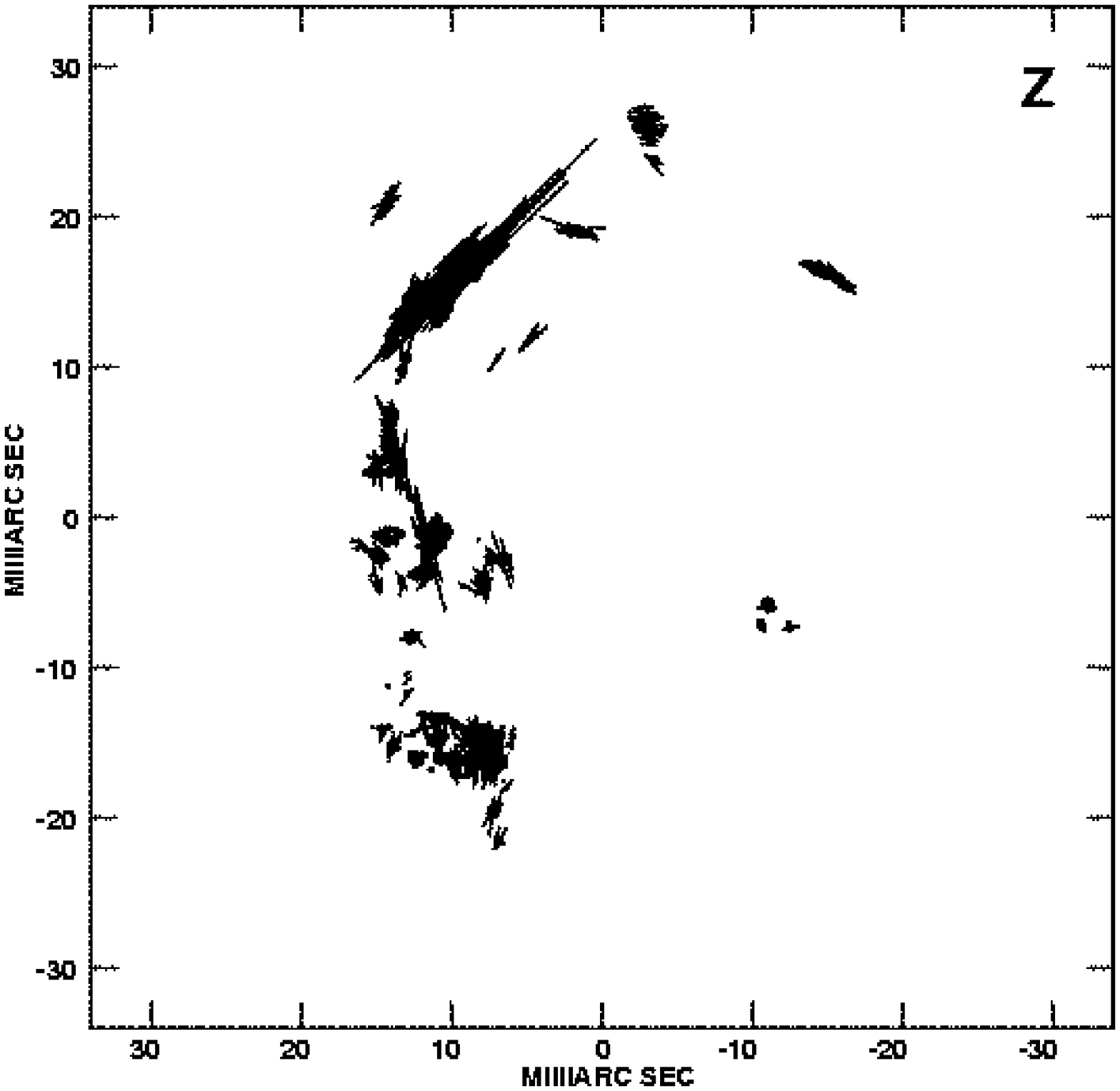} \\ 
\end{array}$ 

\caption{Same as Figure~\ref{fig-pcntr-3}.}

\label{fig-pcntr-6} 
\end{figure}

\begin{figure}[h] 
\advance\leftskip-1cm
\advance\rightskip-1cm
$\begin{array}{cc} 
\includegraphics[width=70mm, height=70mm]{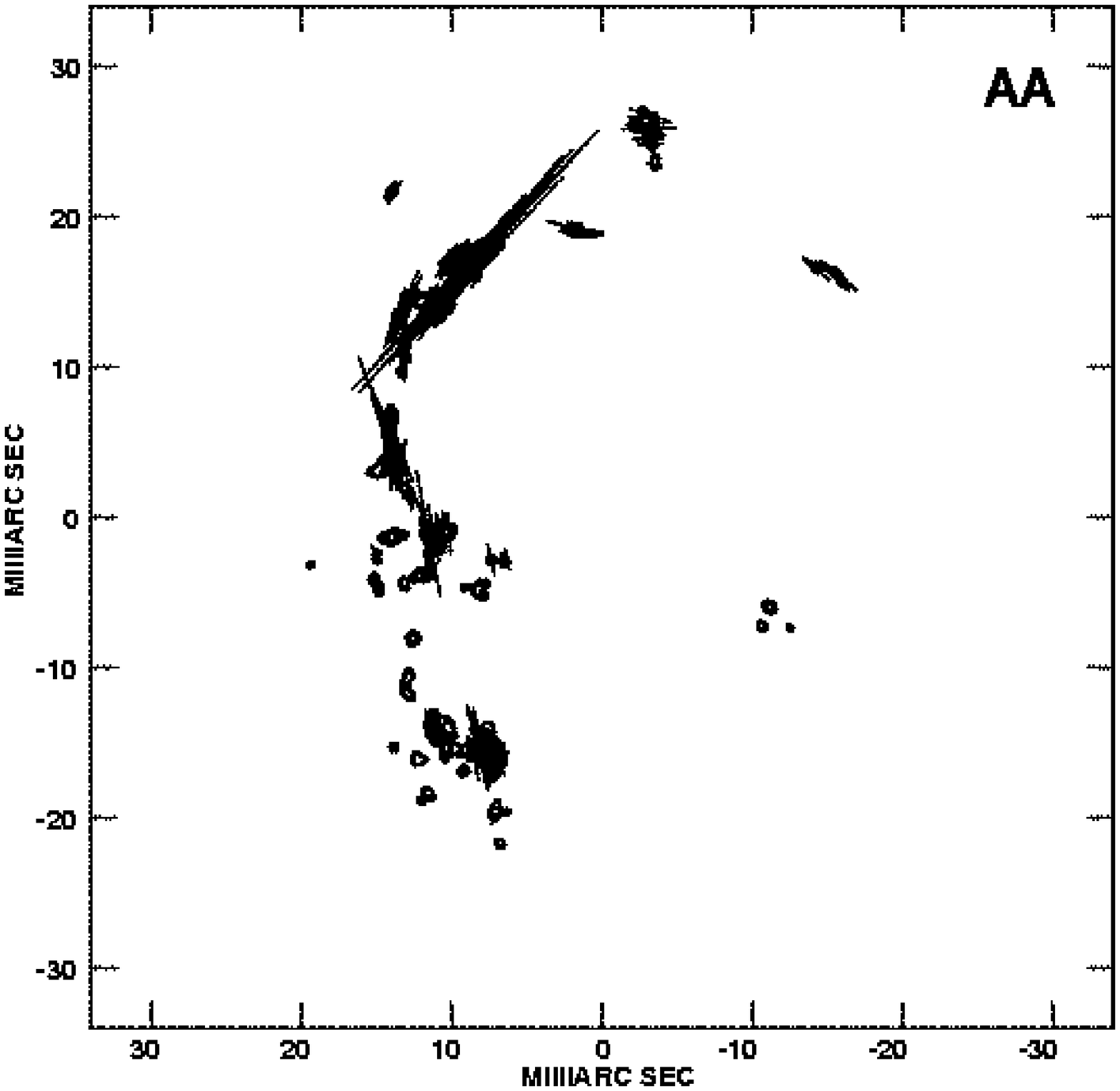} & 
\includegraphics[width=70mm, height=70mm]{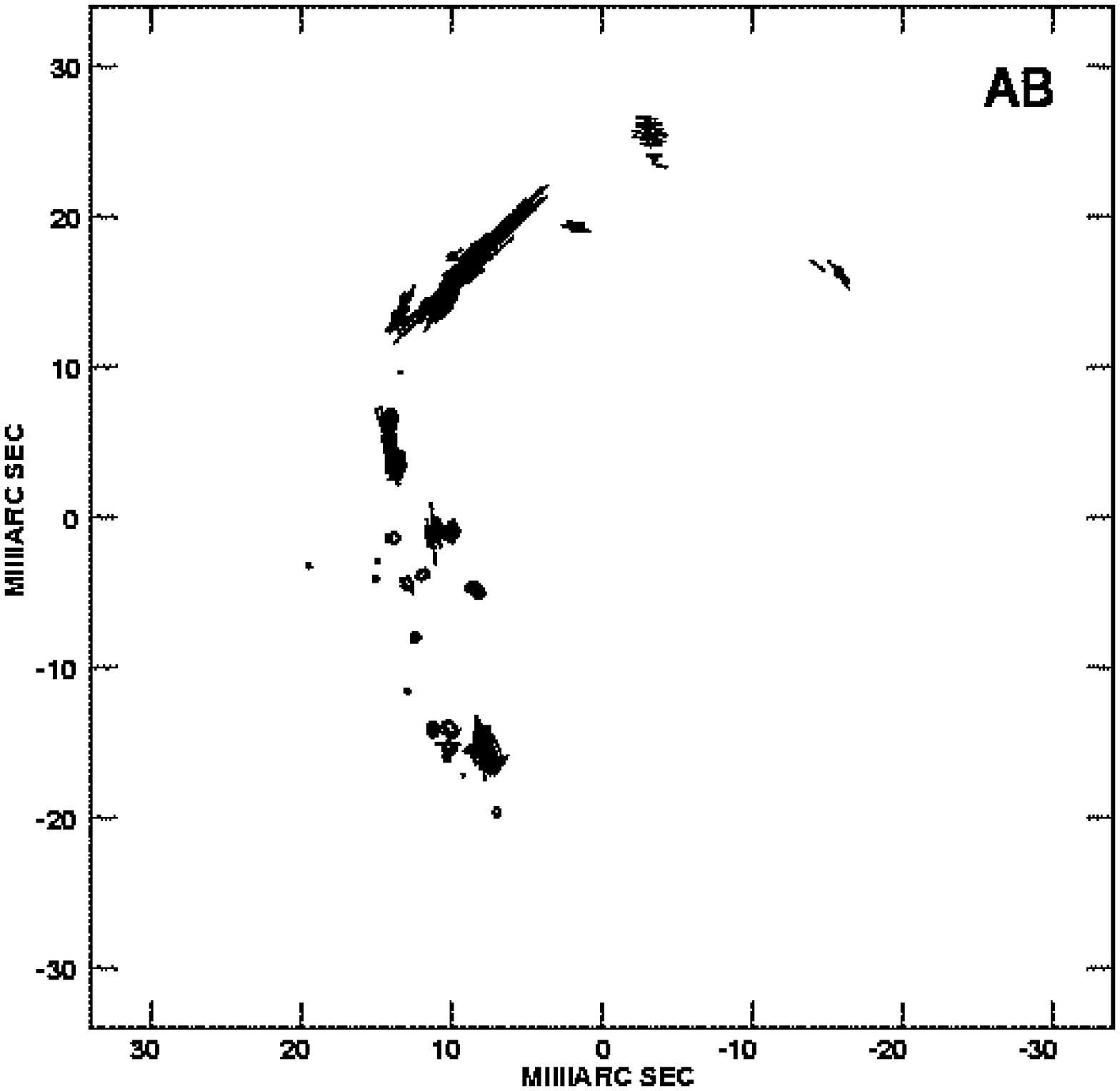} \\ 
\includegraphics[width=70mm, height=70mm]{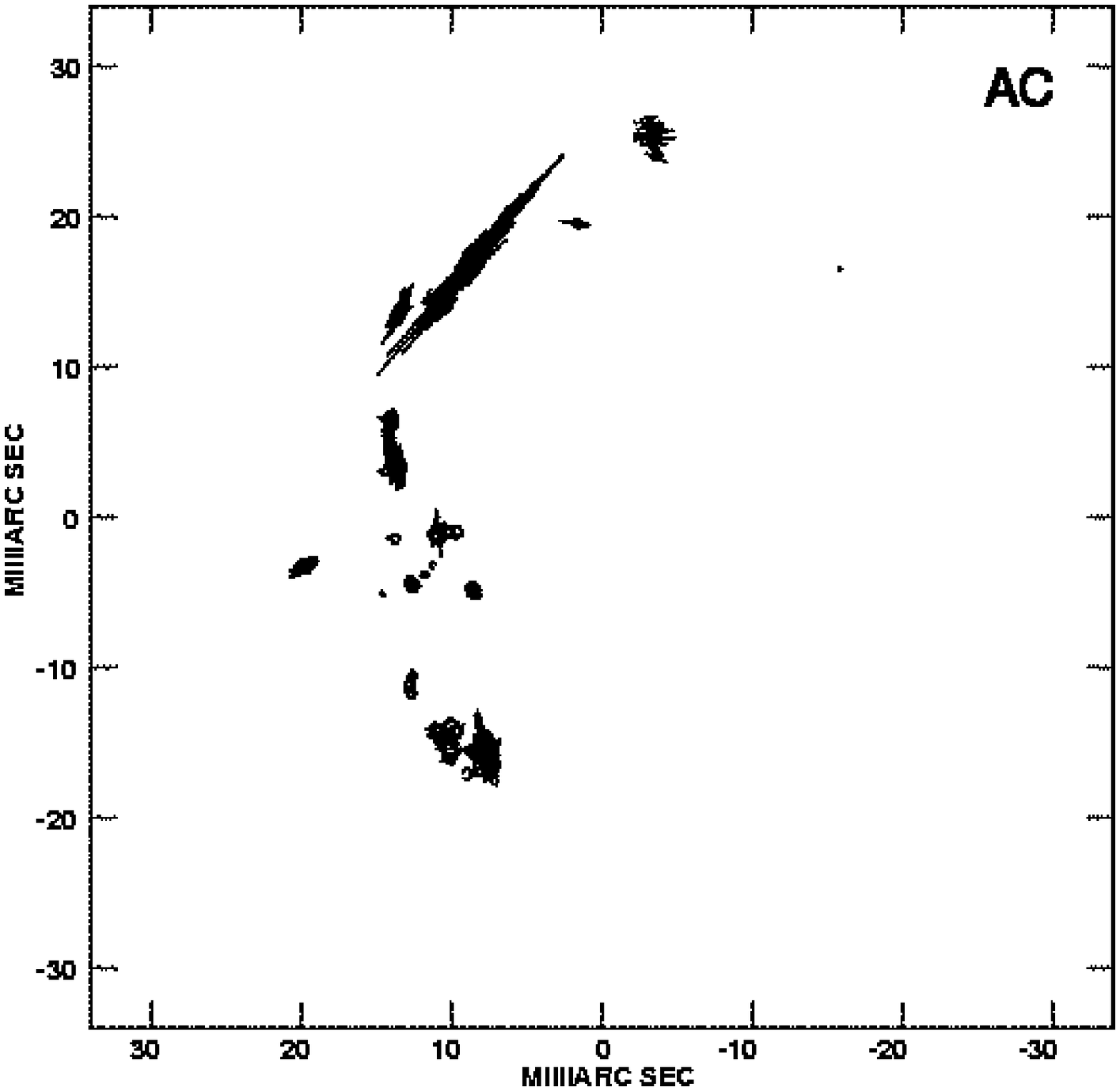} & 
\includegraphics[width=70mm, height=70mm]{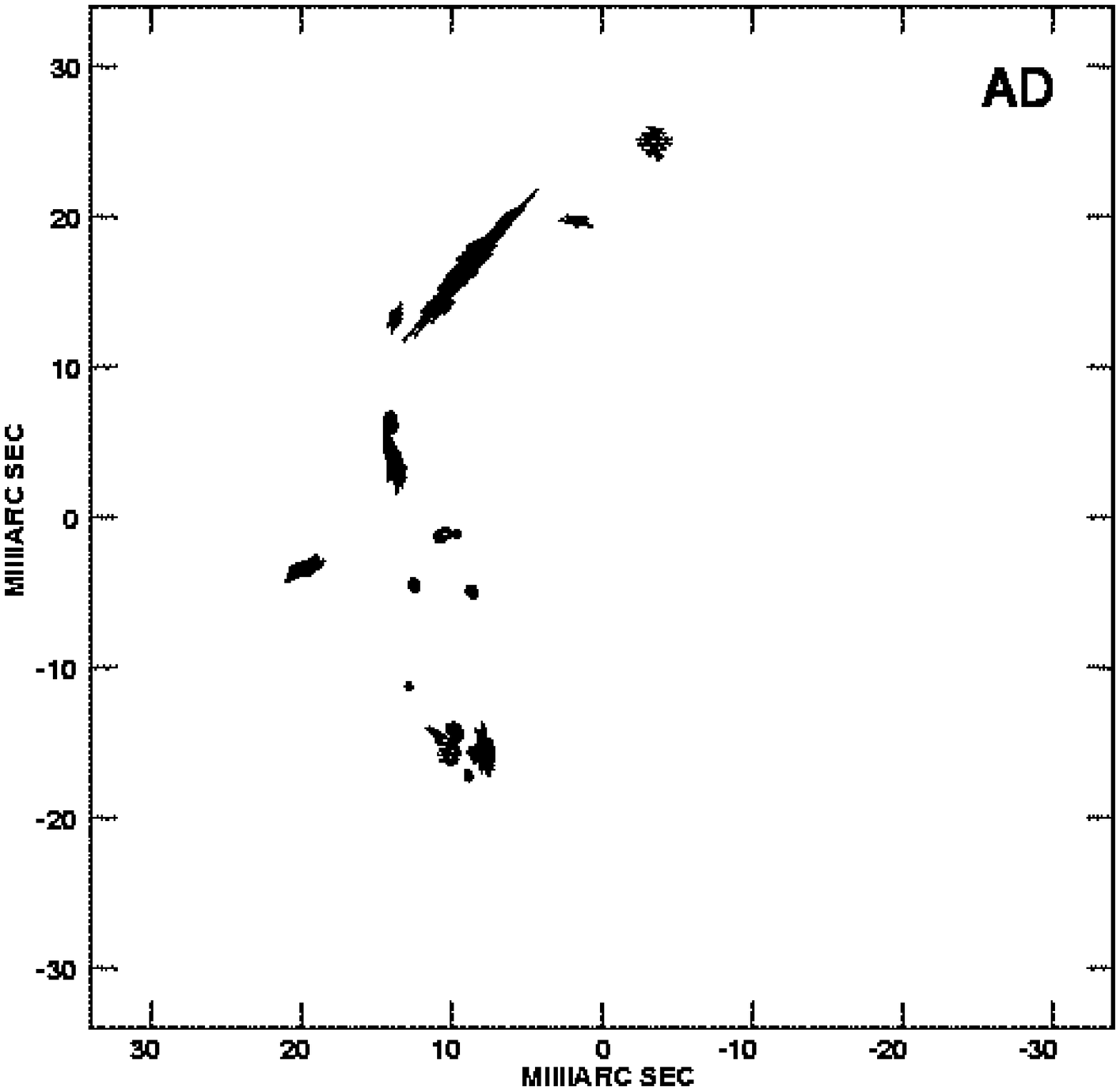} \\ 
\includegraphics[width=70mm, height=70mm]{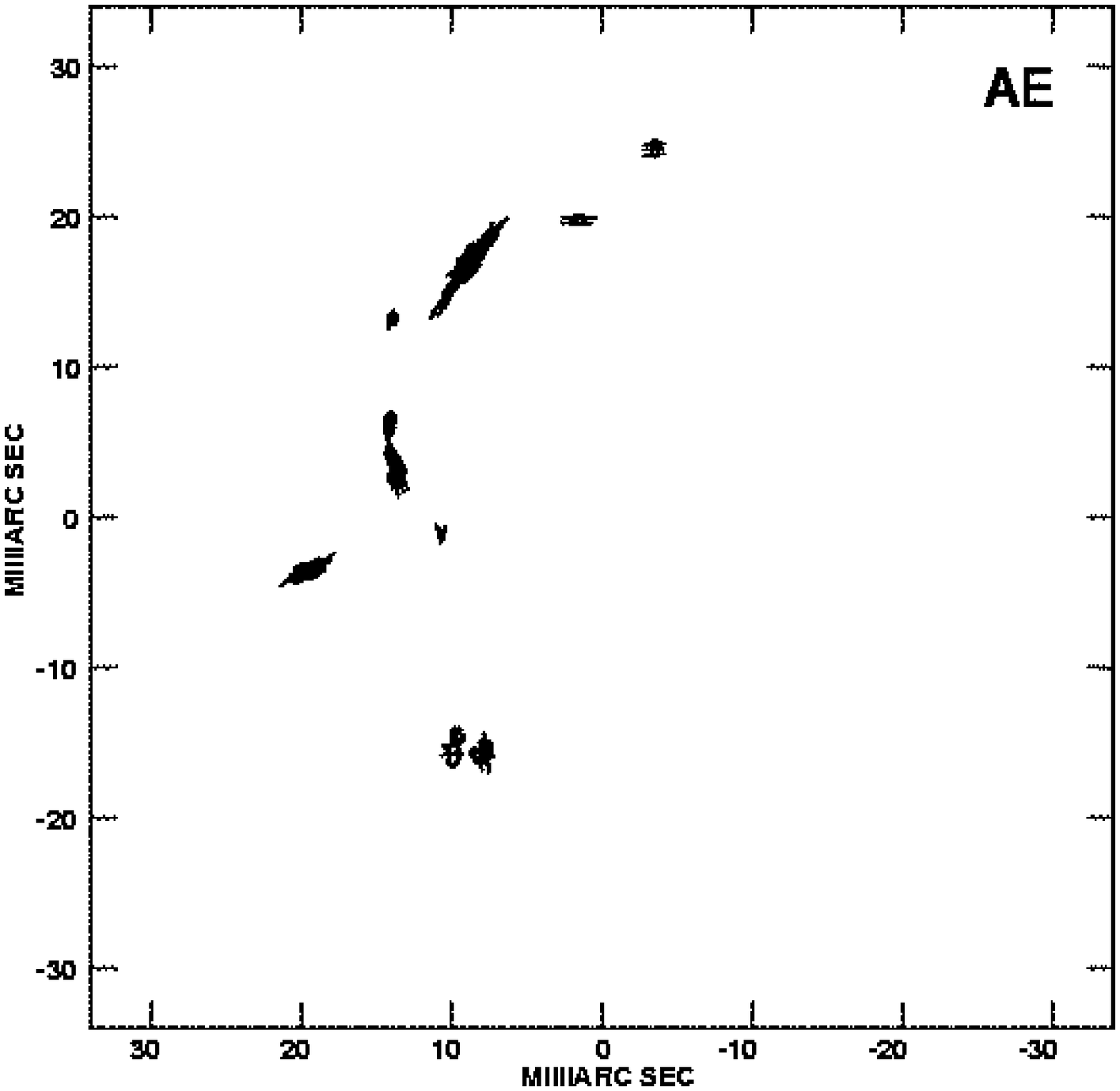} & 
\includegraphics[width=70mm, height=70mm]{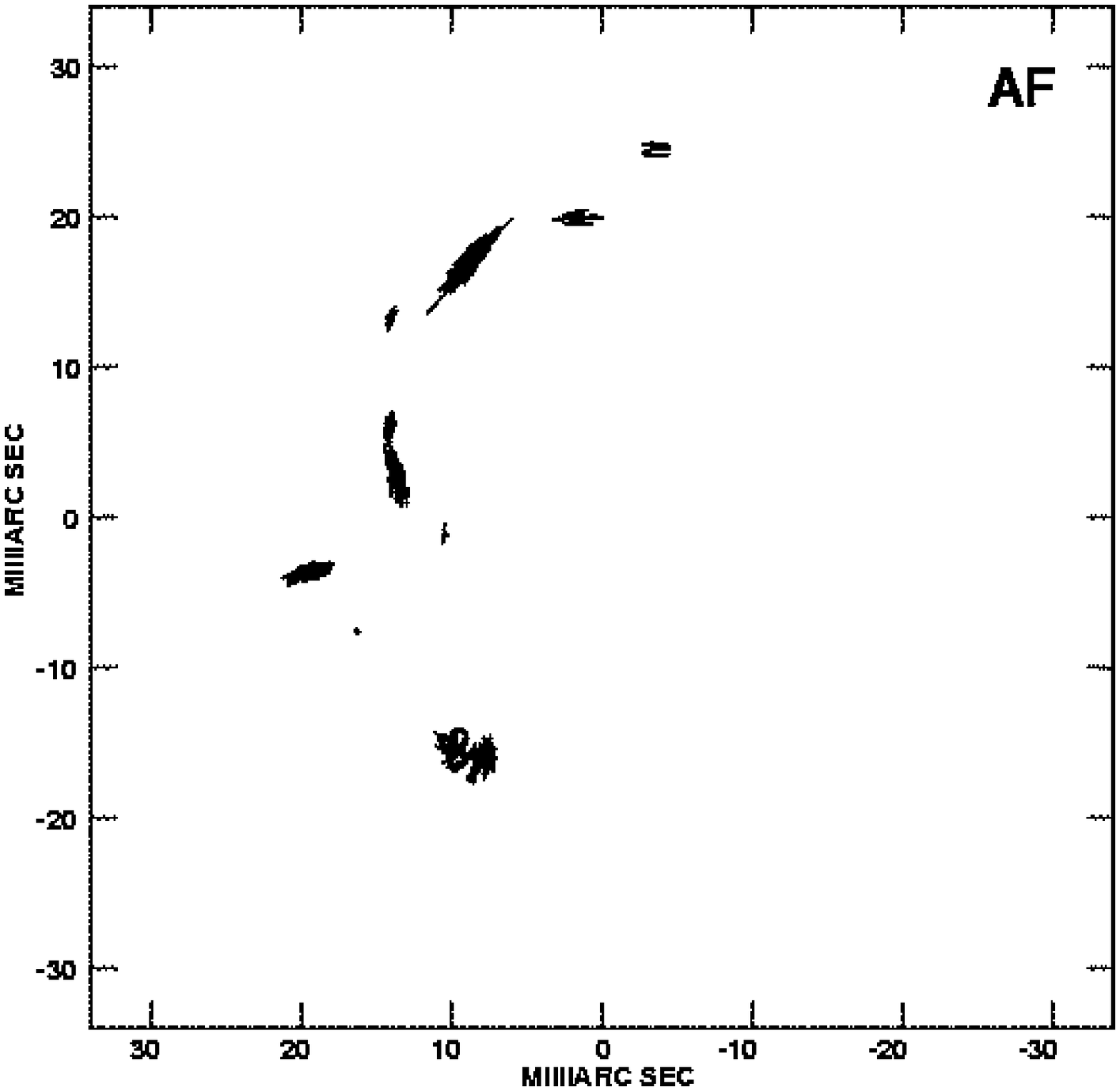} \\ 
\end{array}$ 

\caption{Same as Figure~\ref{fig-pcntr-3}.}

\label{fig-pcntr-7} 
\end{figure}

\begin{figure}[h] 
\advance\leftskip-1cm
\advance\rightskip-1cm
$\begin{array}{cc} 
\includegraphics[width=70mm, height=70mm]{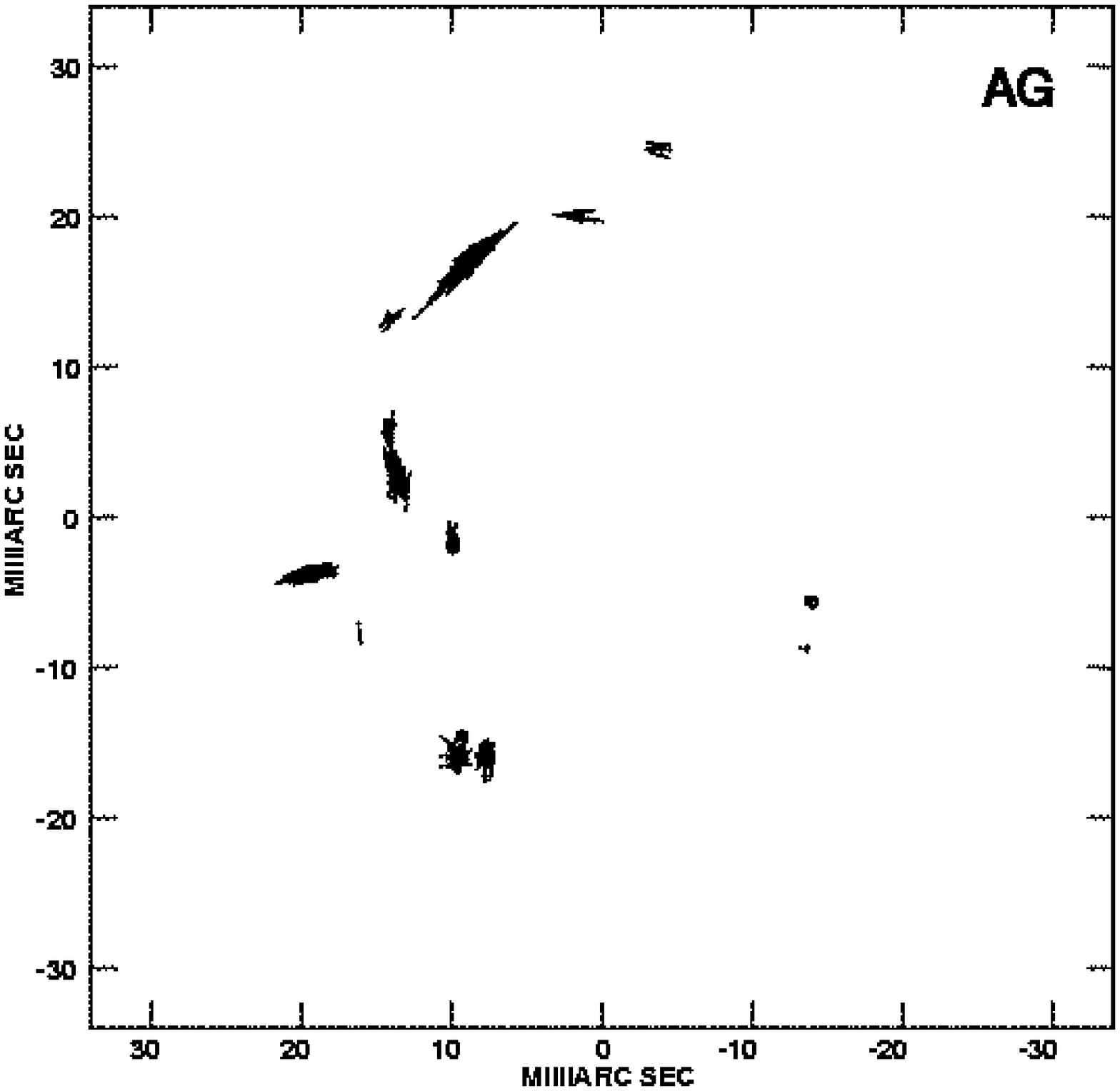} & 
\includegraphics[width=70mm, height=70mm]{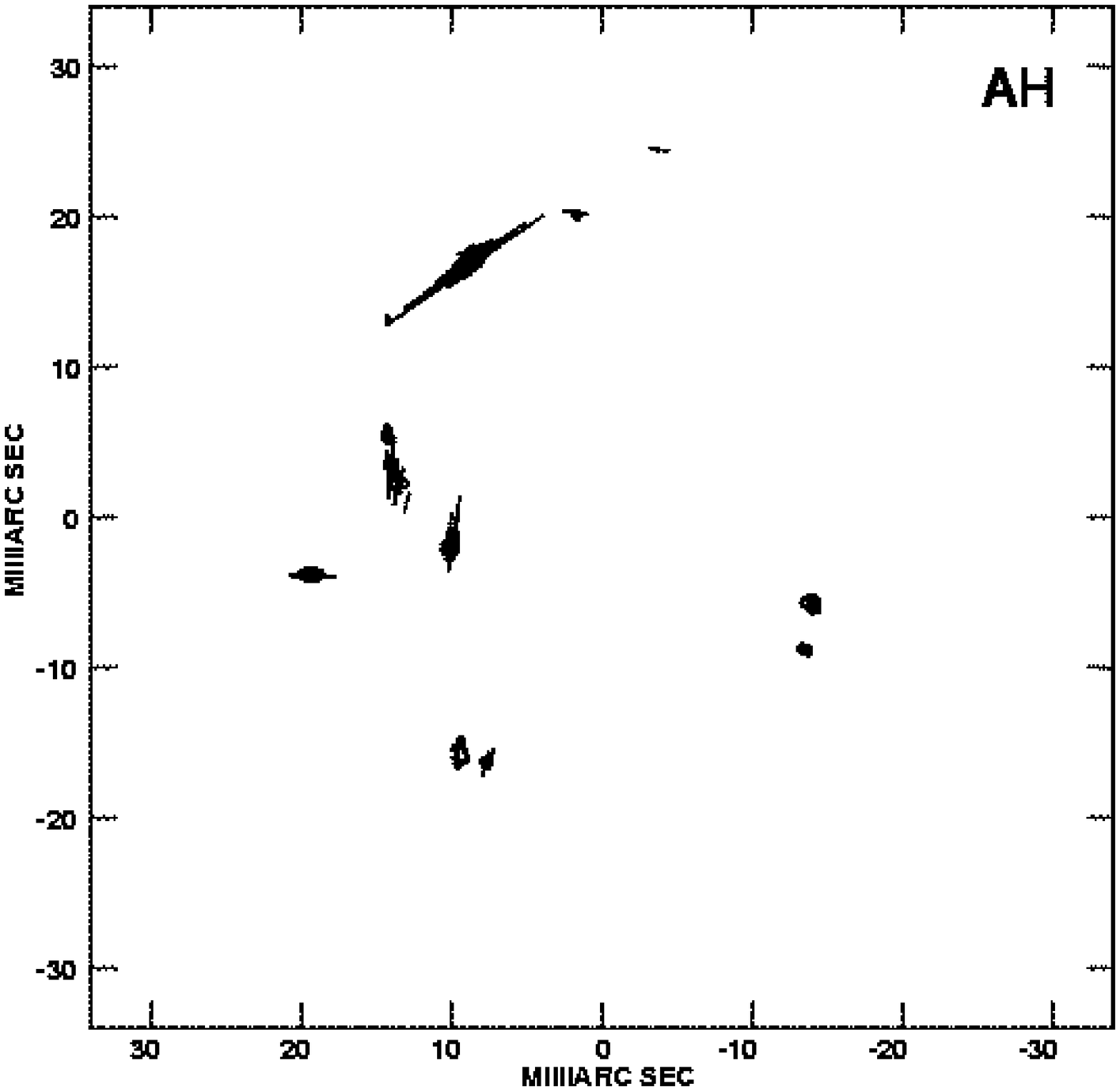} \\ 
\includegraphics[width=70mm, height=70mm]{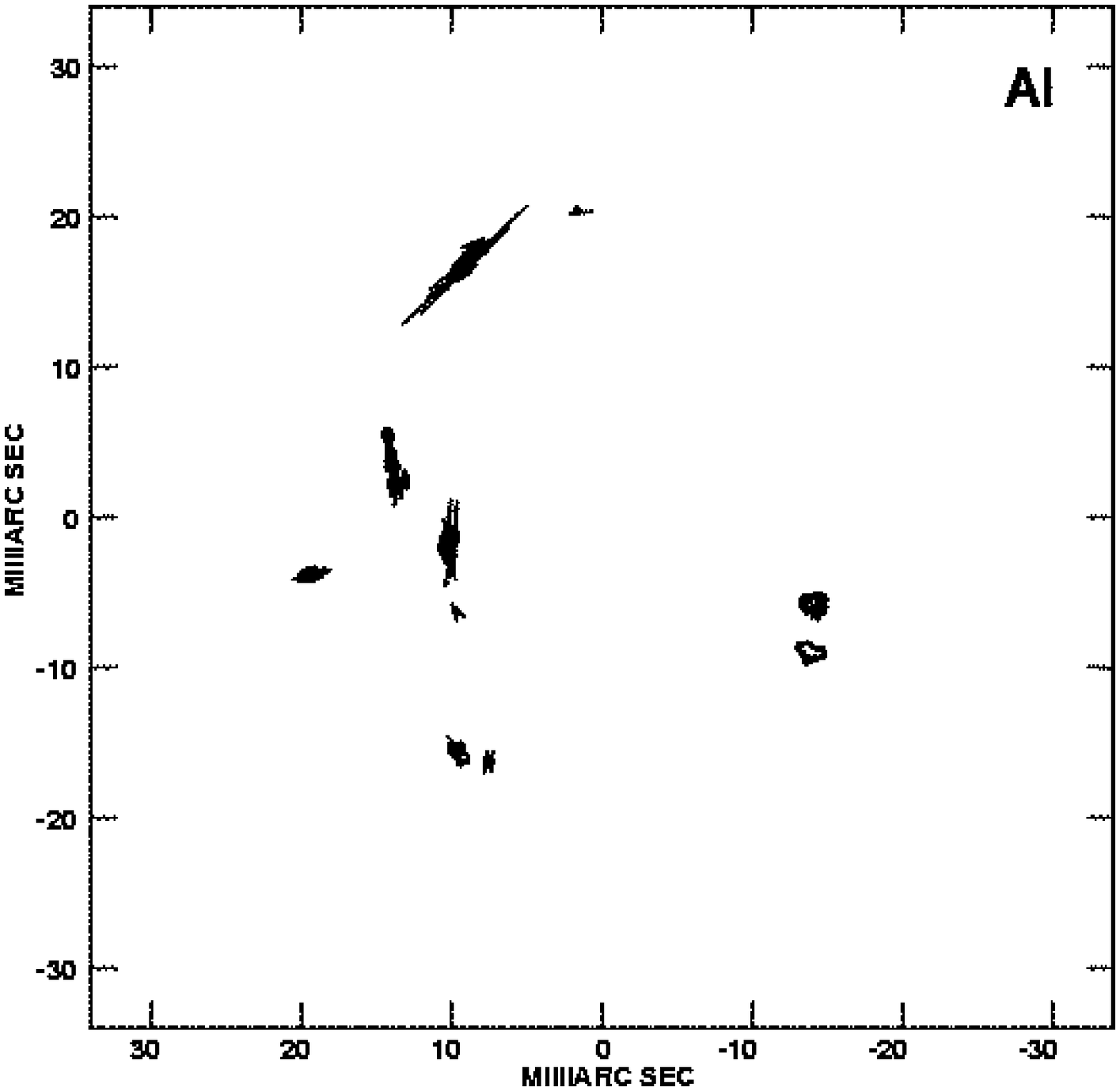} & 
\includegraphics[width=70mm, height=70mm]{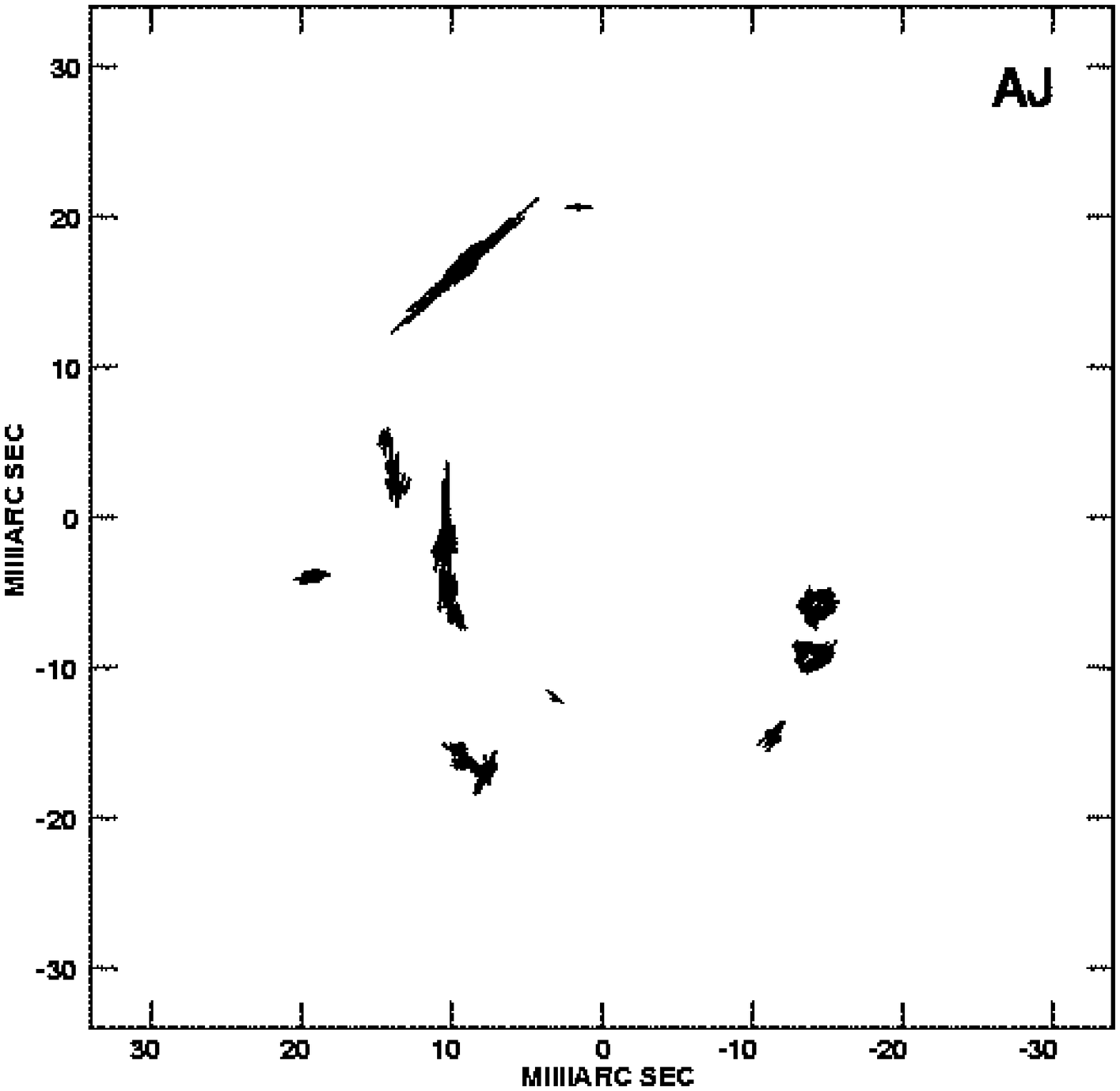} \\ 
\includegraphics[width=70mm, height=70mm]{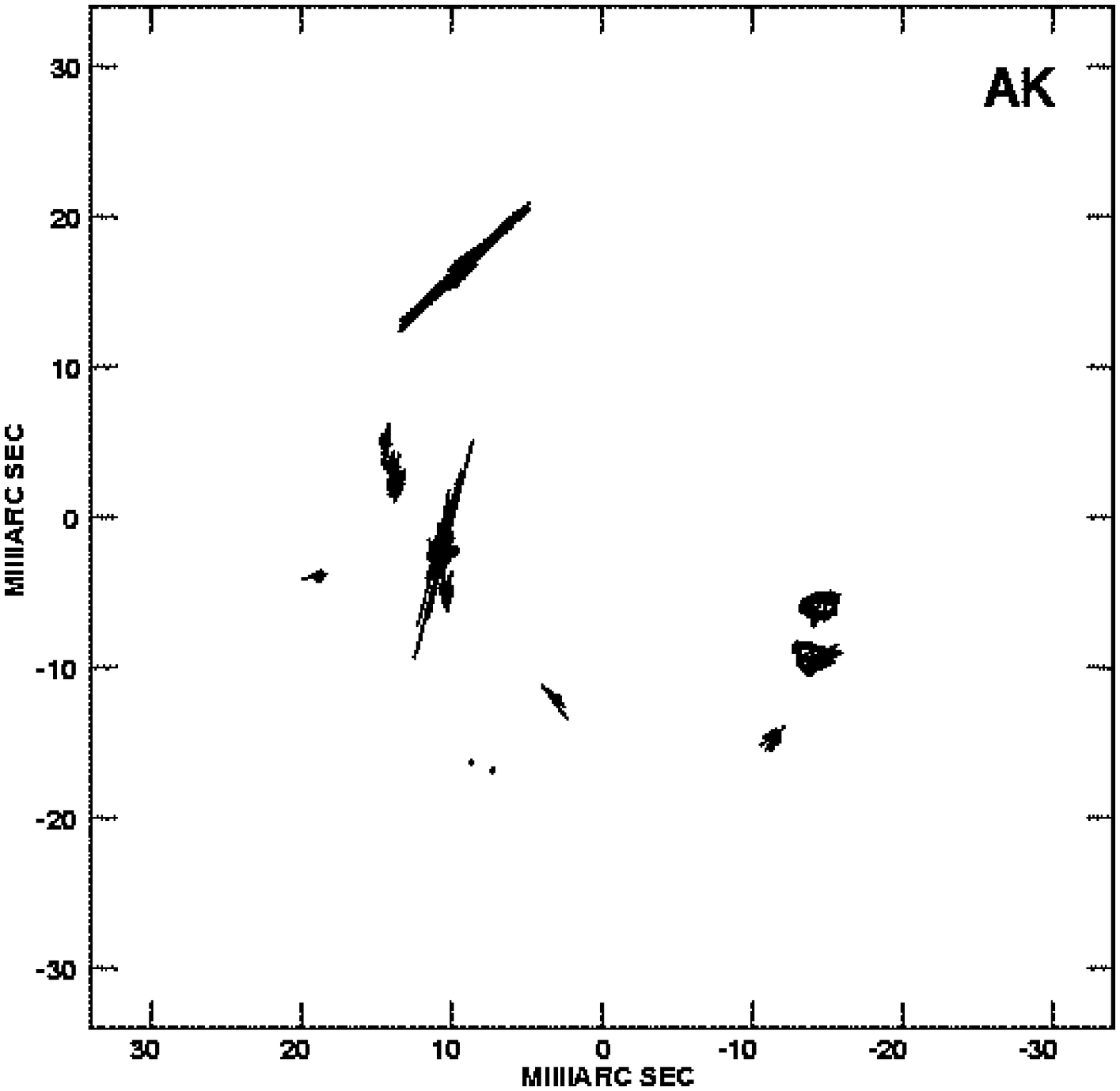} & 
\includegraphics[width=70mm, height=70mm]{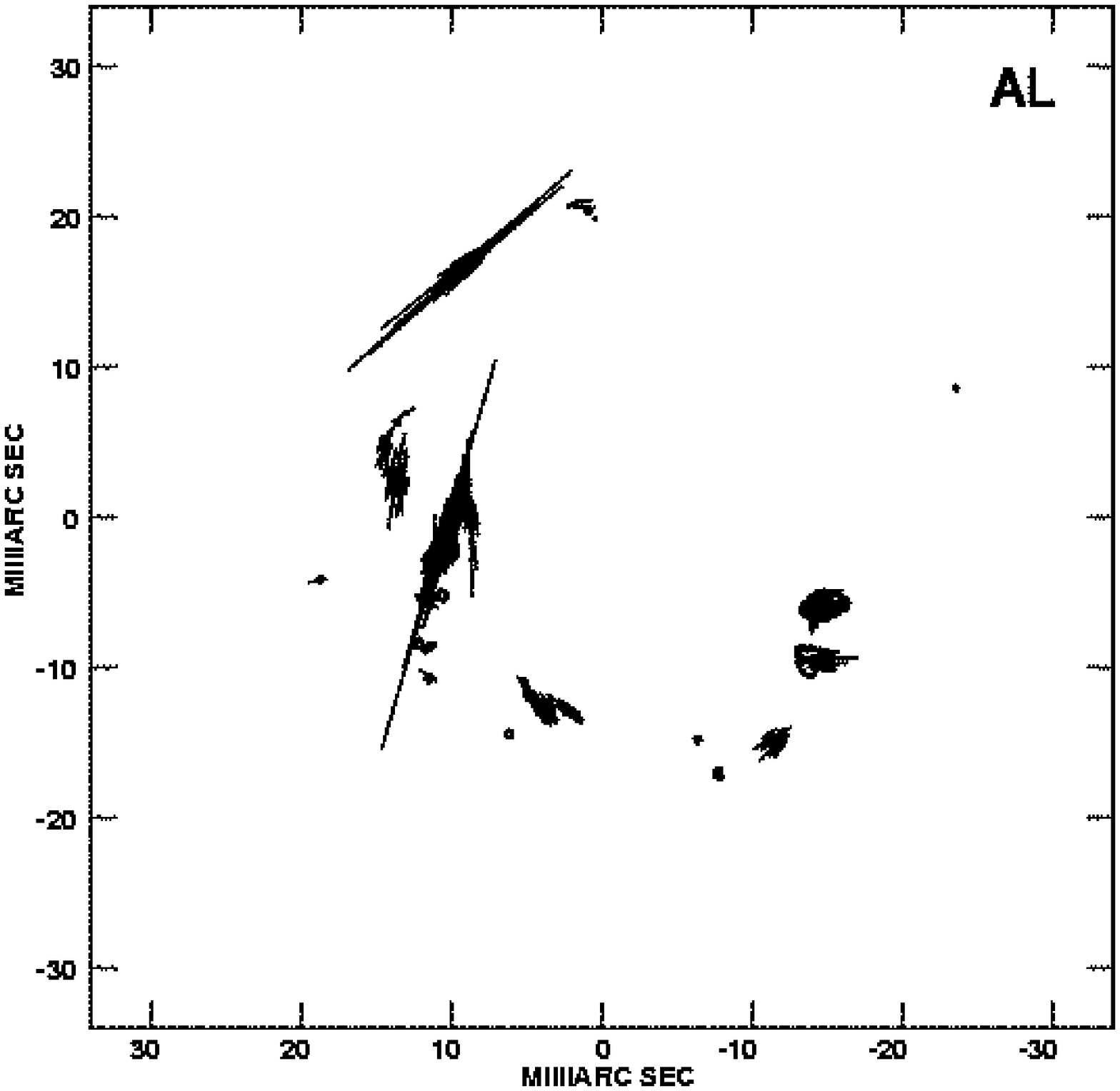} \\ 
\end{array}$ 

\caption{Same as Figure~\ref{fig-pcntr-3}.}

\label{fig-pcntr-8} 
\end{figure}

\begin{figure}[h] 
\advance\leftskip-1cm
\advance\rightskip-1cm
$\begin{array}{cc} 
\includegraphics[width=70mm, height=70mm]{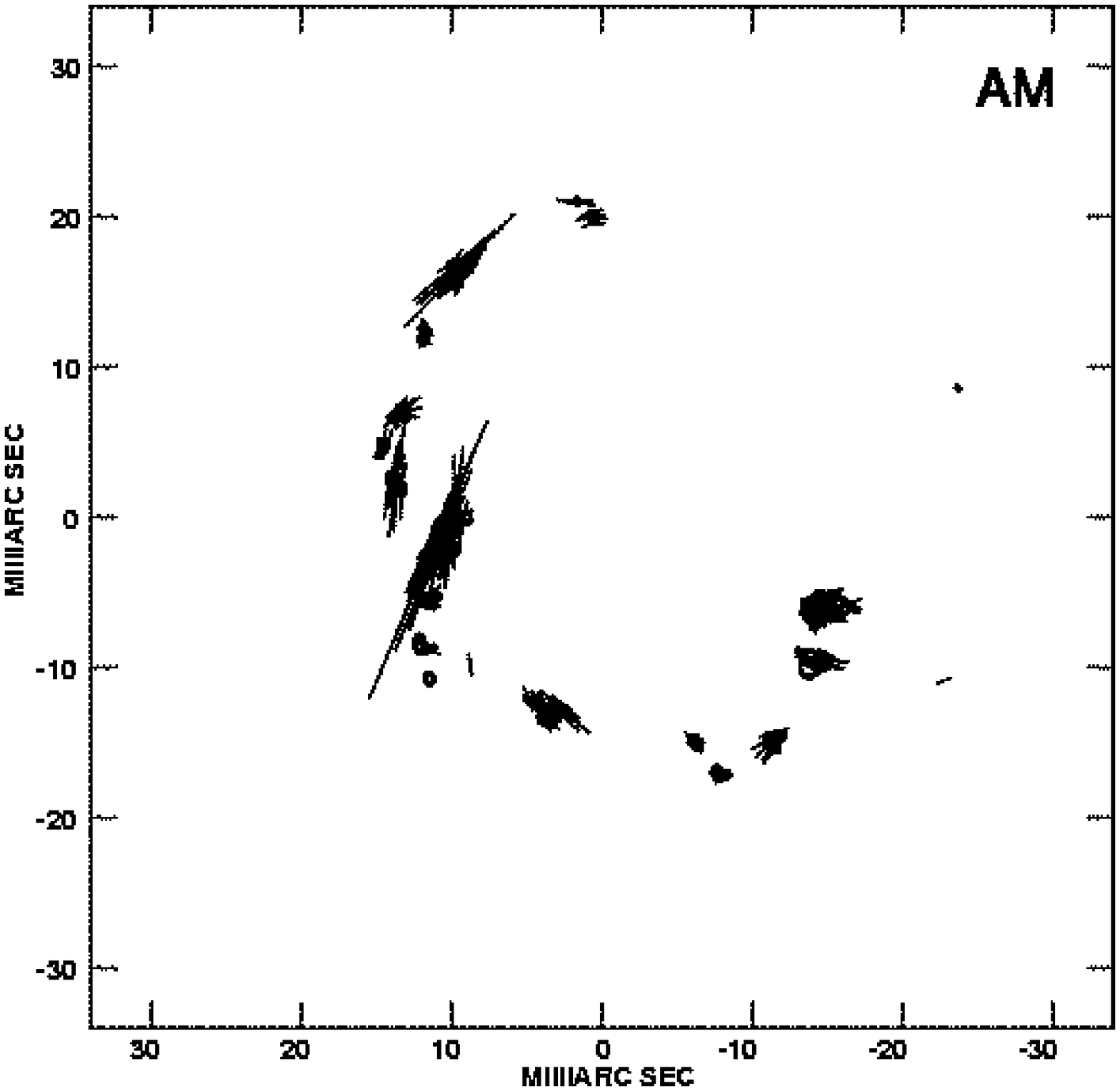} & 
\includegraphics[width=70mm, height=70mm]{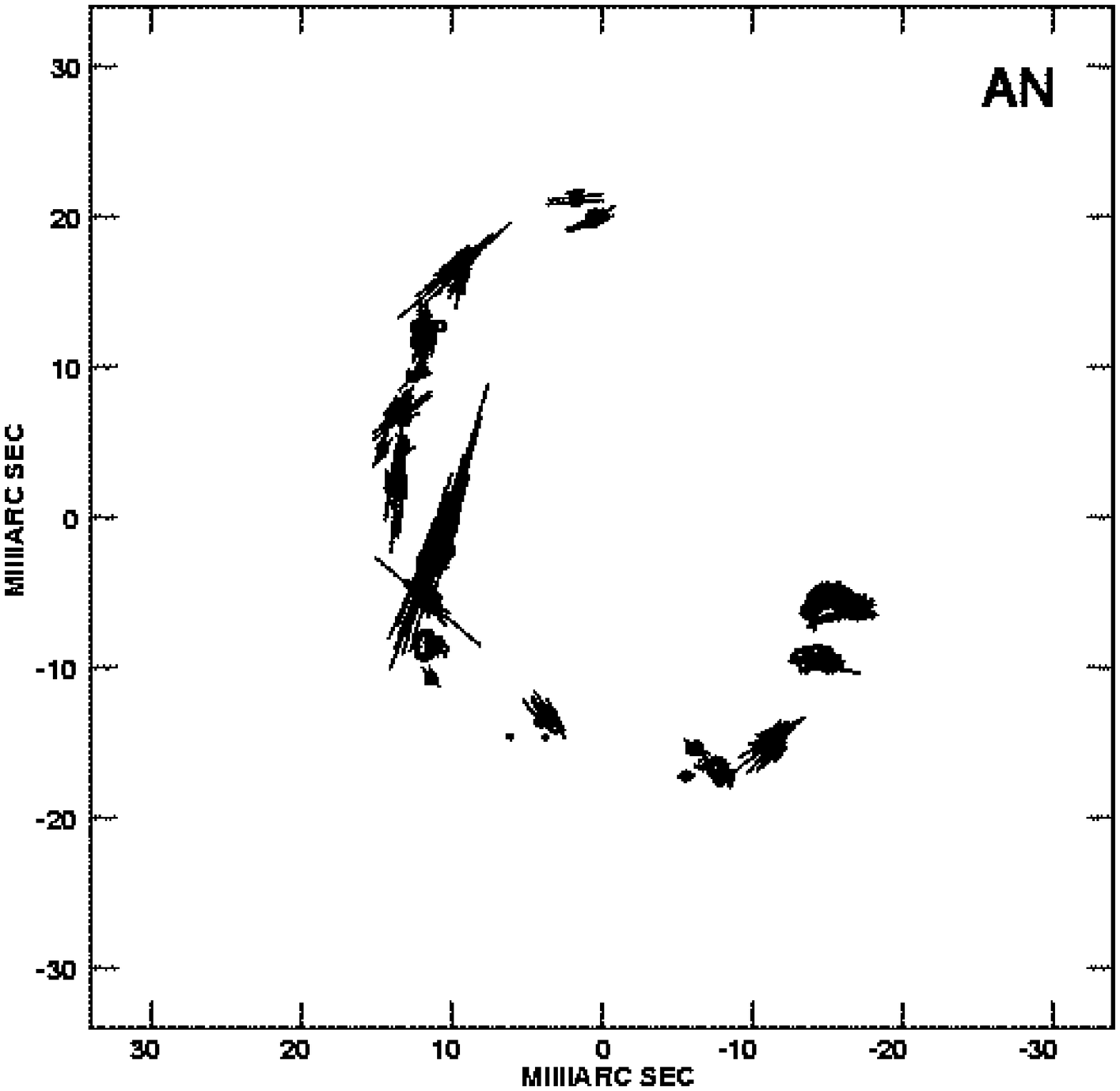} \\ 
\includegraphics[width=70mm, height=70mm]{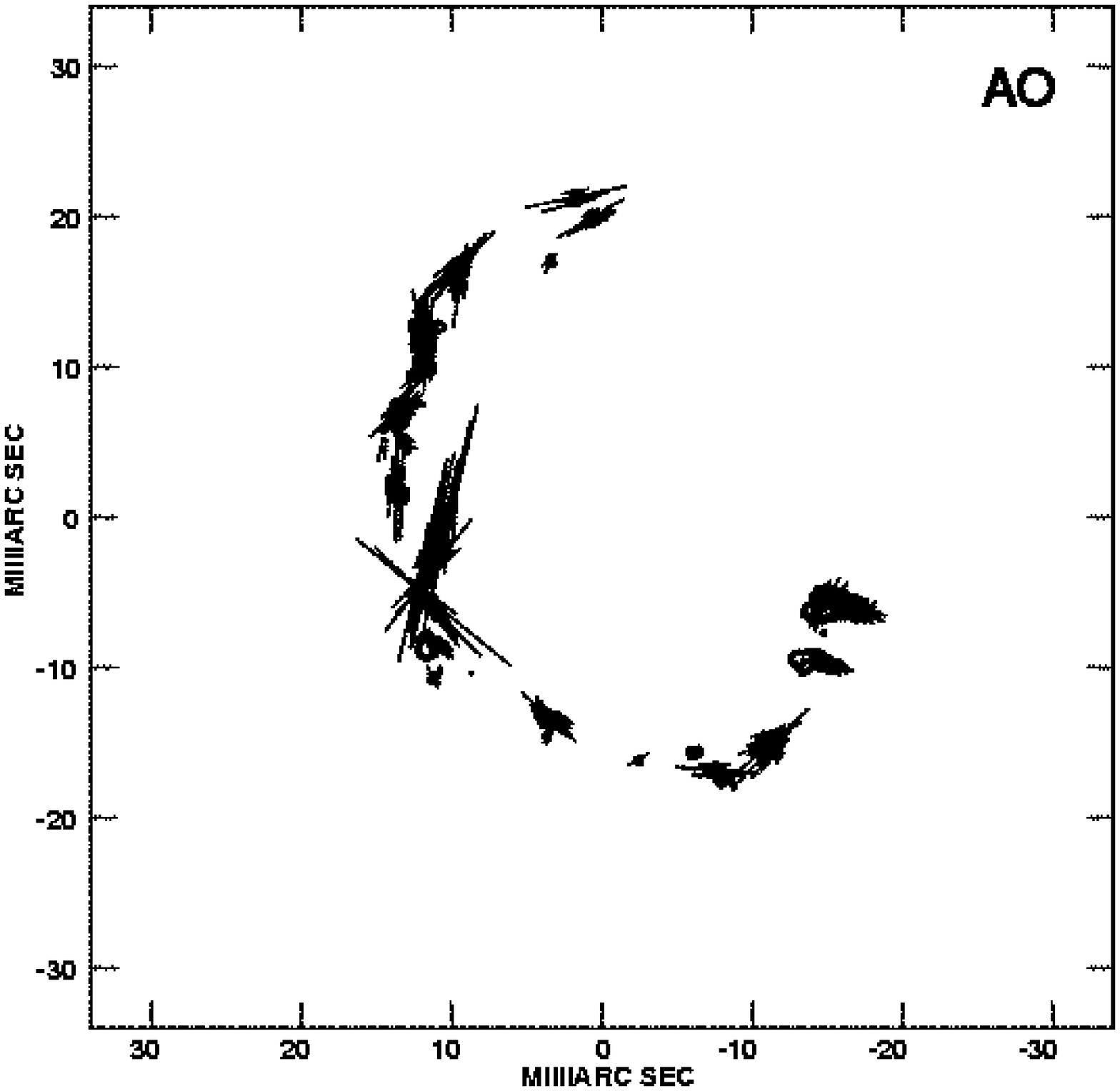} & 
\includegraphics[width=70mm, height=70mm]{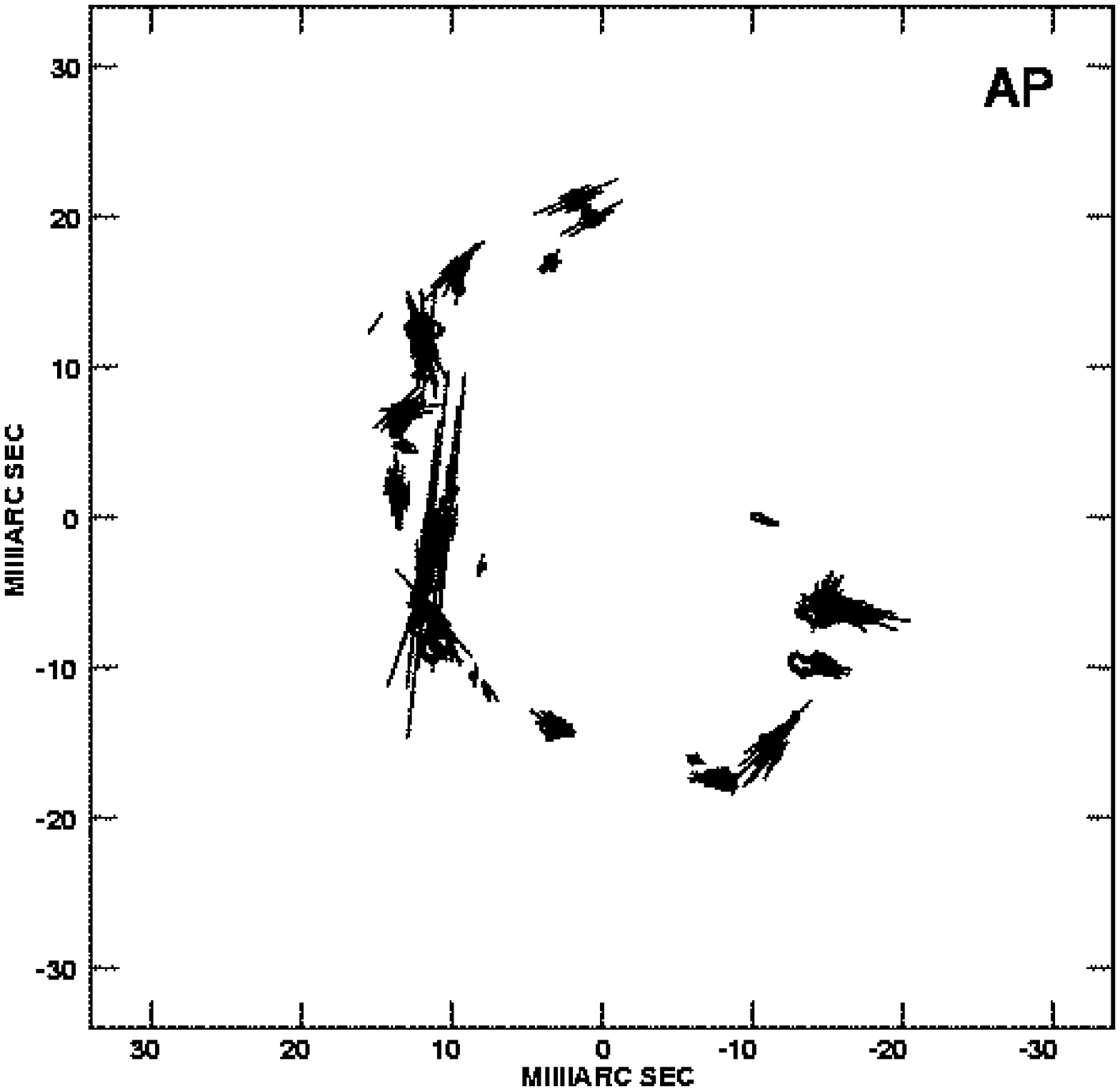} \\ 
\includegraphics[width=70mm, height=70mm]{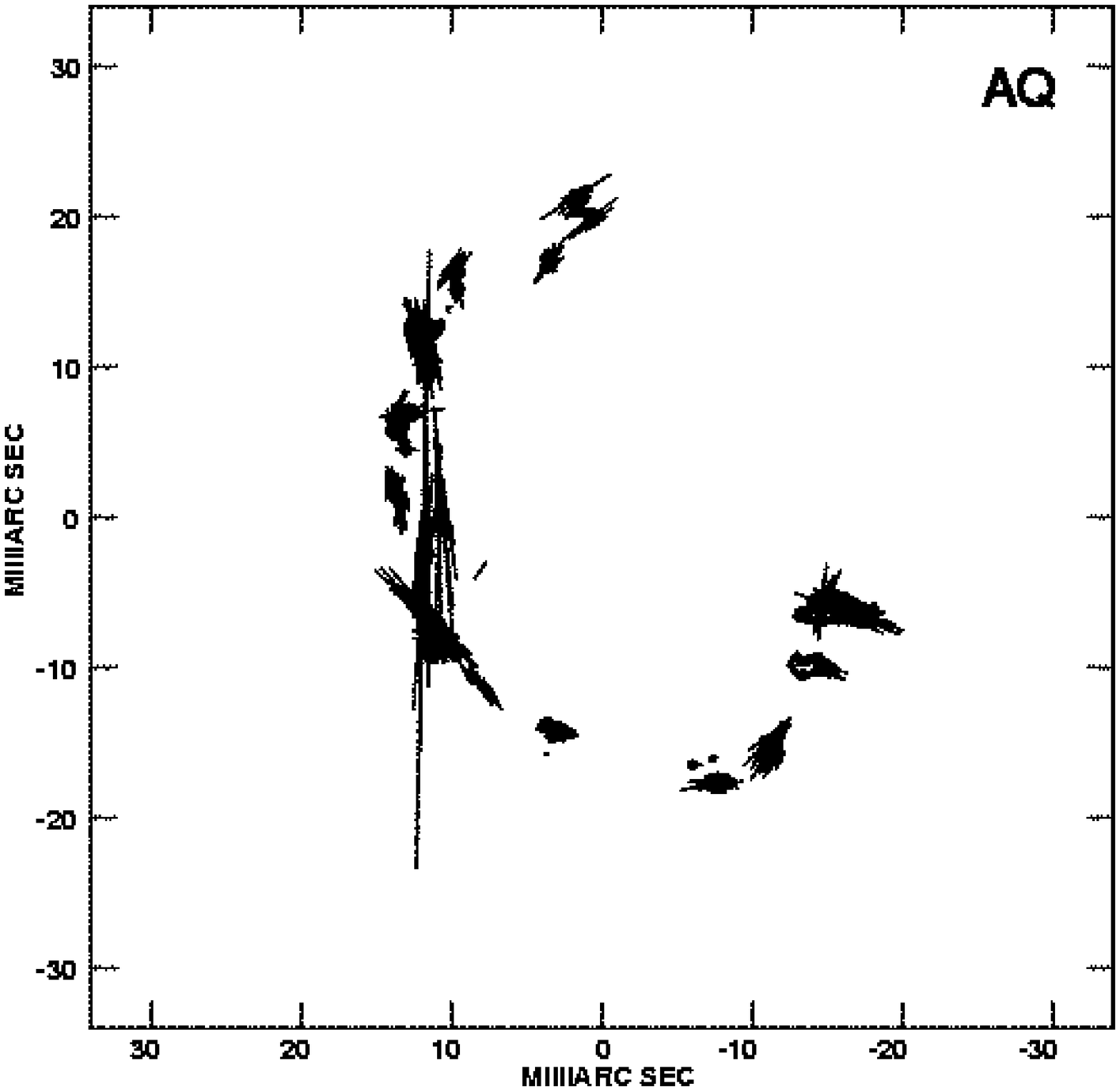} & 
\includegraphics[width=70mm, height=70mm]{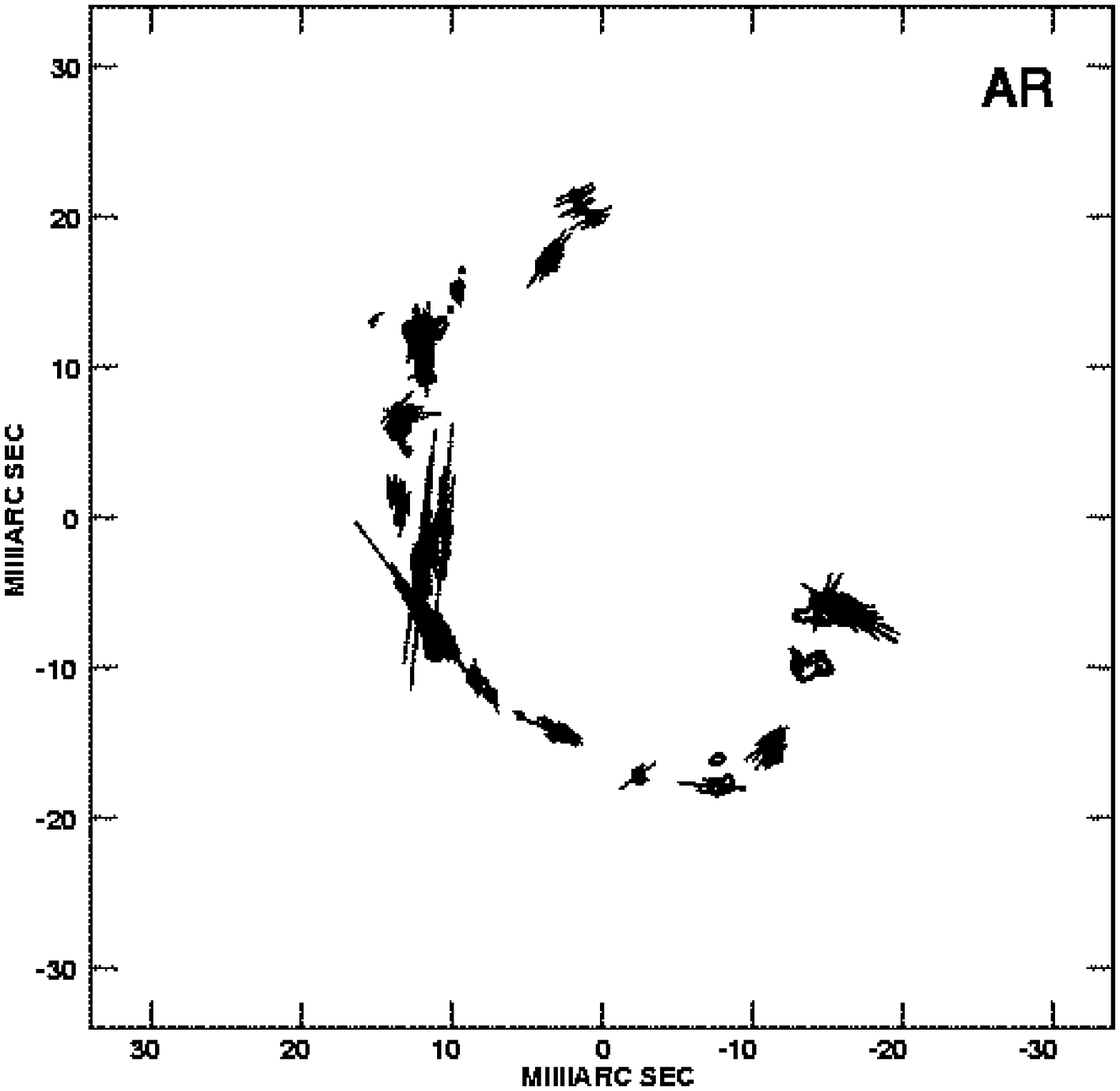} \\ 
\end{array}$ 

\caption{Same as Figure~\ref{fig-pcntr-3}.}

\label{fig-pcntr-9} 
\end{figure}

The total and linearly-polarized intensity images of the $v=1,\ J=1-0$
SiO emission toward TX Cam over the pulsation phase range covered by
this paper are shown in Figures~\ref{fig-pcntr-3} through
\ref{fig-pcntr-9}. In these composite figures, the enhanced-contrast
Stokes I image is plotted as a background contour plot, overlaid by
vectors representing the linearly-polarized emission. The vectors are
drawn with a length proportional to the linearly-polarized intensity
and at the position angle (N through E) corresponding to the EVPA of
the linearly-polarized emission. These figures are the zeroth-moment
(over frequency) of the associated Stokes $(I,Q,U)$ brightness in the
parent image cubes at each epoch.

These polarization images are presented in an animated sequence in the
electronic materials associated with this paper, in the form of a
movie of the linear polarization morphology evolution of the $v=1,\
J=1-0$ SiO maser emission toward TX Cam over the pulsation phase range
$\phi=0.68$ to $\phi=1.82$.

The phase range covered is the first pulsation cycle of our longer
synoptic monitoring campaign. Several striking properties are evident
from this time-series of figures. Over this pulsation phase range, the
linear polarization EVPA is predominantly tangential to the projected
total intensity SiO maser shell. In addition, for the majority of the
SiO maser components, the EVPA of individual features appears to
persist over their component life-times. Finally, the
linearly-polarized intensity of the maser emission is generally
brightest at the inner projected shell boundary and decreases in
magnitude at larger radii.

The dominant tangential linear-polarization morphology is consistent
with that reported by \citet{kemball97} at a single epoch, in the
first VLBI imaging polarimetry of this SiO transition toward TX
Cam. The earlier observations were undertaken on 1 Dec 1994, so
precede the data in the current paper by two pulsation
cycles. \citet{desmurs00} confirmed the tangential linear polarization
morphology toward TX Cam in observations in April 1996. Taken together
with the current data, this suggests that this morphology may be an
inter-cycle property of the SiO maser emission toward this star. It is
not yet clear however, whether this is a generic property of SiO
linear polarization in this transition toward late-type evolved stars
in general. \citet{desmurs00} report a similar tangential morphology
in the SiO maser emission toward IRC+10011. However, for a sample of
Mira variables monitored in $v=1,2,\ J=1-0$ SiO maser emission in a
full-polarization VLBI imaging study, and reported in a series of
papers by \citet[and references therein]{cotton08}, the results are
more mixed. Early images suggested no pervasive linear polarization
pattern \citep{cotton04}; however, later, a bimodal EVPA distribution
(in the sense of being either parallel or perpendicular to the
projected shell) was reported for U Ori and o Cet
\citep{cotton06}. However, not all sources for which images are
shown by \citet{cotton08} show tangential linear polarization
structure. Further observations of larger samples are needed
to resolve this important question.

There are several possible theoretical explanations for tangential
linear polarization structure in SiO maser emission, as we see in the
data presented here. However, these interpretations are conditional on
the theoretical assumptions made concerning the underlying maser
emission; we focus here only on the observational implications of
these theoretical issues in what follows.

Silicon monoxide is a non-paramagnetic molecule and the rotational
transition considered here is in the small-splitting Zeeman
regime. Following \citet{elitzur96}, the ratio $x_B$ of the Zeeman
splitting to the Doppler linewidth $\Delta v_D$ is \citep{kemball97}:

\begin{equation}
x_B = 8.2 \times 10^{-4} \left(\frac{B}{\rm{G}}\right)
\left(\frac{\Delta v_D}{\rm{km.s}^{-1}}\right)^{-1}\label{eqn-xb}
\end{equation}

For the case of $x_B \ll 1$, as equation~\ref{eqn-xb} indicates is
applicable to SiO, competing models for the transport of polarized
maser emission have been presented \citep[and references
therein]{elitzur96, watson02}. Both series of papers extend the
parameter space and generalize the original foundational work of
\citet{goldreich73}. There also remains uncertainty in the literature
as to whether SiO masers are pumped by collisional \citep{elitzur80}
or radiative mechanisms \citep{bujarrabal81}. Finally, an additional
consideration is the degree to which $m-$anisotropic pumping of the
magnetic substates dominates. We neglect for the purposes of this
discussion the role of magnetic or velocity gradients in the masing
regions, and any considerations that apply specifically to transitions
with spins higher than $J=1-0$.

At the outset we need to determine in which theoretical regime of
maser polarization propagation our observational data reside.
Specifically, there are several key parameters that determine this
theoretical interpretive framework. These include the relative
magnitude of the stimulated emission rate, $R$, the cumulative
collisional and radiative decay rate, $\Gamma$, and the Zeeman rate,
$g\Omega$ \citep{goldreich73,watson02}. 

To estimate $R$, we adopt equation 1 of \citet{plambeck03} to derive
the stimulated emission rate of the $v=1,\ J=1-0$ SiO transition as:

\begin{equation}
R=23\ \left(\frac{T_B}{2\times10^{10}\ \rm{K}}\right)\ \left(\frac{d\Omega}{10^{-2}\ \rm{sr}}\right) \rm{s}^{-1}\label{eqn-r}
\end{equation}

where $T_B$ is the maser brightness temperature and $d\Omega$ is the
estimated maser beaming angle. The relative magnitude of $R$ and
$\Gamma$ defines the degree of maser saturation, with saturation
increasing for increasing $\frac{R}{\Gamma}$. The vibrational
radiative transitions $v \to (v-1)$ for low-J rotation levels have
radiative decay rates $\Gamma \sim 5 v$ s$^{-1}$
\citep{kwan74,elitzur92}. These radiative decay rates dominate over SiO-H$_2$
or SiO-H collisional de-excitation. From the collisional rate
equations in \citet{elitzur80}, we calculate $\Gamma < 1$s$^{-1}$ for
collisions, adopting a representative temperature $T$=1300 K and
atomic hydrogen density $n_H = 5 \times 10^{10}$ cm$^{-3}$ in the SiO
maser region, as inferred from \citet{reid97}. The Zeeman rate for SiO
is given by \citet{nedoluha90,plambeck03} as:

\begin{equation}
\frac{g\Omega}{2\pi} \sim 200 \left(\frac{B}{\rm{G}}\right)\ s^{-1}\label{eqn-go}
\end{equation}

In the limiting regime $g\Omega \gg R \gg \Gamma$, and assuming
$m$-isotropic pumping, the \citet{goldreich73} solutions for the
linear polarization are applicable. In this case the fractional linear
polarization depends on the angle between the line of sight and the
magnetic field, and takes the form $q(\theta)=\frac{Q}{I} =
\frac{2-3\sin^2\theta}{3\sin^2\theta}$ for $\theta \ge \theta_B$, and
$q(\theta) = 1$ for $\theta \le \theta_B$, where $\theta_B \approx
35^\circ$ \citep{goldreich73,elitzur92}. At a break angle, $\theta_F
\approx 55^\circ$, defined as the point where
$\sin^2\theta_F=\frac{2}{3}$, and Stokes $Q$ changes sign, the
observed EVPA on the plane of the sky switches from parallel to the
projected magnetic field ($\theta < \theta_F)$ to perpendicular to the
projected magnetic field ($\theta > \theta_F)$
\citep{goldreich73,elitzur92}. We note that this change of EVPA over
90$^\circ$ has been detected observationally by \citet{vlemmings06a}
for a water maser component in the source W43A.

The degree of saturation $\frac{R}{\Gamma}$ required to achieve the
the limiting magnitude of fractional linear polarization differs
however between the two theoretical models of maser polarization
propagation \citep{nedoluha90,elitzur96}. A saturation of
$\frac{R}{\Gamma} \approx 30$ is required to achieve 70\% linear
polarization in the models summarized by \citet{watson01}. In the
model of \citet{elitzur96}, however, the full polarization solution is
realized well before saturation.

In addition, the observational interpretation of linearly-polarized
SiO maser emission is further influenced by whether the maser
inversion is assumed to be due to collisional or radiative pumping. If
the pumping is assumed to be radiative, then the effects of
$m$-anisotropic pumping of the magnetic sublevels $m$ of the
rotational transitions needs to be considered \citep{bujarrabal81}
along with related radiative effects \citep{ramos05}. In the case of
stellar SiO masers, there is an inherent anisotropy in the angular
distribution of the photons from the central star incident on the
masing region, and, in this model, $m$-anisotropic pumping can produce
large linear polarization magnitudes, oriented tangentially to the
incident radial stellar photons \citep{western83,desmurs00}. For
$m$-anisotropic pumping, the observed EVPA is still likely to be
either parallel or perpendicular to the projected magnetic field on
the plane of the sky, if the conditions $g\Omega > R$ and $g\Omega >
\Gamma$ hold, even weakly \citep{watson02}, however the shape of the
function $q(\theta)$ and the break angle $\theta_F$ are likely to
differ from the standard Zeeman case above. If the magnetic field is
sufficiently weak so that $g\Omega \approx R$ holds, then the EVPA
need no longer be strictly parallel or perpendicular to the magnetic
field in this model \citep{watson02}. Given the degree of order we
observe in the linear polarization morphology, the data suggest that
$g\Omega \neq R$ in the current study.

Both theoretical models of maser polarization predict circular
polarization that is broadly proportional to the magnetic field
strength \citep{watson01,elitzur96}, but with differences in predicted
line-shape and dependence on saturation as well as the angle $\theta$
between the magnetic field and the line of sight. Given $x_B$ for SiO
masers, circular polarization at the several percent level predicts
Gauss-level magnetic fields \citep{barvainis87,kemball97}. In the
models summarized by \citet{watson02}, circular polarization can also
arise from non-Zeeman effects caused by a difference between the
optical axes along the maser propagation path and the direction of
linear polarization. Circular polarization can be produced in this
model when $R \approx g\Omega$ \citep{nedoluha90} or by turbulent
rotation of the magnetic field along the line of sight in the regime
$g\Omega \gg \{R,\Gamma\}$ \citep{wiebe98}.

To constrain the theoretical regime in which our current data reside,
it is first useful to estimate the relative magnitude of $g\Omega$,
$R$, and $\Gamma$. There are two main sources of uncertainty in this
estimation, namely the strength of the magnetic field, $B$, and the
unknown maser beaming angle $d\Omega$ used to calculate the stimulated
emission rate in equation~\ref{eqn-r} (equivalently, the degree of
saturation).  We make an initial estimate of $B$ by comparing energy
densities. A magnetic field $B$ implies a magnetic energy density:

\begin{equation}
\frac{B^2}{2\mu_0} \sim 4 \times 10^{-3} \left(\frac{B}{\rm{G}}\right)^2 \rm{J.m}^{-3}
\end{equation}

For our adopted $n_H = 5 \times 10^{10}$ cm$^{-3}$ and $T=1300$ K, the
thermal energy density is $\frac{3}{2}n_HkT \sim 1.3 \times 10^{-3}$
J.m$^{-3}$, and the bulk kinetic energy density is $\frac{1}{2}\rho
v^2 \sim 2.0 \times 10^{-3}$ J.m$^{-3}$, where $\rho$ is density and
we adopt a bulk motion $v=7$ \kms from Paper I. Equating these energy
densities suggests a zeroth-order magnetic field estimate in the range
$\sim 580-720$ mG. For $B\sim0.6$ G, equation~\ref{eqn-go} predicts
$\frac{g\Omega}{2\pi} \approx 120$ s$^{-1}$. Tangential shock
compression driven by stellar pulsation could increase the magnetic
field strength by an order of magnitude
\citep{soker02a,hartquist97}. Adopting $R \sim 23$ s$^{-1}$, $\Gamma
\sim 5$ s$^{-1}$, and $g\Omega \geq 10^{3-4}$, would place us in the
regime: $g\Omega \gg R \gg \Gamma$ or $g\Omega > R \gg \Gamma$. In
this regime, and absent significant Faraday rotation, the standard
Zeeman interpretation of the parallel or perpendicular orientation of
the EVPA relative to the projected magnetic field would apply, and the
linearly-polarized morphology could either be tracing a
globally-organized stellar magnetic field, or a sparse strong field
that is only dynamically significant on local scales. There are
arguments against a globally organized stellar magnetic field in
isolated (non-binary) TP-AGB stars, primarily the likely low rotation
velocity at this point in their evolutionary path in the H-R diagram,
and the difficulties of sustaining a regular $\alpha \omega$ dynamo in
stars with such an extended mantle, amongst other considerations
\citep{soker06}.However, as noted above, there are several strands of
observational evidence supporting surface magnetic fields of order
$10^{2-3}$ G in Mira variables \citep{vlemmings05} and the central
stars of PNe \citep{jordan05} respectively, as well as magnetic jet
collimation in post-AGB transition objects \citep{vlemmings06}.

A sparse field in an AGB extended atmosphere could arise from an
$\alpha^2 \omega$ convective dynamo, as noted above, and might include
local loops rising above the photosphere, other flaring events, and
magnetic cool spots \citep{soker99,soker03}. These localized fields
are also expected to be shock-compressed tangentially due to the
outwardly-propagating shock waves from the pulsating central star,
introducing a preferred geometry for field enhancement. This is a
complex three-dimensional picture, but it is also important to
remember that SiO masers sample only those regions of the NCSE that
lie relatively close to the orthogonal plane to the line of
sight. This constraint arises from the observed tangential
amplification of the total intensity emission.

If however, the masers are pumped radiatively and the radiative
pumping is $m$-anisotropic, then the tangential linear polarization
morphology could originate from that mechanism \citep{western84}, even
in the non-magnetic case $g\Omega \sim 0$ \citep{ramos05}. For the
magnetic case, $g\Omega > 0$, the Hanle effect can, in certain cases,
produce an EVPA rotation exceeding 45 $\deg$ relative to the tangential
direction \citep{ramos05}. In the case of $m$-anisotropic pumping in
general, we would also expect to measure a functional shape for
$q(\theta)$, and associated break angle $\theta_F$, that differs from
the standard Zeeman form noted above \citep{western84}, and a higher
degree of linear polarization magnitude \citep{watson02}.

\begin{figure}
\epsscale{0.9}
\plotone{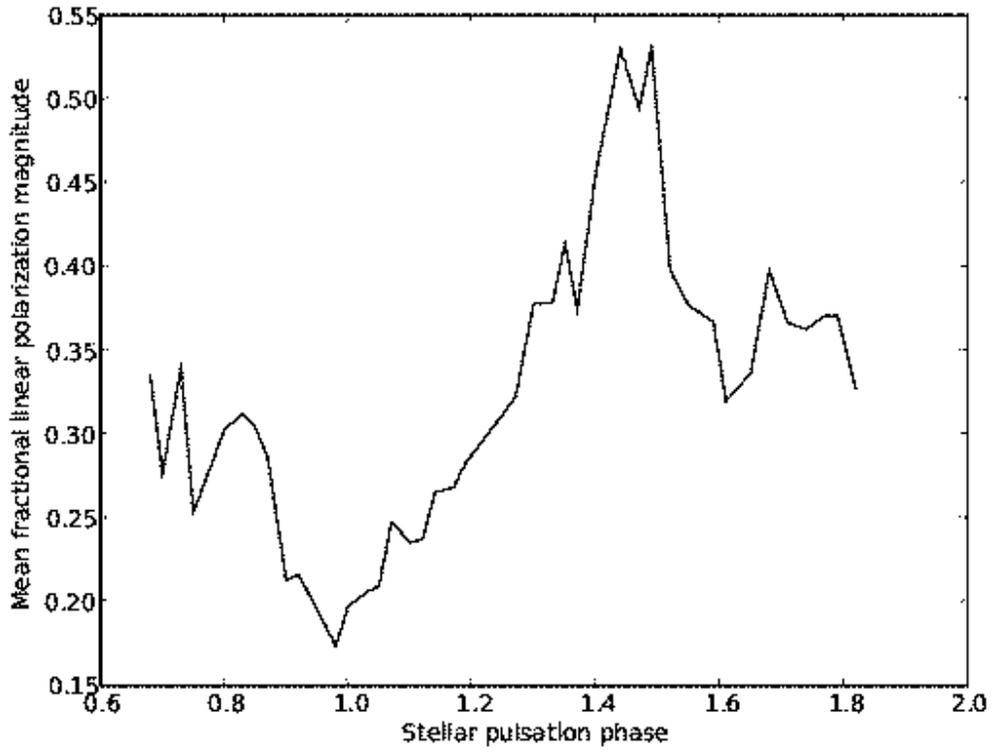}

\caption{The mean fractional linear polarization magnitude as a
function of stellar pulsation phase.}

\label{fig-ml}
\end{figure}

In Figure~\ref{fig-ml} we examine the variation of mean fractional
linear polarization magnitude ${\overline m_l}(\phi)$ against pulsation
phase. This value was computed at each epoch from the sum of Stokes
$I$ and $P=\sqrt{Q^2 + U^2}$, using a pixel mask defined as the set of
all pixels with a total intensity exceeding three times the Stokes $I$
image rms, and corrected for statistical bias in $P$. The value of
${\overline m_l}(\phi)$ has a minimum near $\phi \sim 1$ and a maximum near
$\phi \sim 1.5$. It has higher values when the inner edge of the shell
is well-defined and dominates the overall total intensity
morphology. Unfortunately this result can be interpreted either as
evidence for linear polarization driven primarily by magnetic field
enhancement through shock compression, or is supportive of
$m$-anisotropic pumping as the dominant origin of linear
polarization: both are expected to be stronger in the inner shell
boundary. We note that the mean fractional linear polarization is
broadly inversely correlated with the shell width. 

We identify several key uncertainties that need to be further
constrained by observations before the remaining theoretical
polarization transport uncertainties can be resolved, and that are
amenable to observational    tests: i) the statistical incidence of
tangential linear polarization morphology in Mira variables as a
class; ii) independent measurement of the exact degree of saturation
of stellar SiO masers; and iii) detailed measurements of the
functional form of $q(\theta)$ in maser regions where this is
possible. We plan to address these issues in future papers. As yet,
there is insufficient observational data to distinguish the relative
influence of $m$-anisotropic pumping on the magnitude or orientation
of the linearly-polarized SiO emission

The linear polarization maps we present here have significant
fine-scale, locally-ordered, polarization structure, as noted in the
first SiO polarization observations of TX Cam by \citet{kemball97}.
Recently, in their sample of O-rich AGB stars, \citet{cotton08} have
cited locally-ordered polarization morphologies along jet-like
features pointing radially toward the central star. Due to the
persistence of the SiO linear polarization properties over a
sufficient fraction of the component lifetime in the polarization
movie presented here, we are able to explore the time evolution of the
polarization morphology for individual maser components. In this paper
we undertake a preliminary analysis confined to two isolated
components selected subject to the criteria that they: i) follow continuous
trajectories across a sufficient range of epochs; ii) are located at a
larger projected separation from the inner shell boundary, so as to
minimize the possible impact of strong anisotropic pumping effects;
and, iii) have compact structure and high SNR.

\begin{figure}[h] 
\advance\leftskip-1cm
\advance\rightskip-1cm
$\begin{array}{cccc} 
\includegraphics[width=34mm, height=48mm]{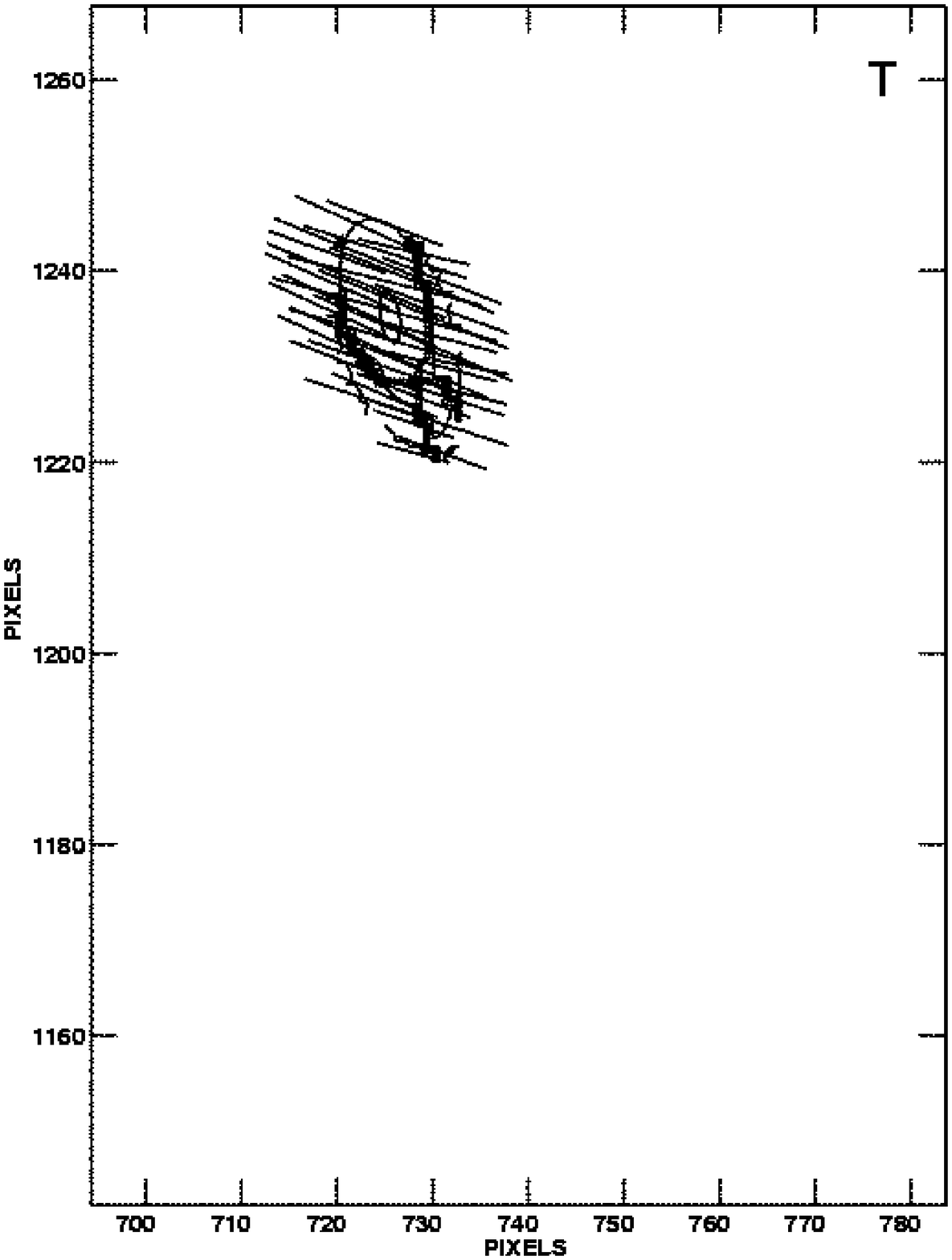} & 
\includegraphics[width=34mm, height=48mm]{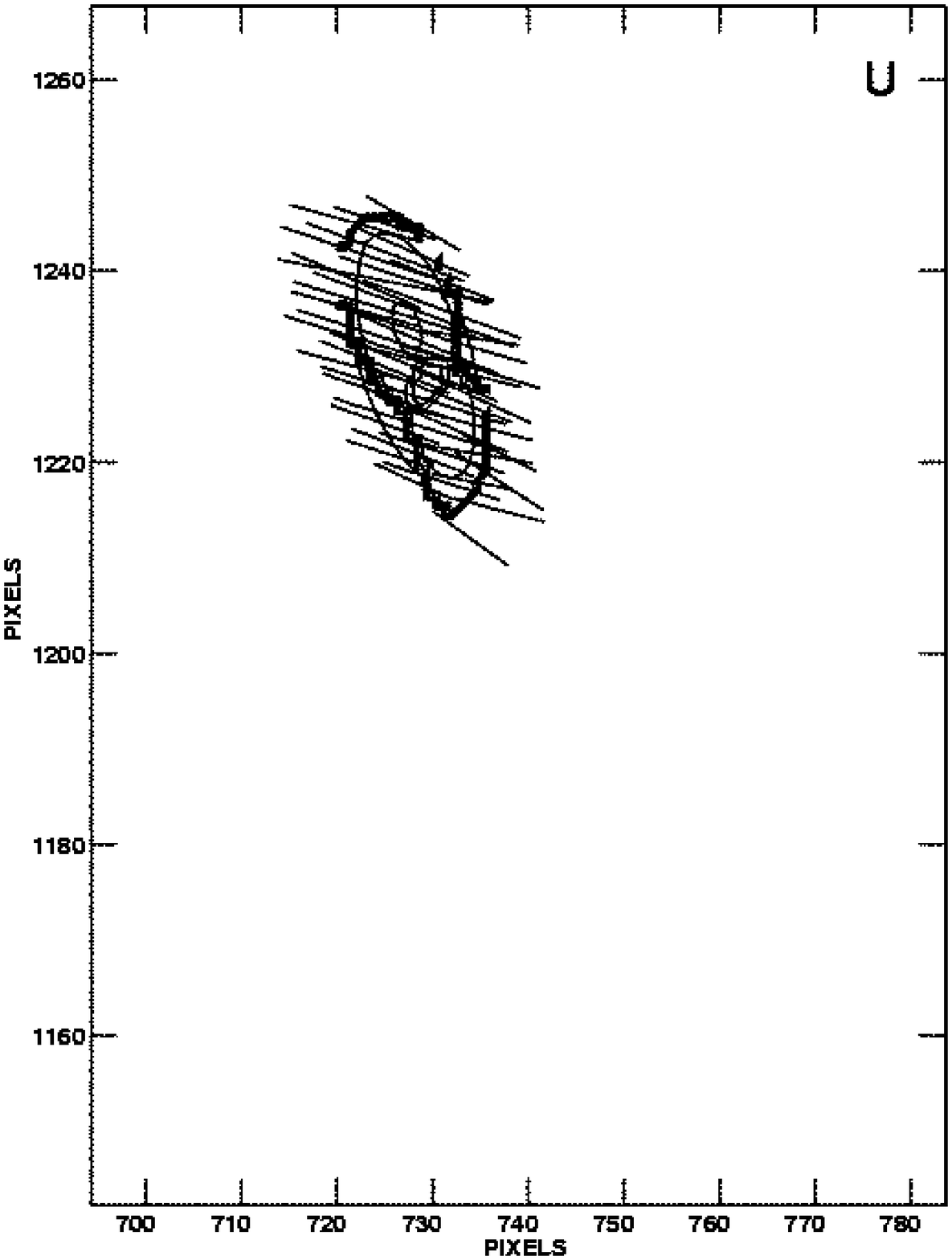} &
\includegraphics[width=34mm, height=48mm]{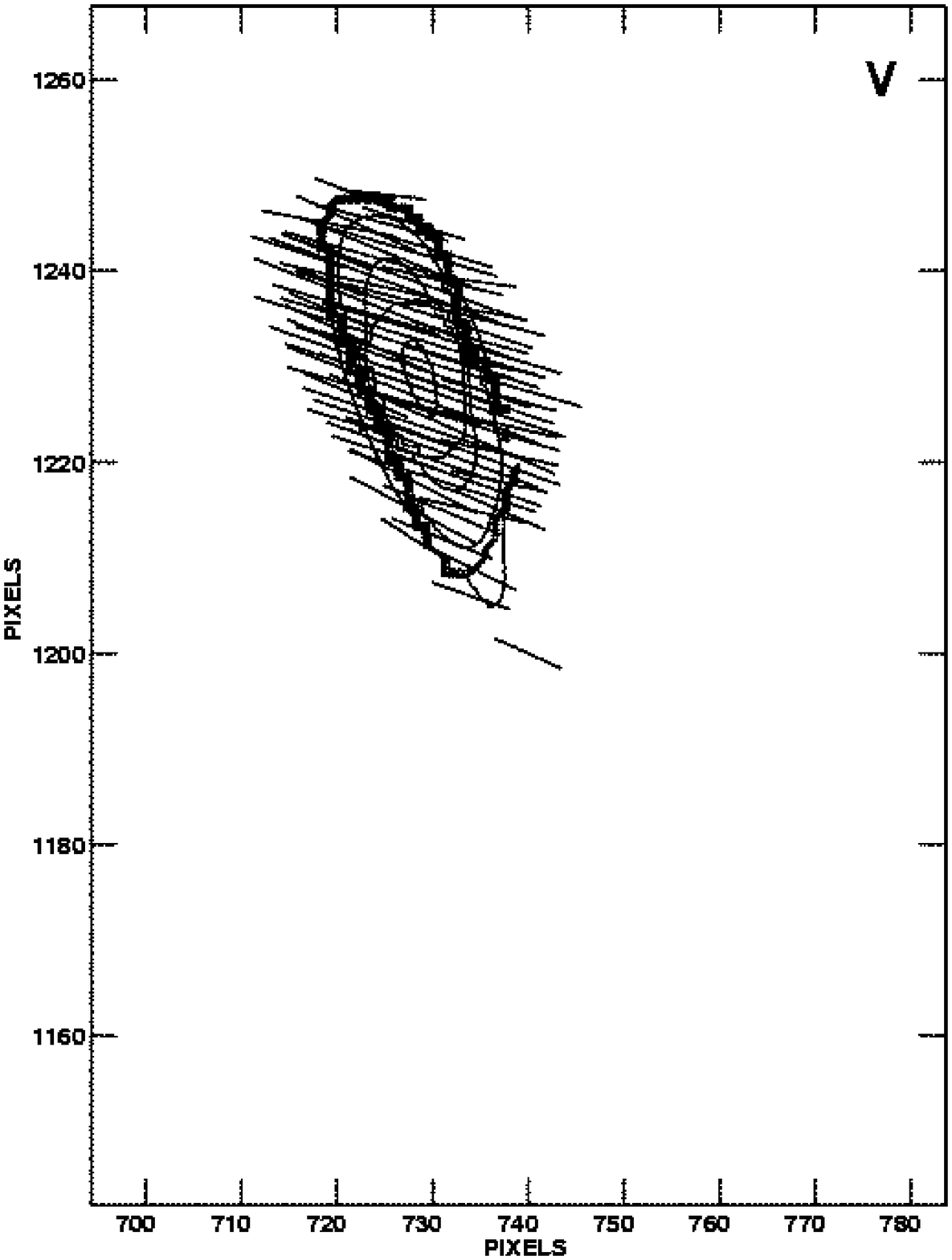} &
\includegraphics[width=34mm, height=48mm]{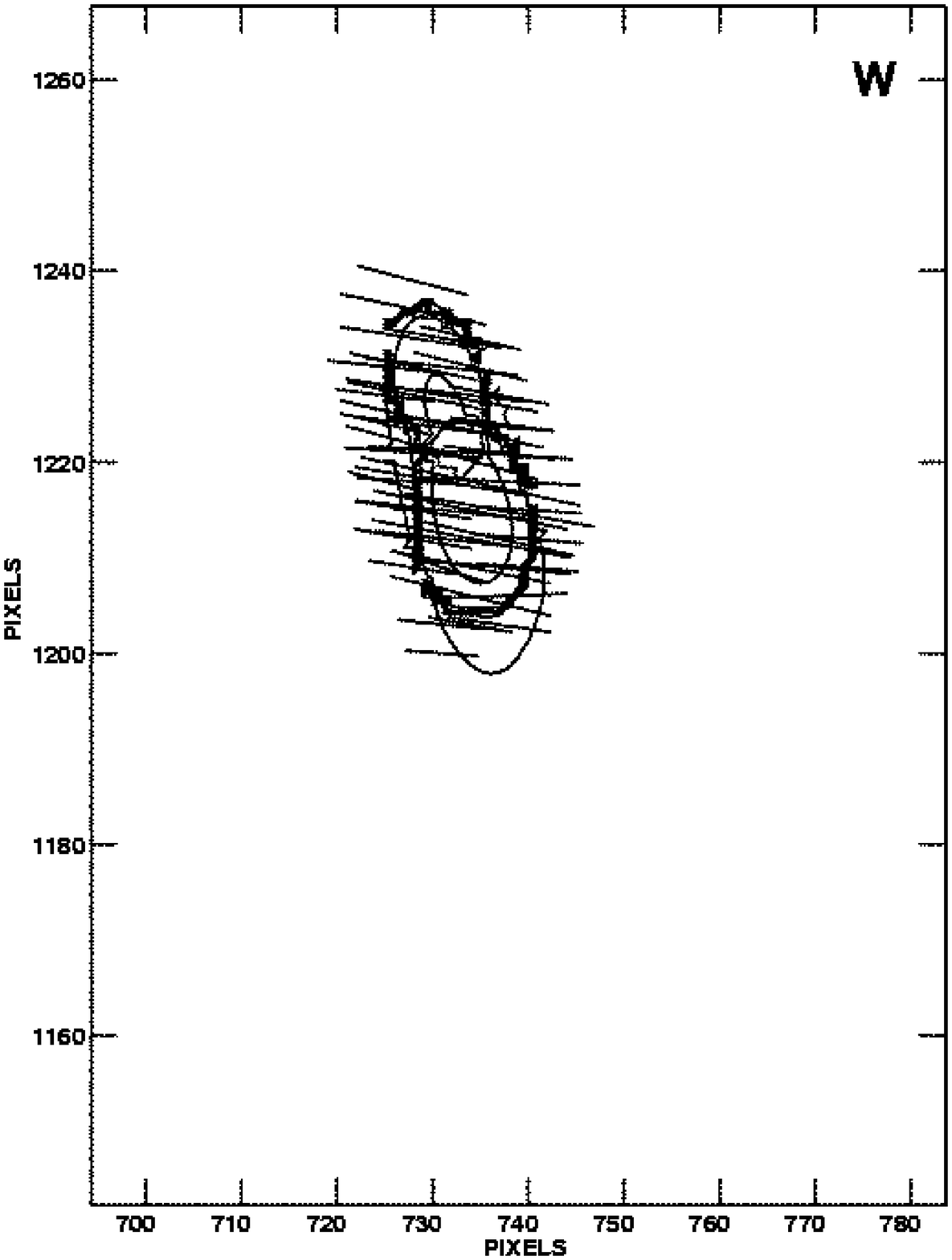} \\ 
\includegraphics[width=34mm, height=48mm]{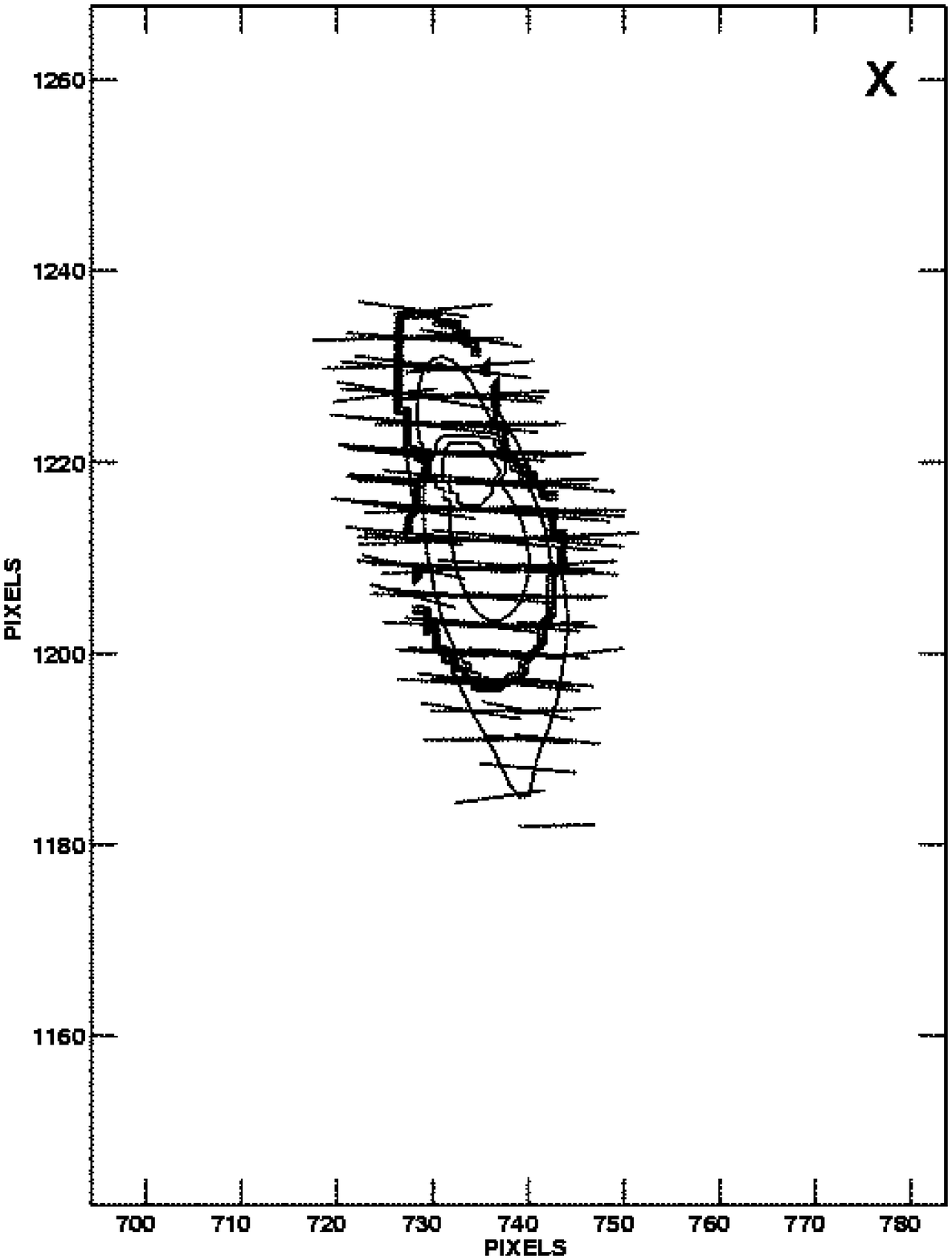} & 
\includegraphics[width=34mm, height=48mm]{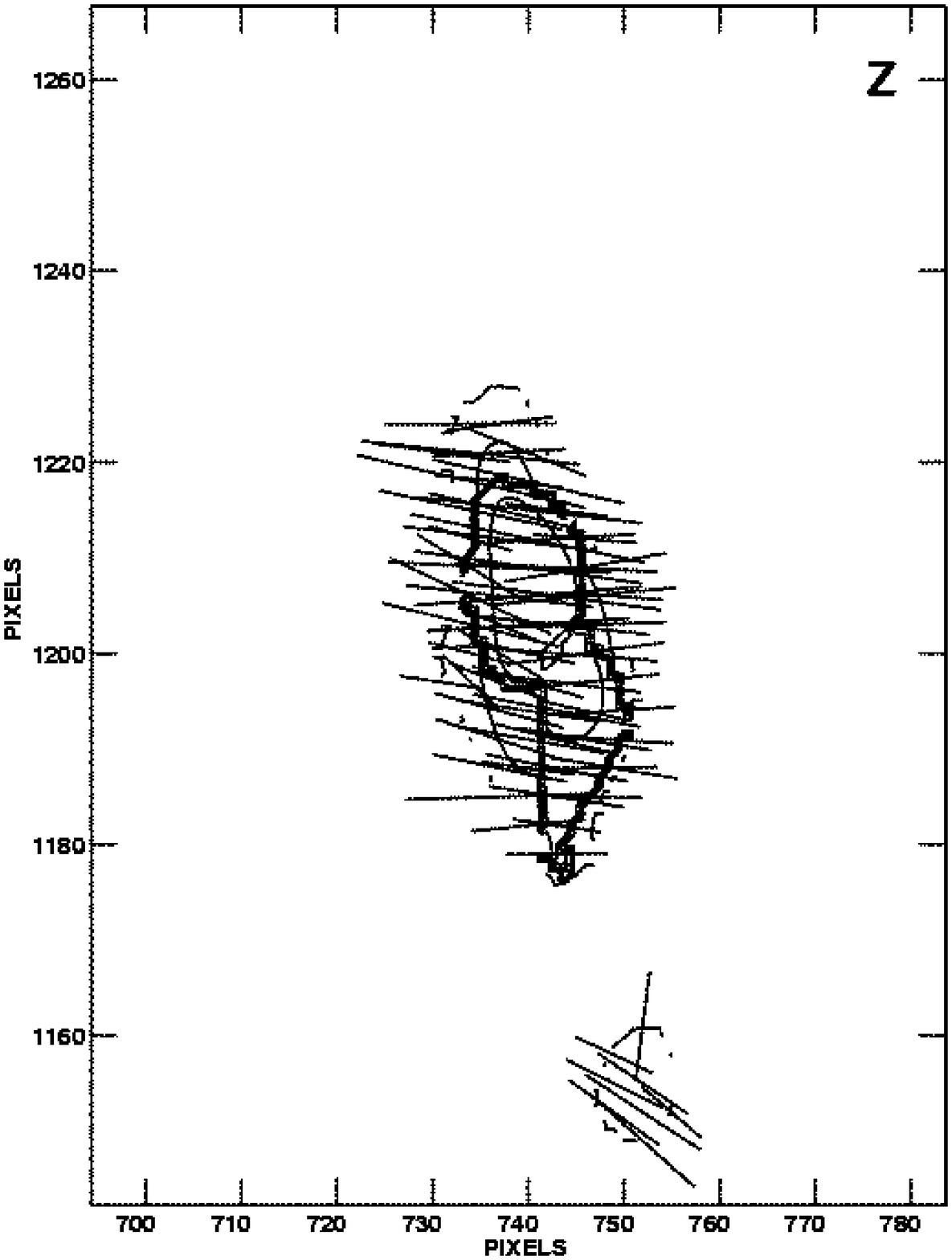} &
\includegraphics[width=34mm, height=48mm]{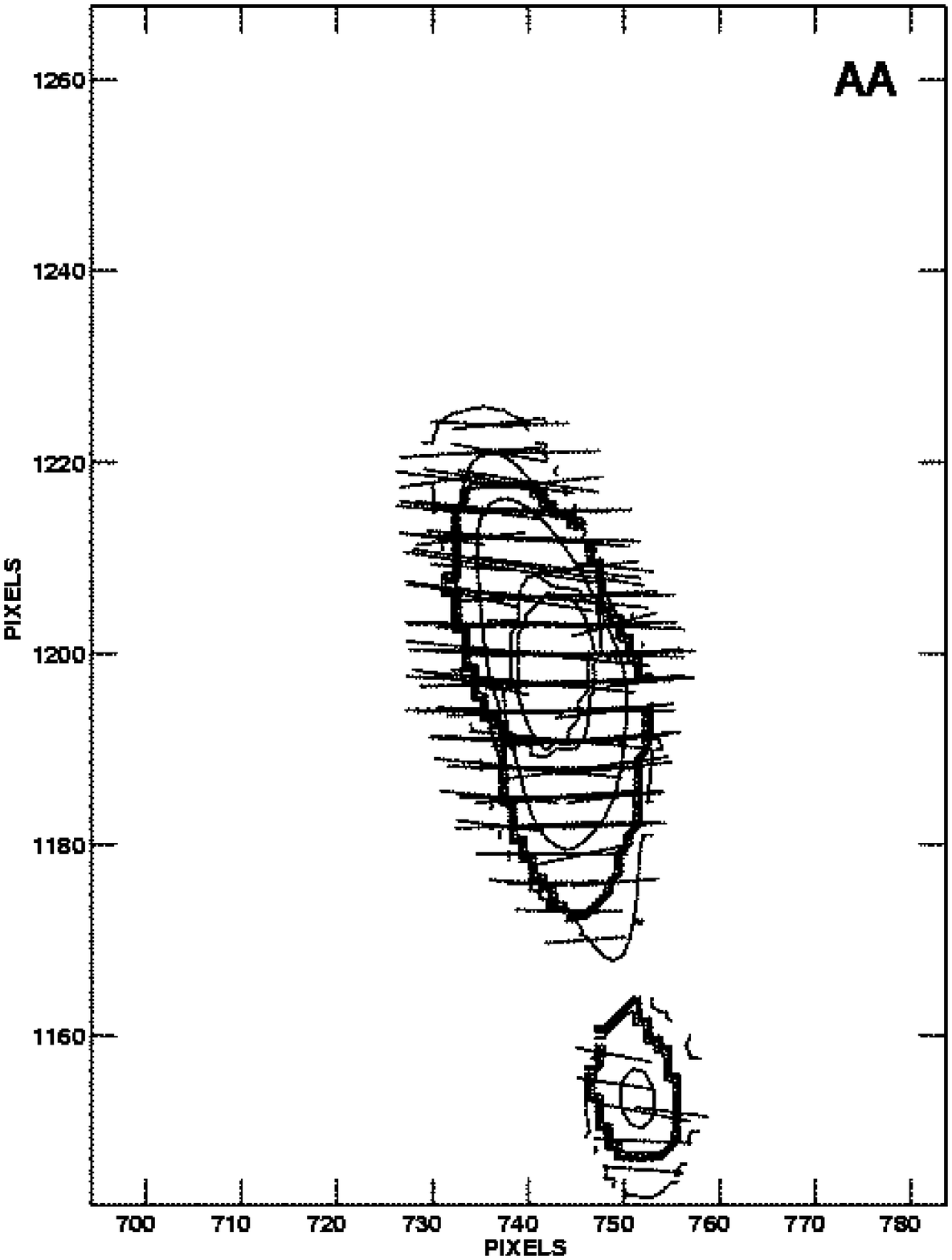} & 
\includegraphics[width=34mm, height=48mm]{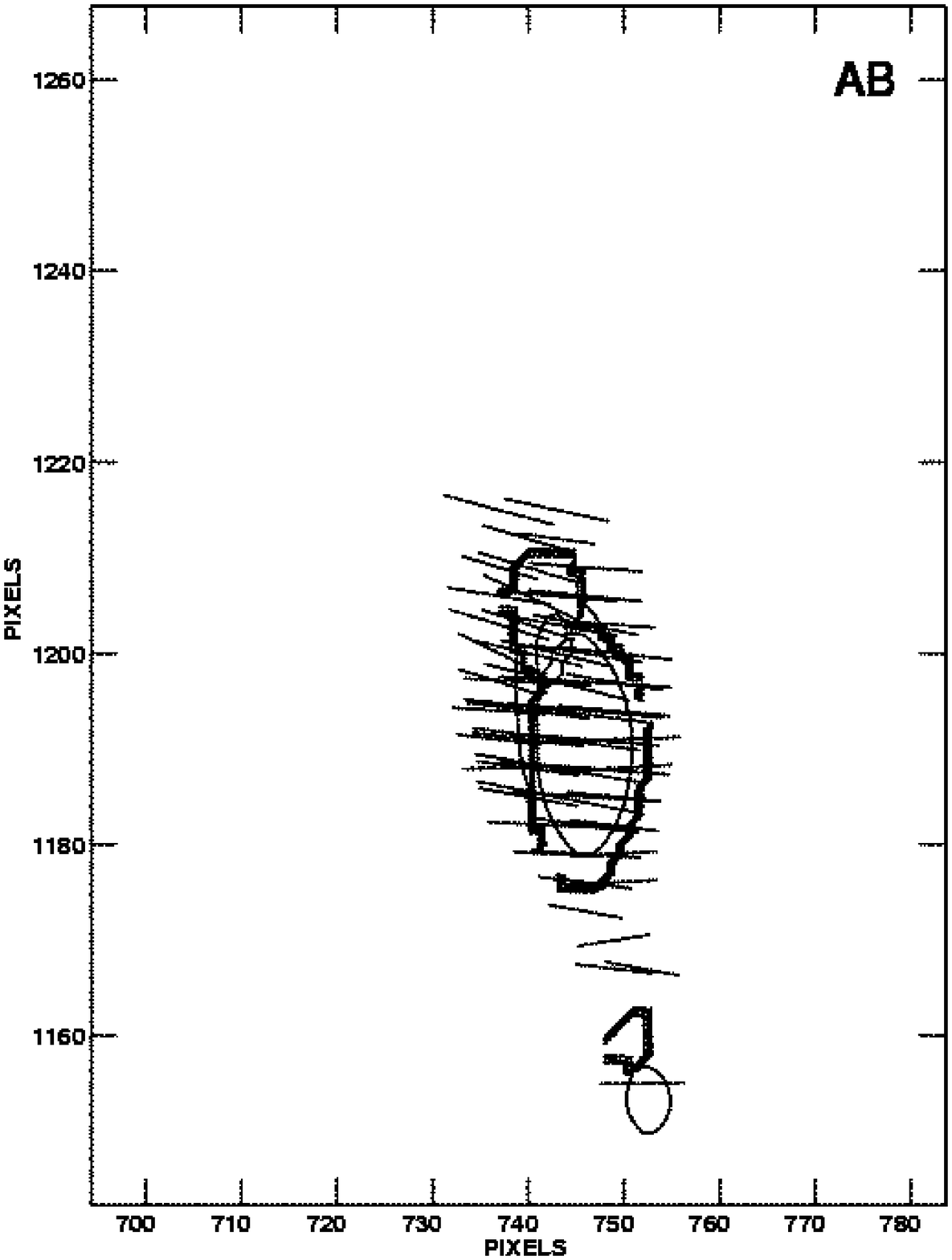} \\
\includegraphics[width=34mm, height=48mm]{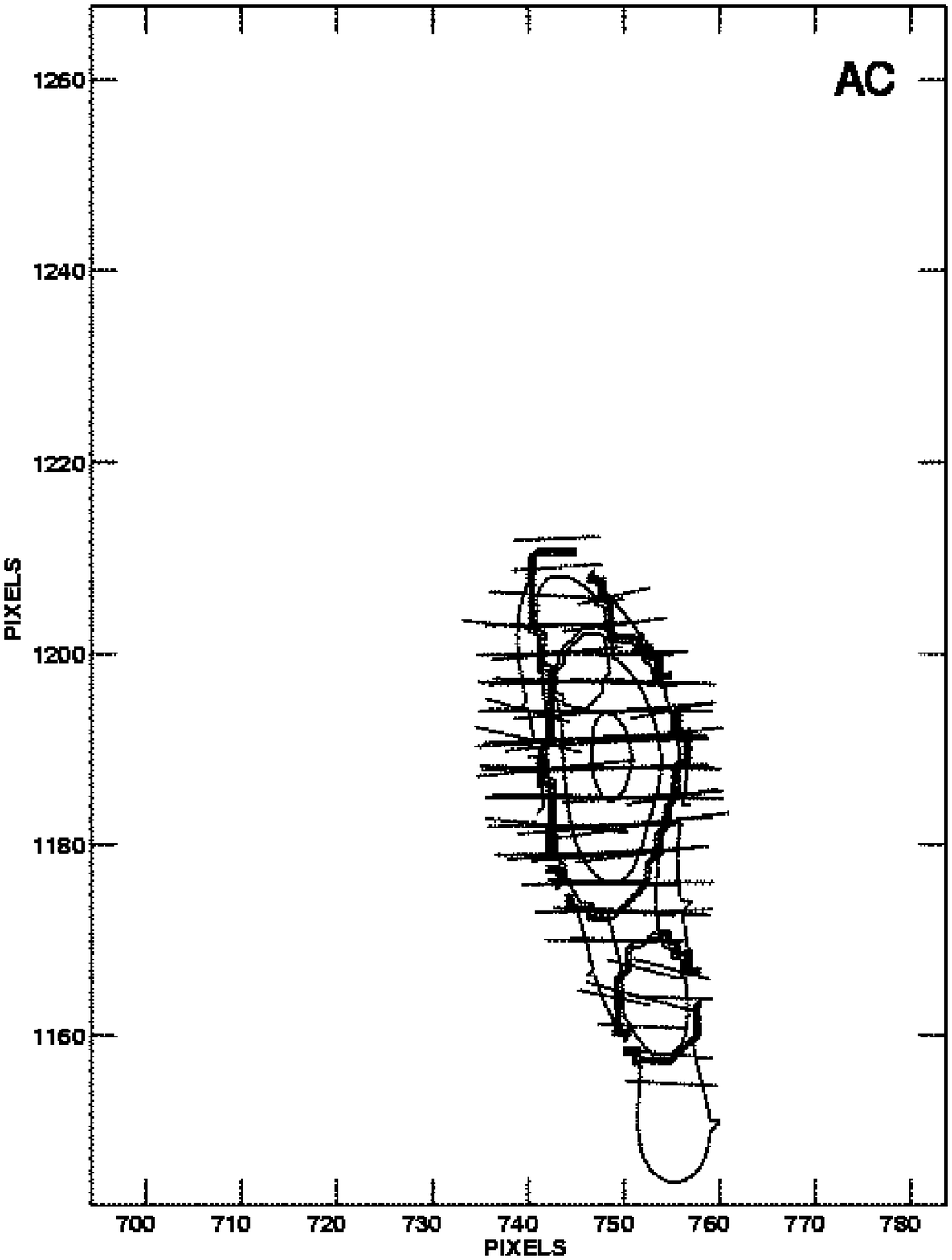} & 
\includegraphics[width=34mm, height=48mm]{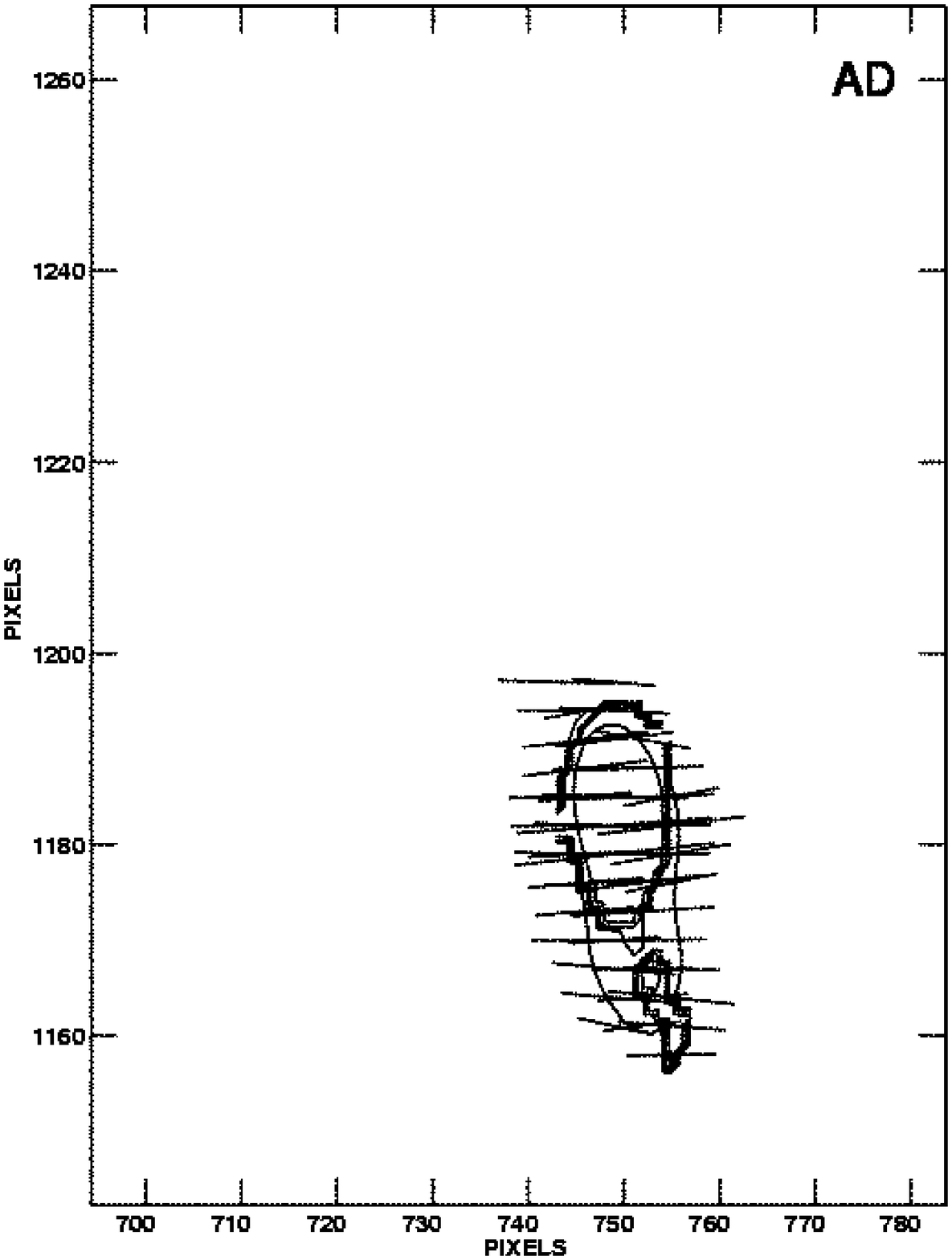} &
\includegraphics[width=34mm, height=48mm]{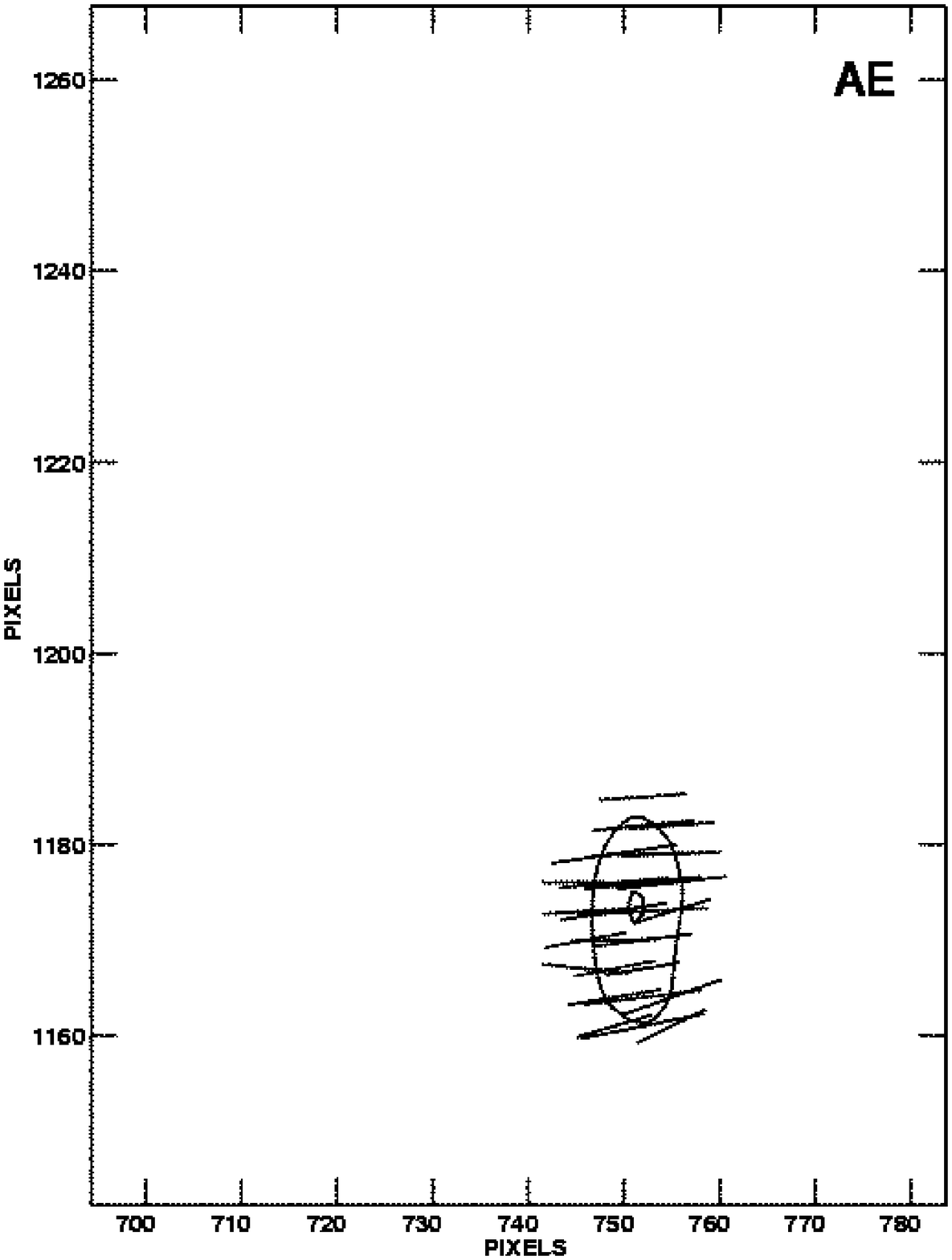} & 
\includegraphics[width=34mm, height=48mm]{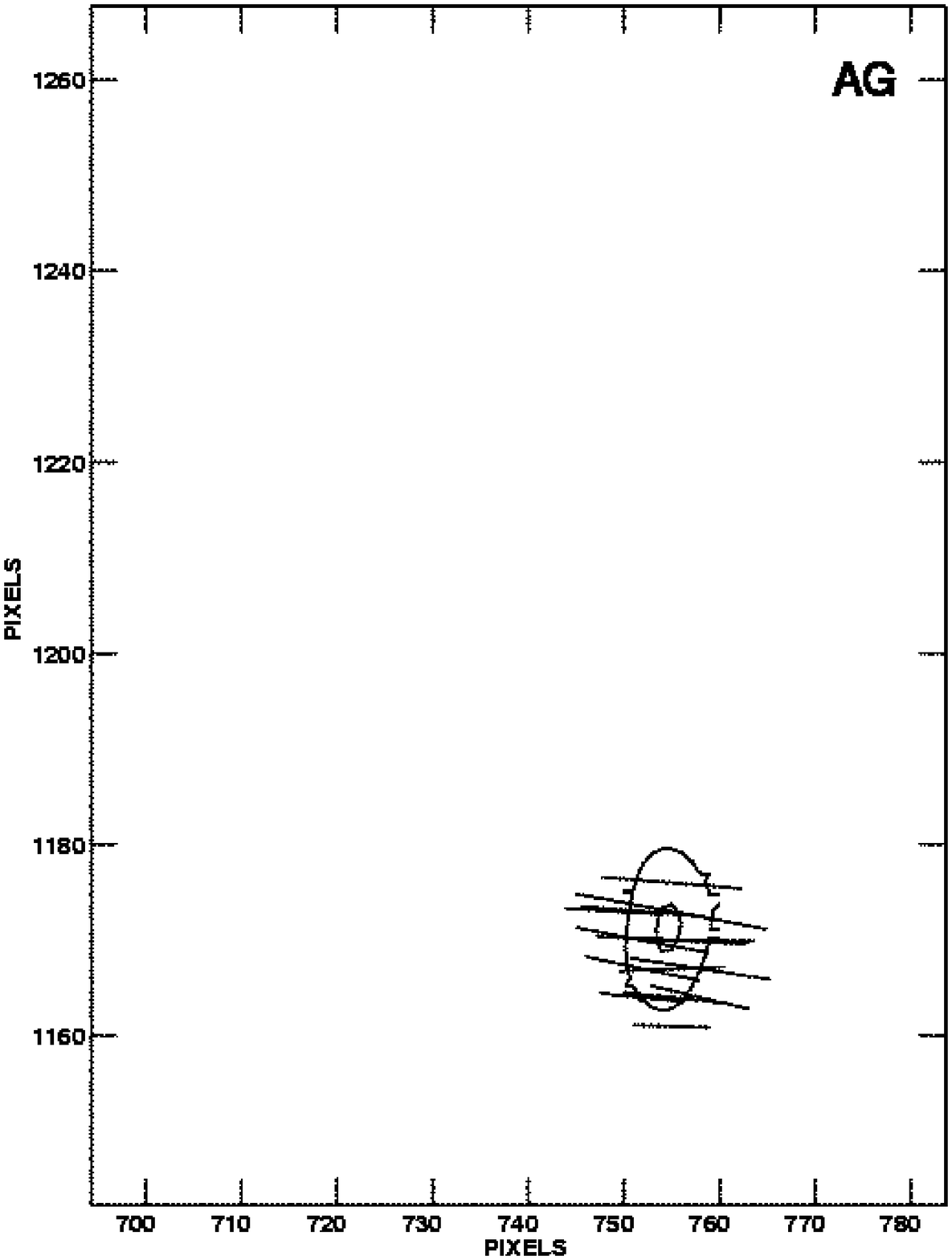} \\
\end{array}$ 

\caption{A time-series of expanded sub-images of the isolated SiO
maser component visible in the far north of images between epochs
\{S-AH\}, falling in toward the northern shell. The plot parameters
are the same as for Figure~\ref{fig-pcntr-3}, except for a more
finely-sampled set of enhanced contrast Stokes $I$ contour levels at
($10^{-1 + 0.1k}\ $ for k=1,30.)}

\label{fig-pcntr-n} 
\end{figure}

\begin{figure}[h] 
\advance\leftskip-1cm \advance\rightskip-1cm

\plotone{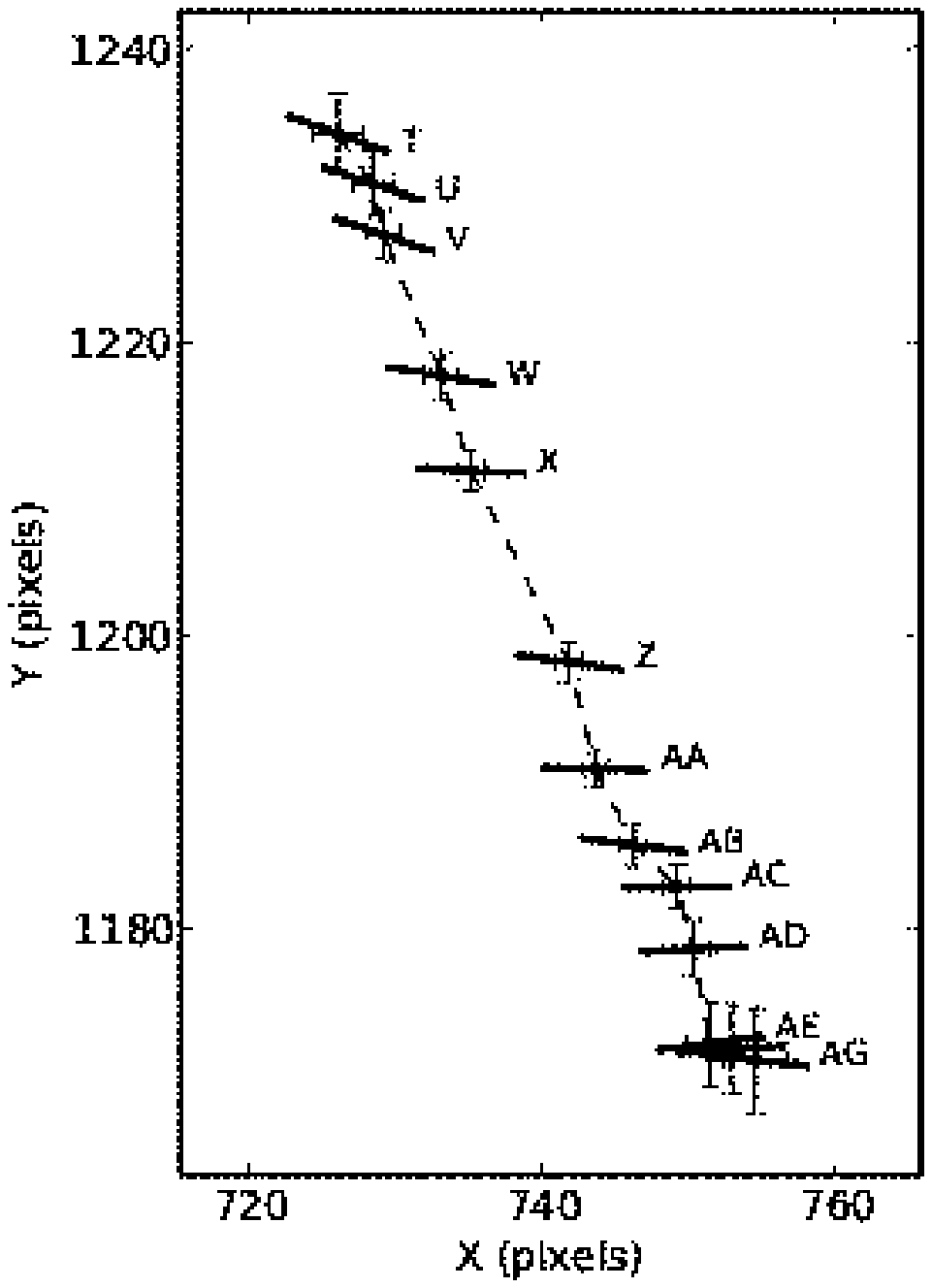}

\caption{A sequence of weighted-mean component positions for the
northern component trajectory shown in Figure
~\ref{fig-pcntr-n}. Error bars in x and y are shown for each
weighted-mean position, and the EVPA is drawn at each epoch as a bold
vector, of uniform, constant length over epoch.}

\label{fig-traj-n} 
\end{figure}

\begin{figure}[h] 
\advance\leftskip-1cm \advance\rightskip-1cm

\plotone{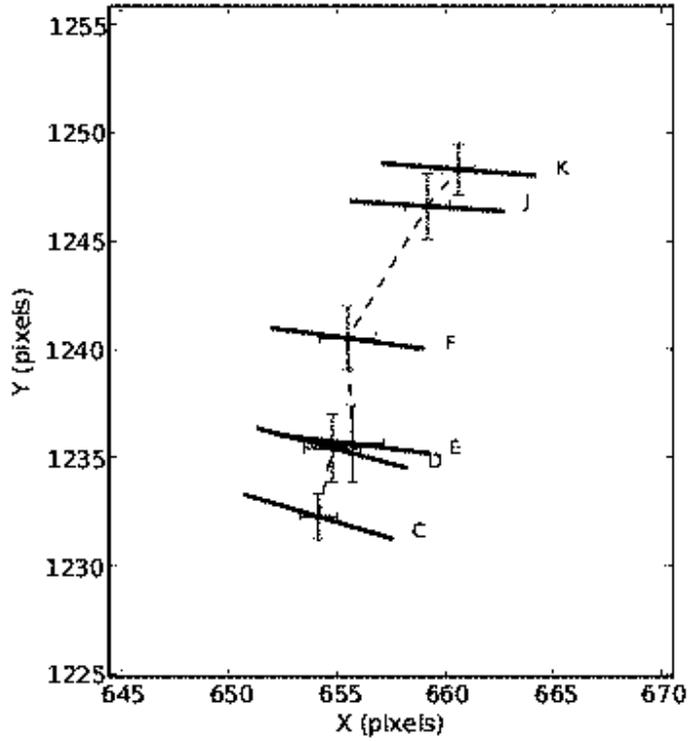}

\caption{A sequence of weighted-mean component positions for the
northern-eastern component discussed in the text. Error bars in x and
y are shown for each weighted-mean position, and the EVPA is drawn at
each epoch as a bold vector, of uniform, constant length over epoch.}

\label{fig-traj-ne} 
\end{figure}

\begin{figure}[h] 
\advance\leftskip-1cm \advance\rightskip-1cm

\plotone{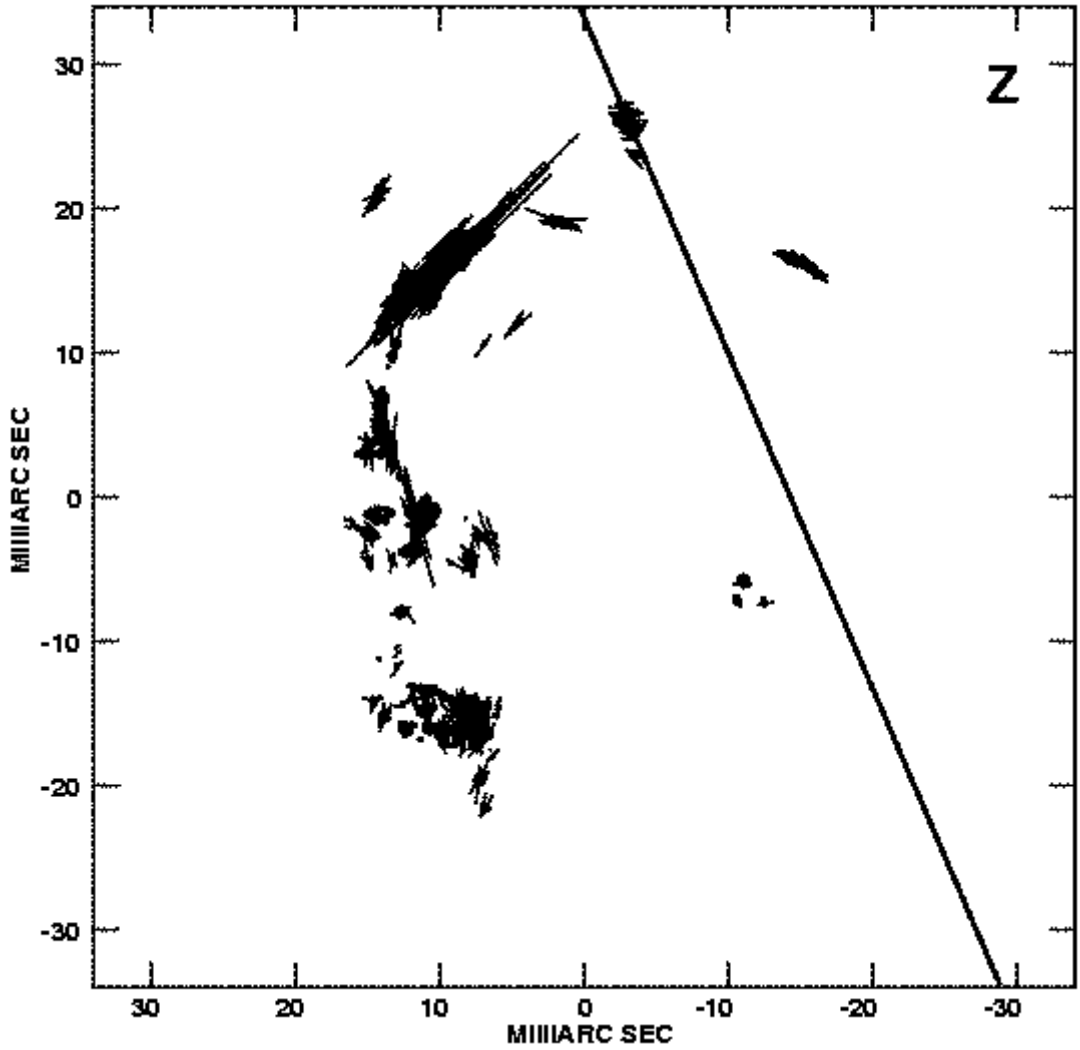}

\caption{The best-fit linear projected trajectory to the northern
component positions in Figure ~\ref{fig-traj-n}, extrapolated across
the full diameter of the SiO maser shell at epoch \{Z\}, here chosen
as the mid-point epoch of the infalling component trajectory.}

\label{fig-nr-n} 
\end{figure}

In Figure~\ref{fig-pcntr-n} we show this analysis for a prominent
infalling component at the center north of the projected shell, that
is visible over epochs \{S-AH\}. In Figure~\ref{fig-pcntr-n}, the
linear polarization structure of this northern component is shown at
the twelve epochs \{T,U,V,W,X,Z,AA,AB,AC,AD,AE,AG\} spaced along the
trajectory over time, using the same composite representation of
linearly-polarized emission as used in Figures~\ref{fig-pcntr-3} to
\ref{fig-pcntr-9}, but with a more finely-sampled total intensity
contour scale, described in further detail in the figure caption. The
outermost epochs \{S,AH\} of the component trajectory are omitted due
to the low SNR when the component appears and disappears. In common
with Figures~\ref{fig-pcntr-3} to \ref{fig-pcntr-9}, each panel in
Figure~\ref{fig-pcntr-n} is a zeroth-moment image over frequency; the
pronounced component elongation visible in these zeroth-moment
component images at several epochs along the trajectory is most
readily explained in terms of a velocity gradient along the elongation
major axis. These velocity gradients are most directly assumed to
arise from linear acceleration along a three-dimensional component
trajectory inclined to the line of sight to the observer. The
weighted-mean component positions, with associated error bars, are
plotted in Figure~\ref{fig-traj-n} for each epoch along the
trajectory. At each component position in this figure, a bold vector
of uniform length (over epoch) is drawn at the orientation of the
integrated component EVPA. A least-squares fit to the component
positions, incorporating x- and y-errors, was performed using software
described by \citet{weiner06}. The fitted mean projected trajectory is
extrapolated in Figure~\ref{fig-nr-n} across the full diameter of the
SiO maser shell at epoch \{Z\}, here chosen as an epoch near the
mid-point of the northern component trajectory shown in
Figure~\ref{fig-traj-n}. Figure~\ref{fig-nr-n} shows that the
projected component trajectory is non-radial with respect to the
likely stellar position relative to the SiO shell distribution; this
qualitative assertion remains valid within the statistical errors of
the projected linear fit parameters. If dynamically-significant gas
pressure gradients are not present, then purely gravitational infall
would be radial toward the central star. We note that this component
has an EVPA oriented predominantly orthogonal to the projected proper
motion vector. As noted above, the location of this component is at a
significantly larger radius than the inner shell boundary, and
therefore minimizes the possibility of strong anisotropic pumping
effects. Over parts of the earlier epoch range \{C-N\}, a component is
visible approximately 3.25 mas to the east of the northern component
considered here. This north-eastern component is at a significantly
lower SNR, and is not clearly detected in linear polarization at
epochs \{G-I,M-N\} as a result. Given the proximinity to the northern
component, however, we plot the counterpart of Figure~\ref{fig-traj-n}
for this component as Figure~\ref{fig-traj-ne}. This shows the
north-eastern component moving outward over these epochs with an EVPA again
predominantly orthogonal to the projected proper motion vector.

\begin{figure}[h] 
\advance\leftskip-1cm
\advance\rightskip-1cm
$\begin{array}{cccc} 
\includegraphics[width=34mm, height=48mm]{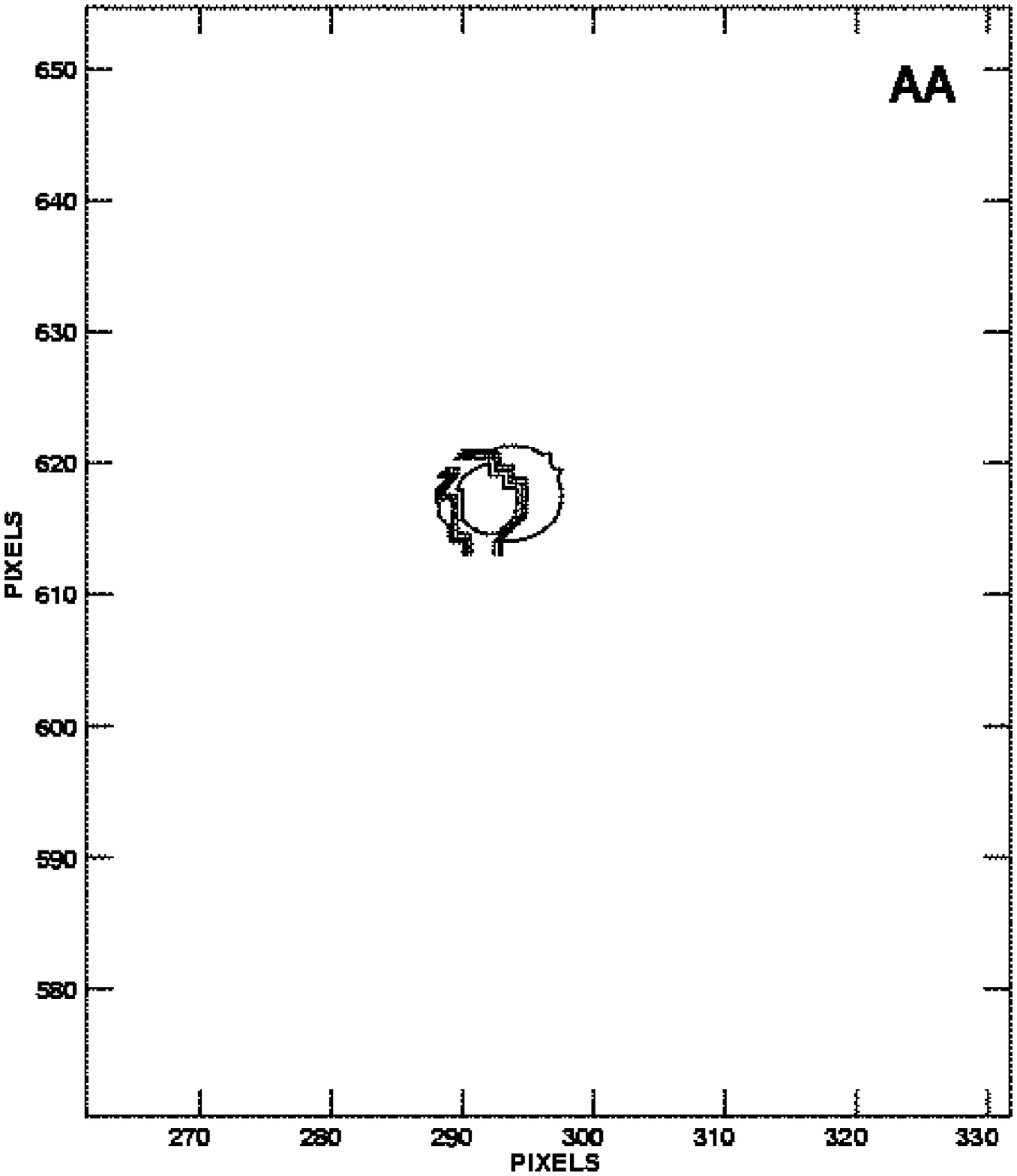} & 
\includegraphics[width=34mm, height=48mm]{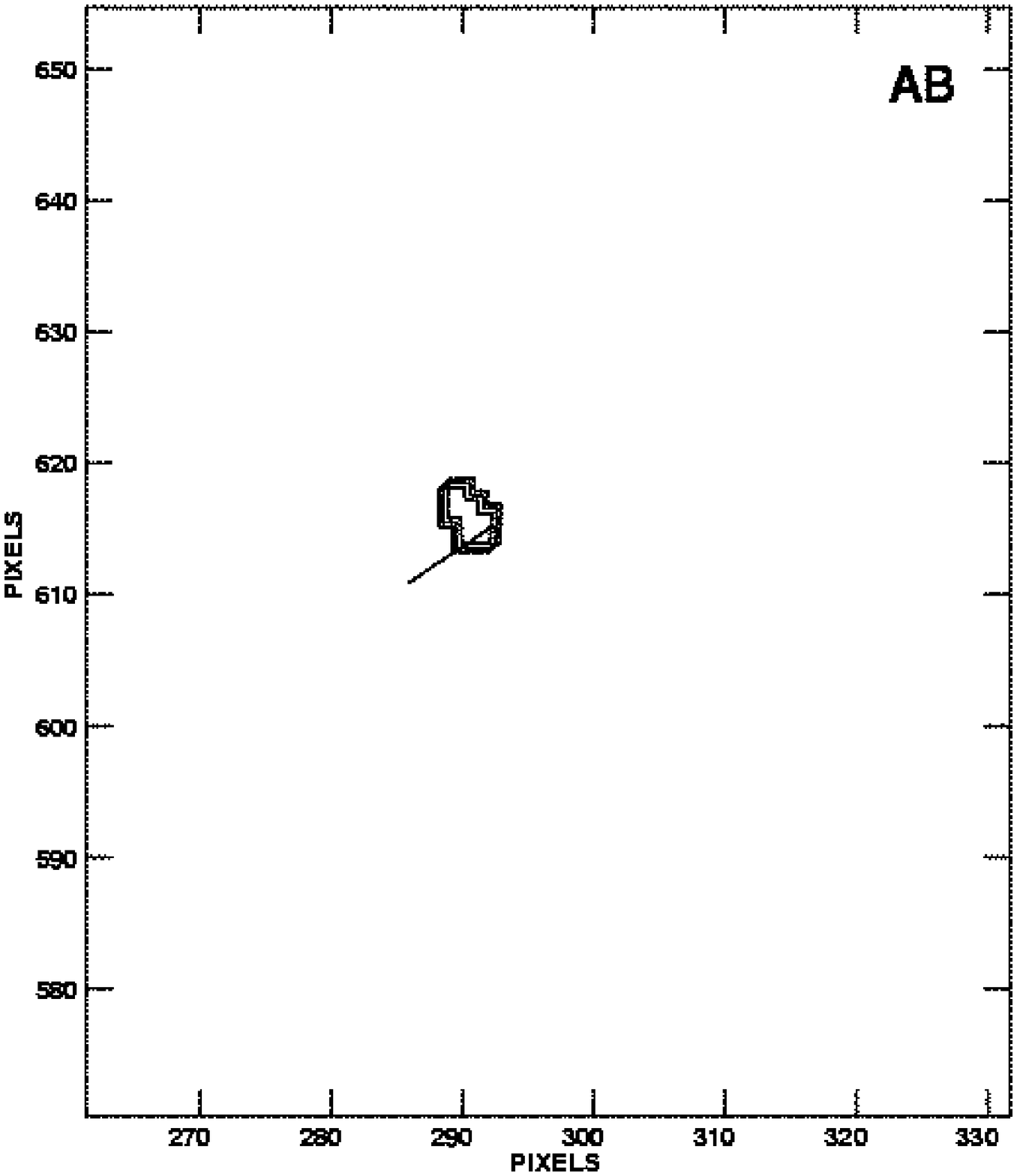} &
\includegraphics[width=34mm, height=48mm]{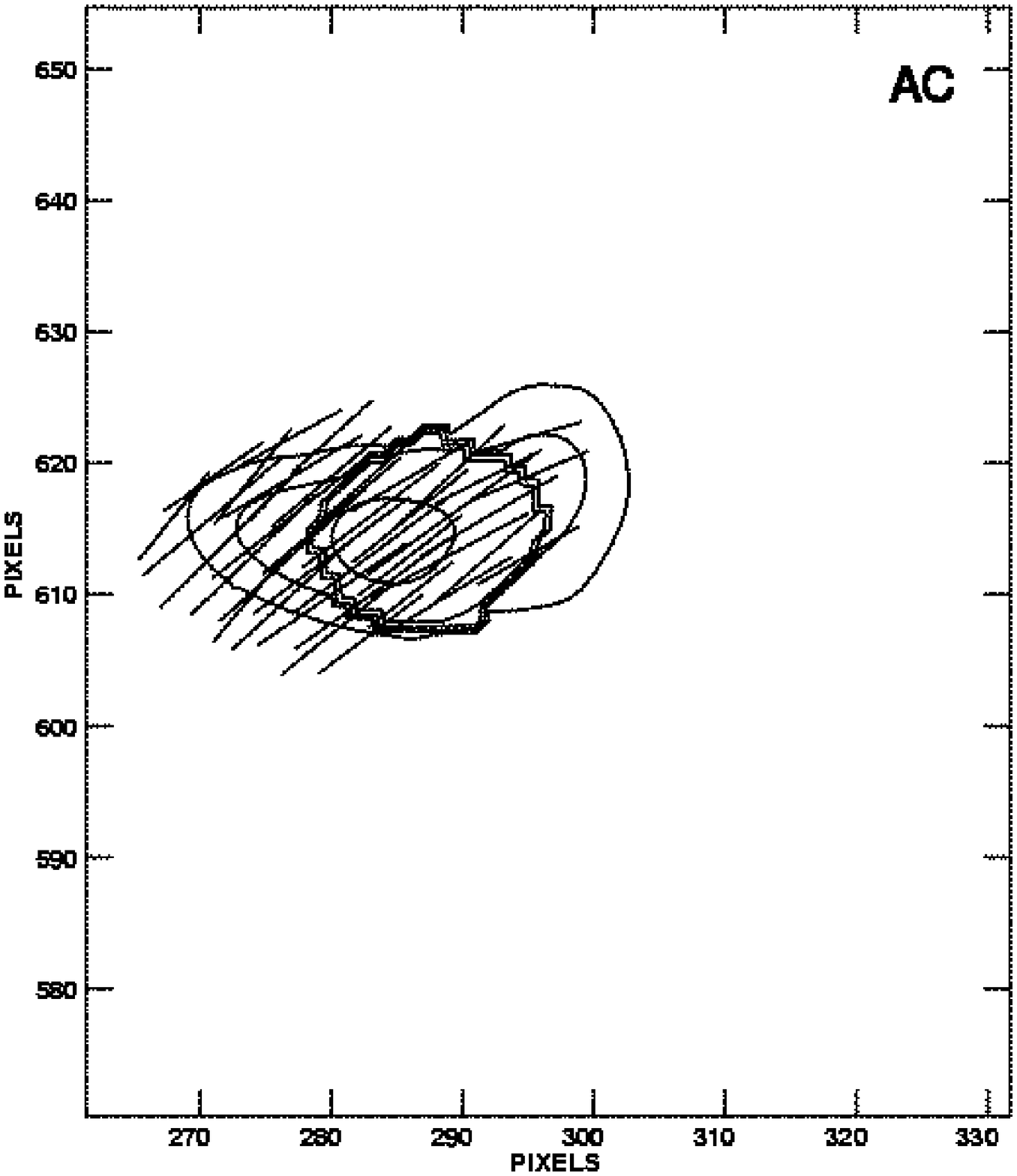} &
\includegraphics[width=34mm, height=48mm]{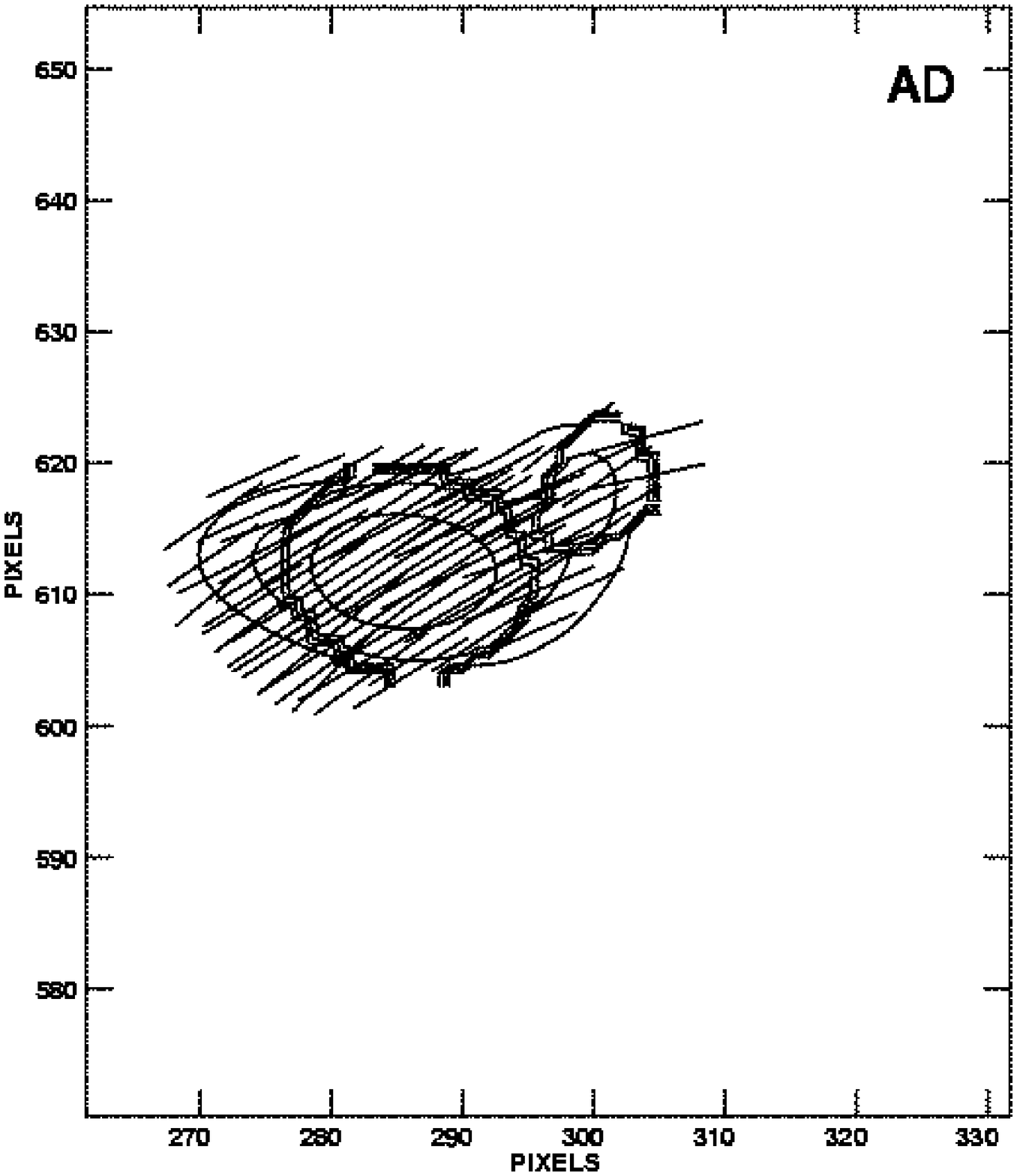} \\ 
\includegraphics[width=34mm, height=48mm]{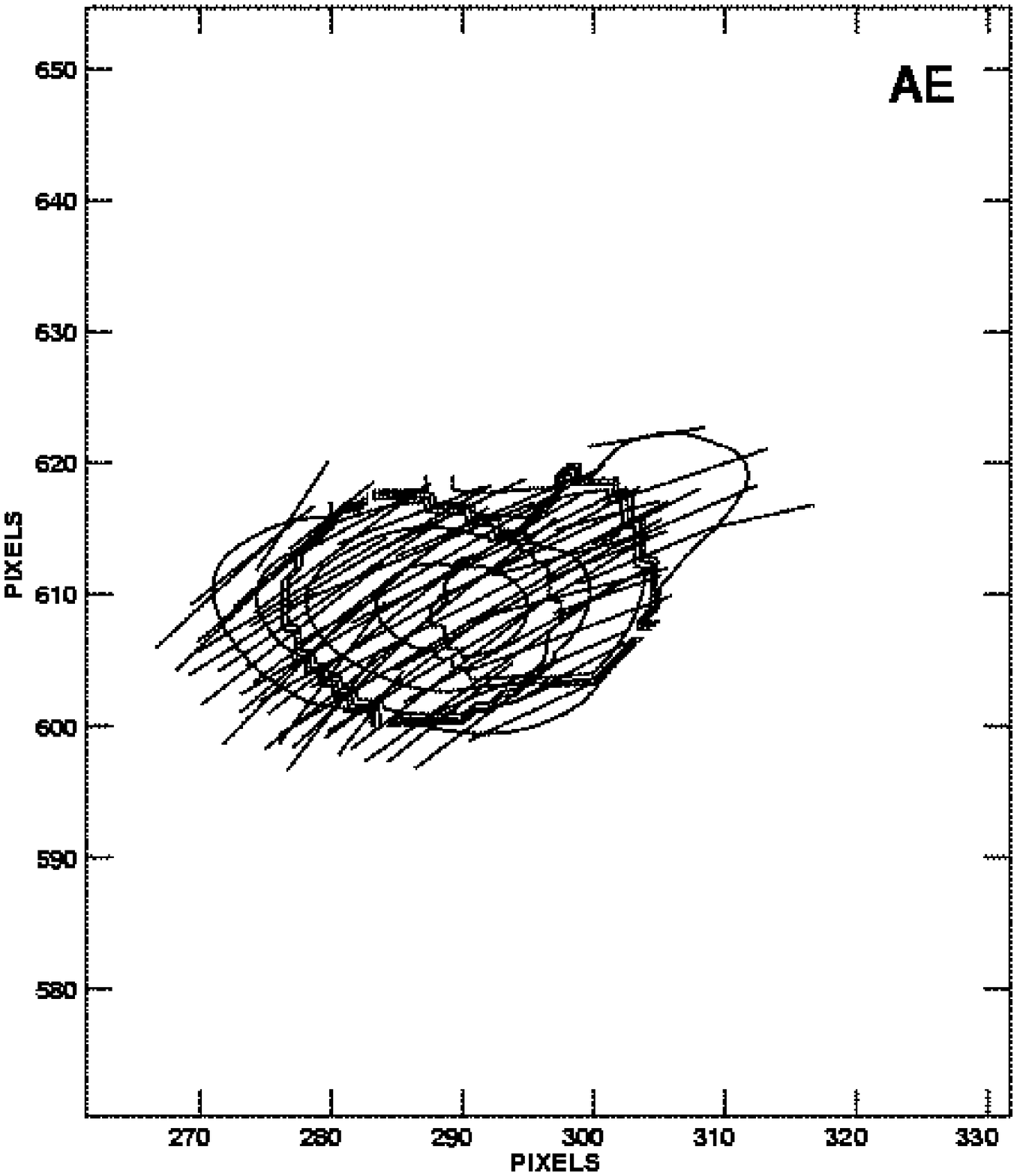} & 
\includegraphics[width=34mm, height=48mm]{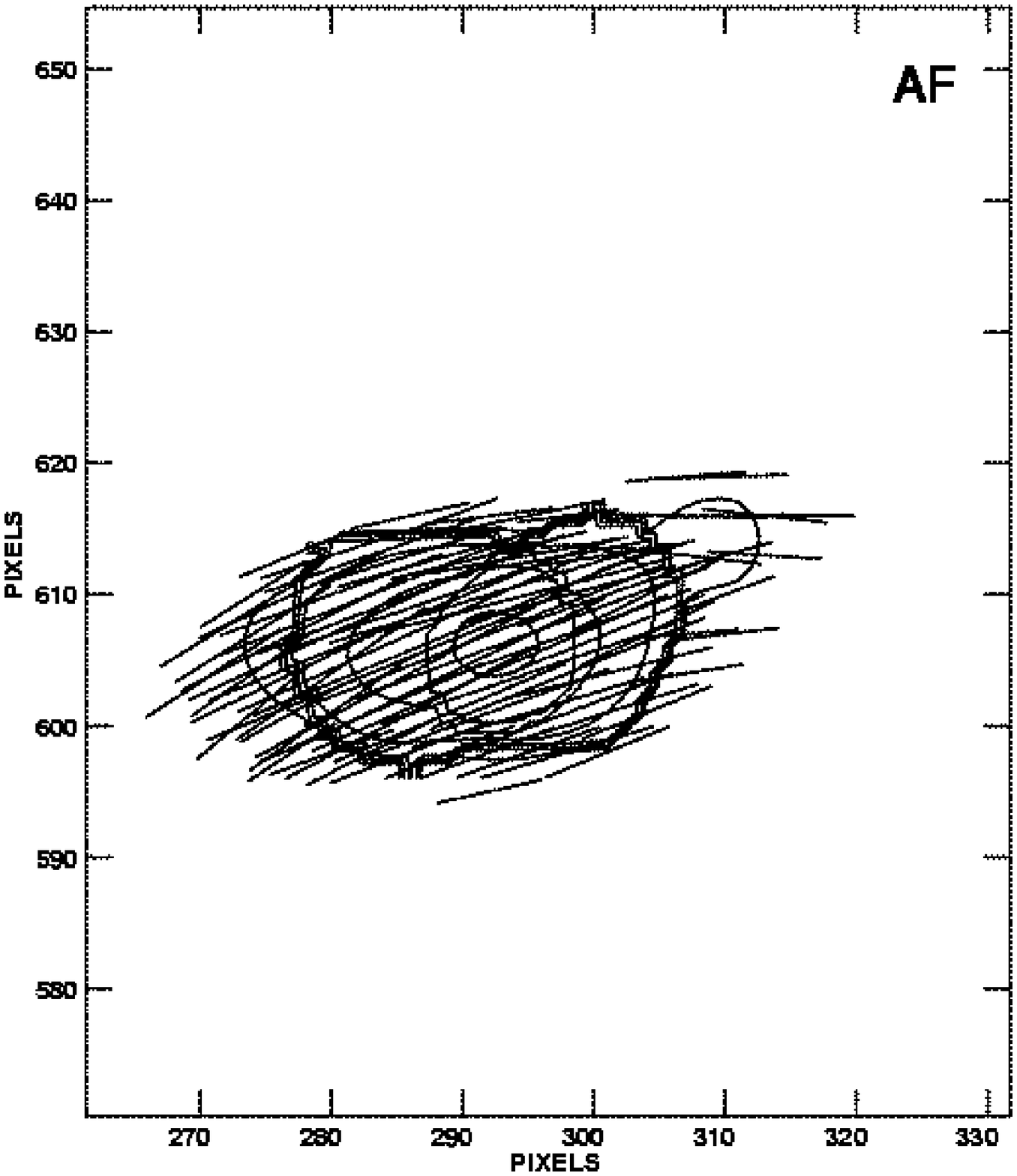} &
\includegraphics[width=34mm, height=48mm]{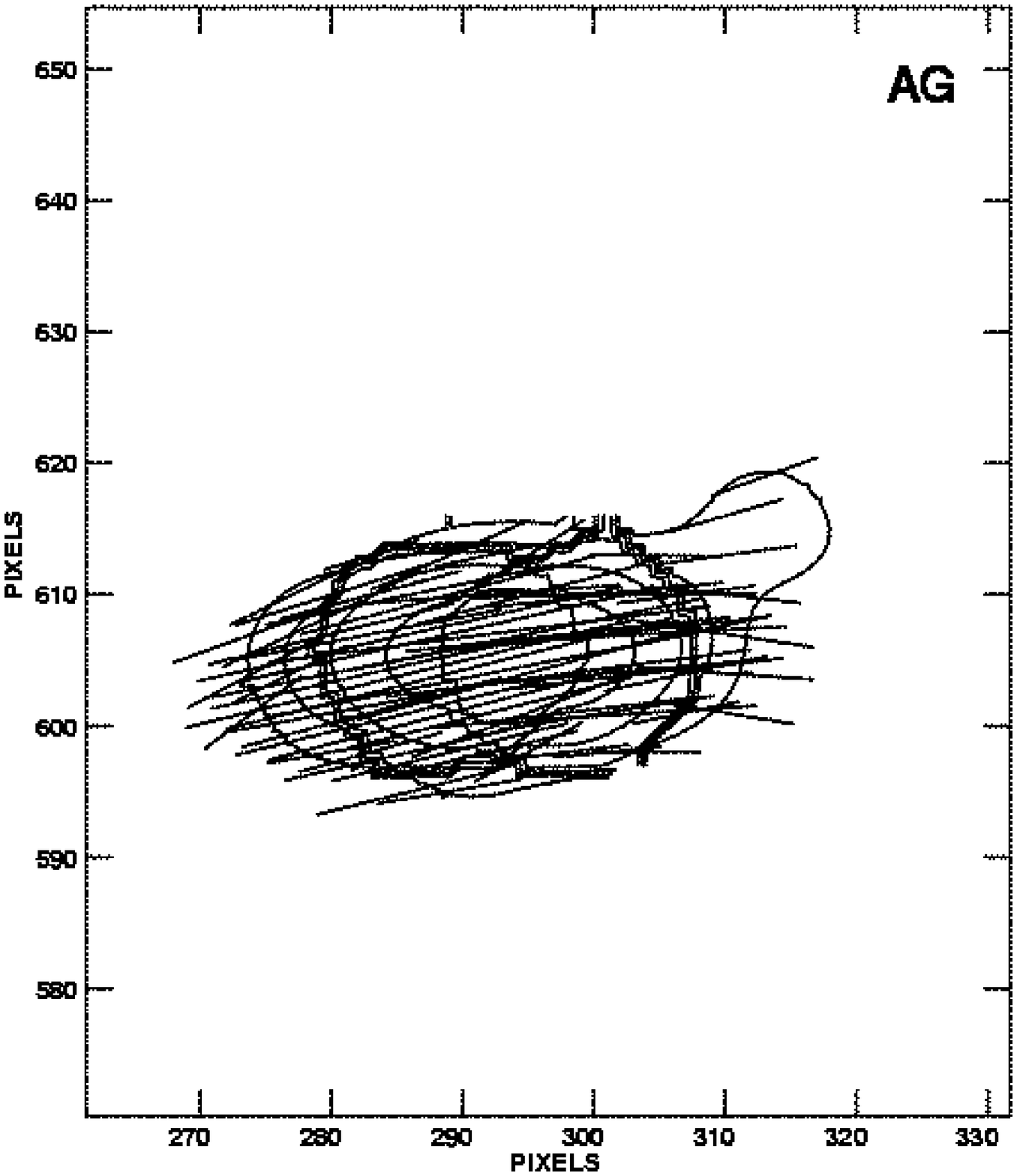} & 
\includegraphics[width=34mm, height=48mm]{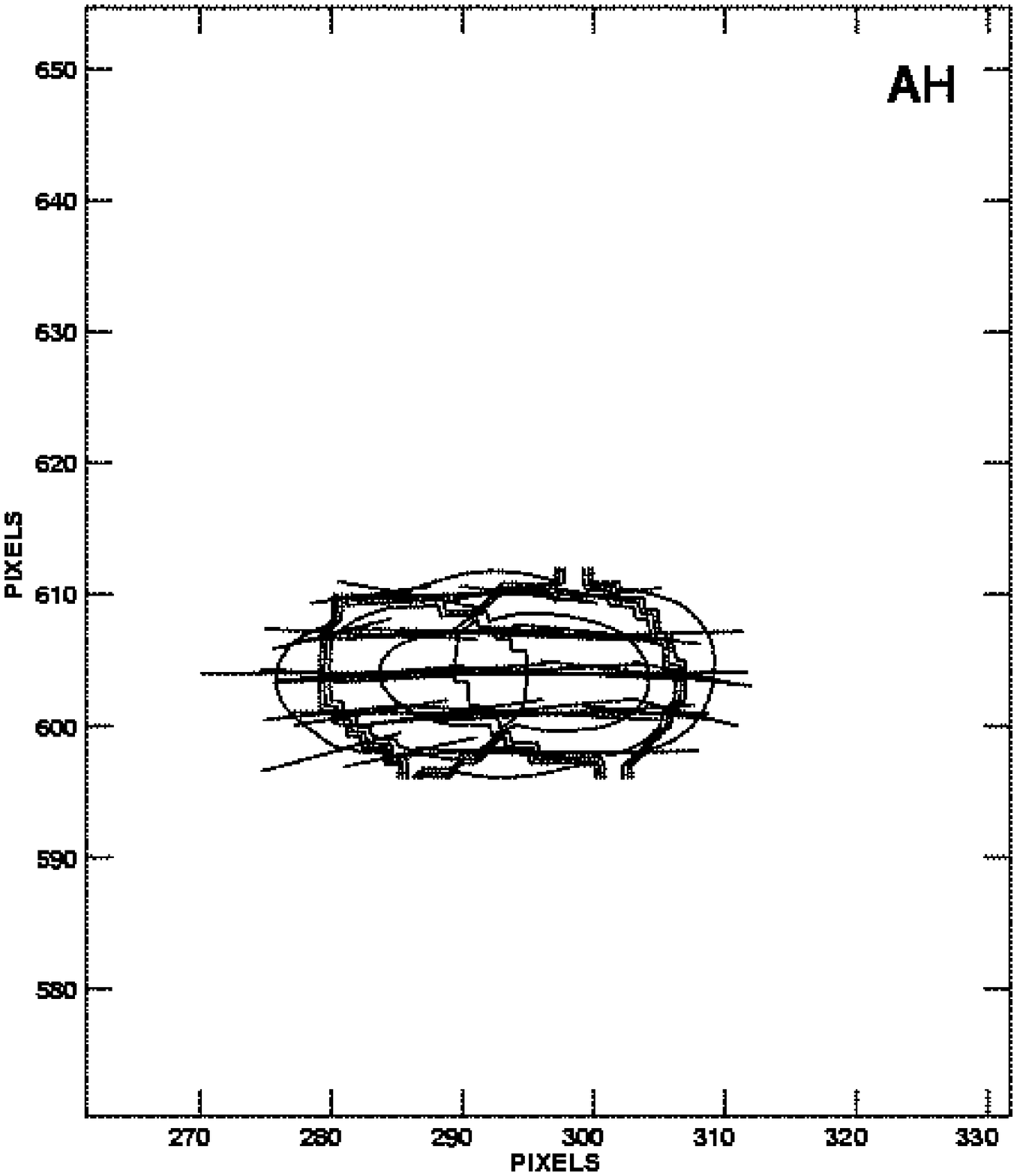} \\
\includegraphics[width=34mm, height=48mm]{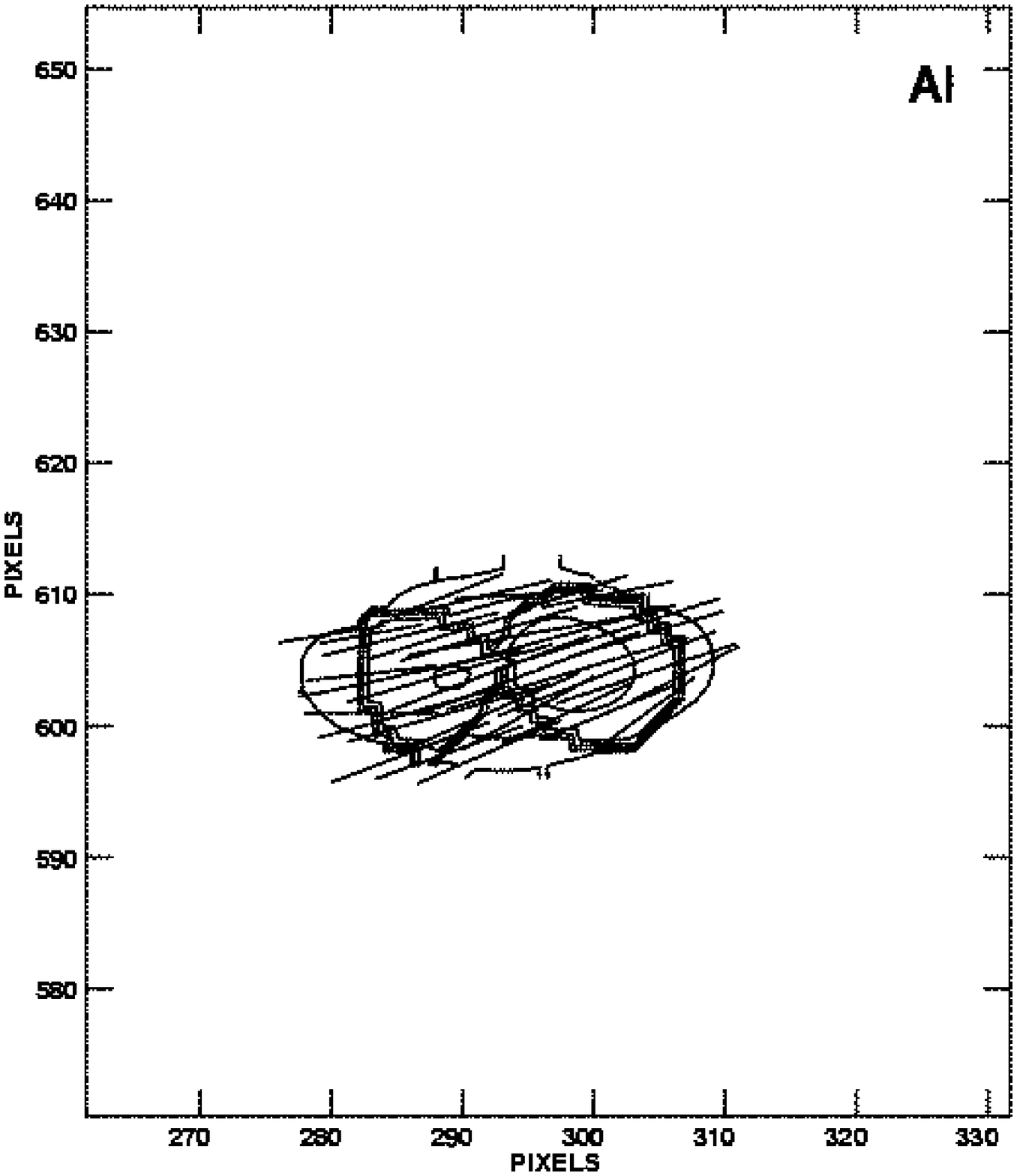} & 
\includegraphics[width=34mm, height=48mm]{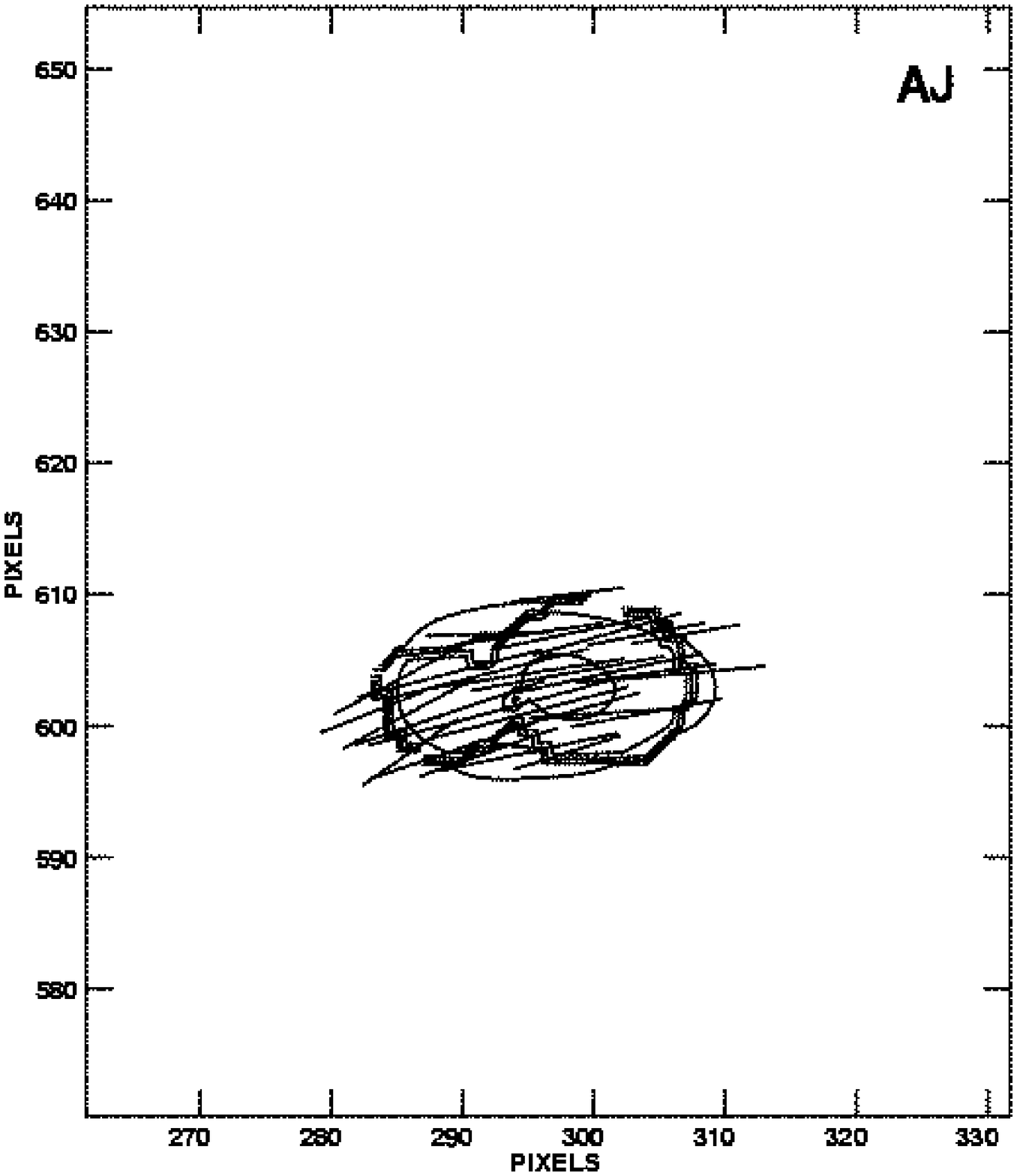} &
\includegraphics[width=34mm, height=48mm]{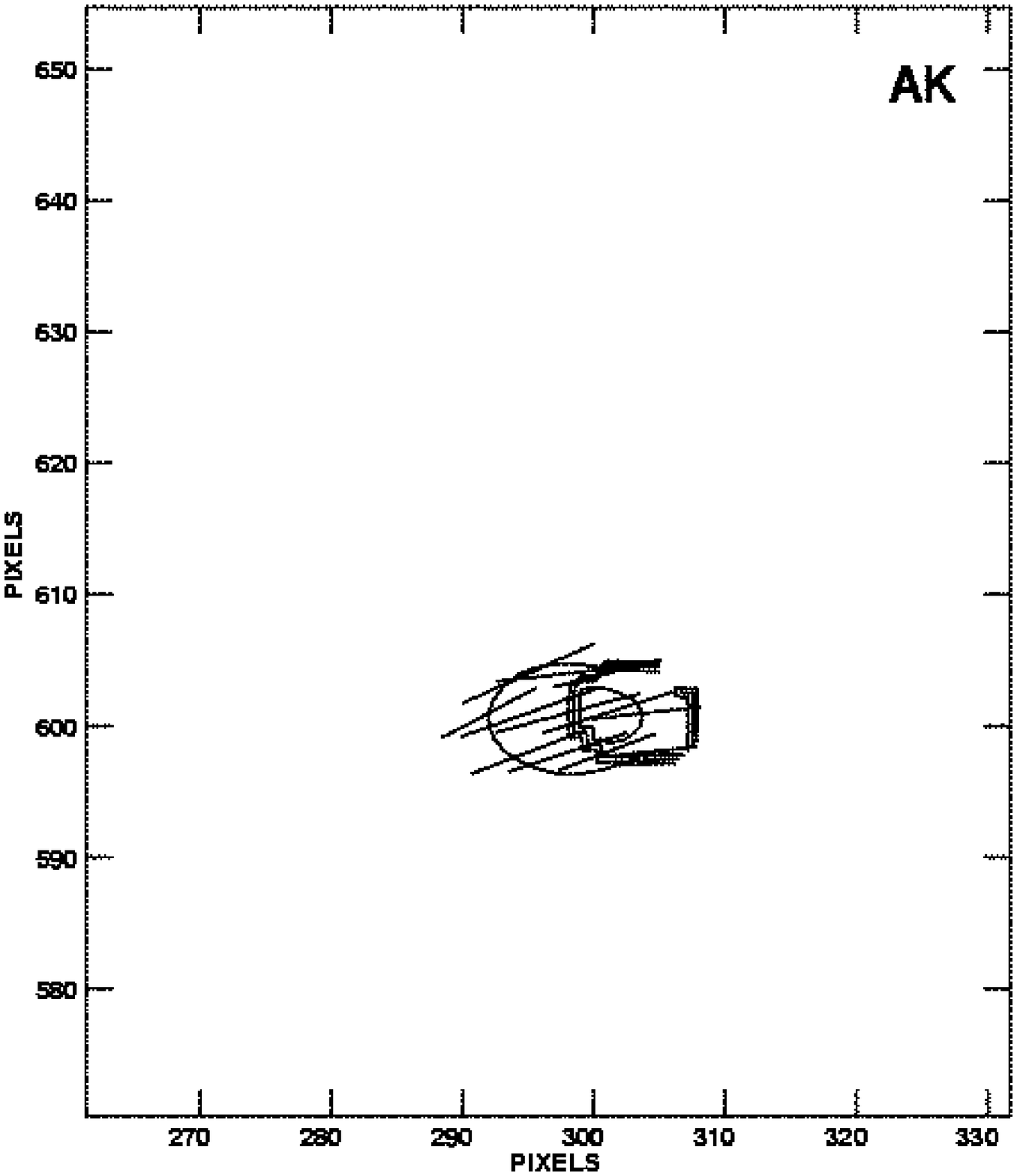} & 
\includegraphics[width=34mm, height=48mm]{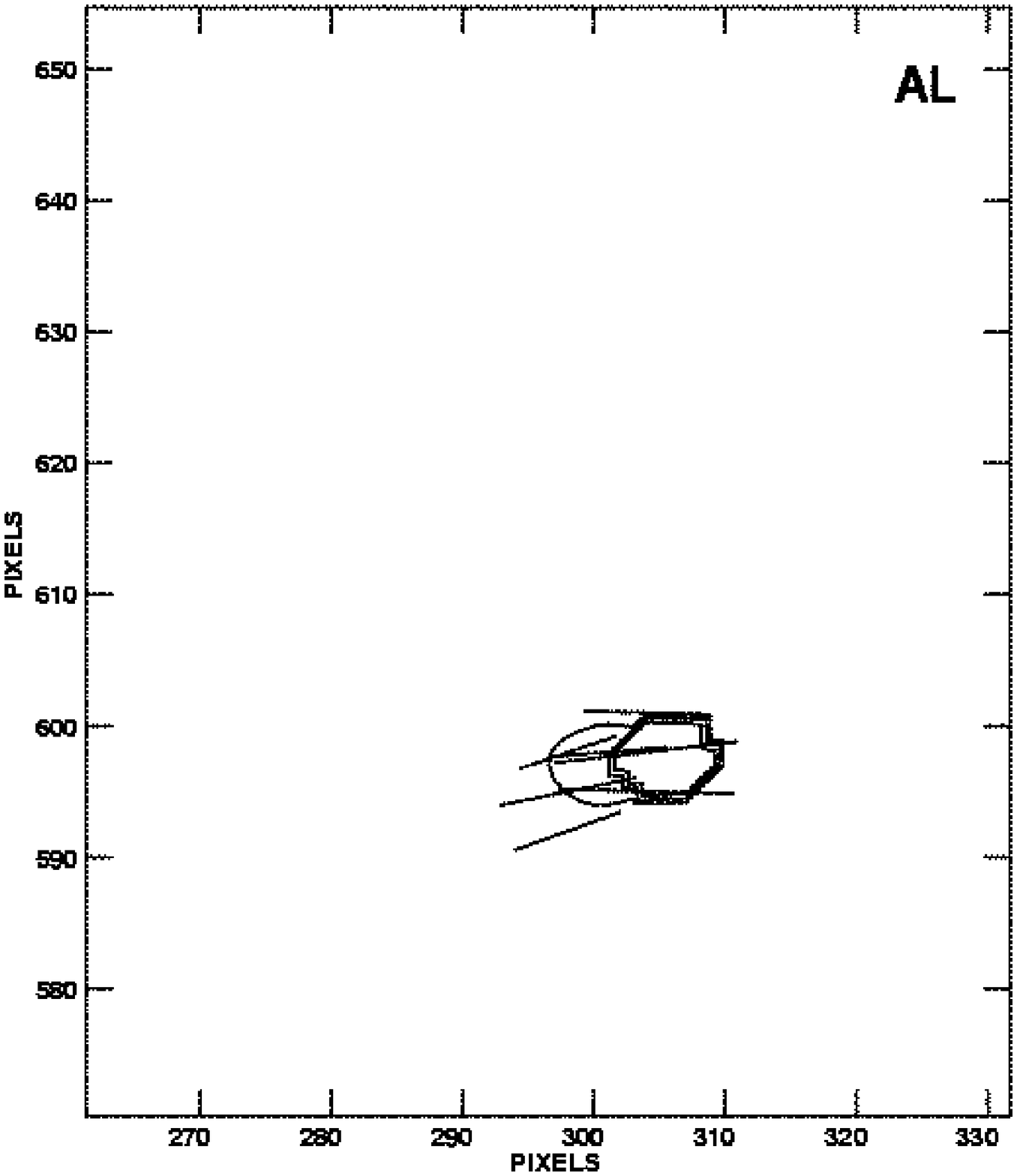} \\
\end{array}$ 

\caption{A time-series of expanded sub-images of the isolated SiO
maser component visible in the eastern part of images between epochs
\{AA-AL\}, falling in toward the eastern shell. The plot parameters
are the same as Figure~\ref{fig-pcntr-n}.}

\label{fig-pcntr-e} 
\end{figure}

\begin{figure}[h] 
\advance\leftskip-1cm \advance\rightskip-1cm

\plotone{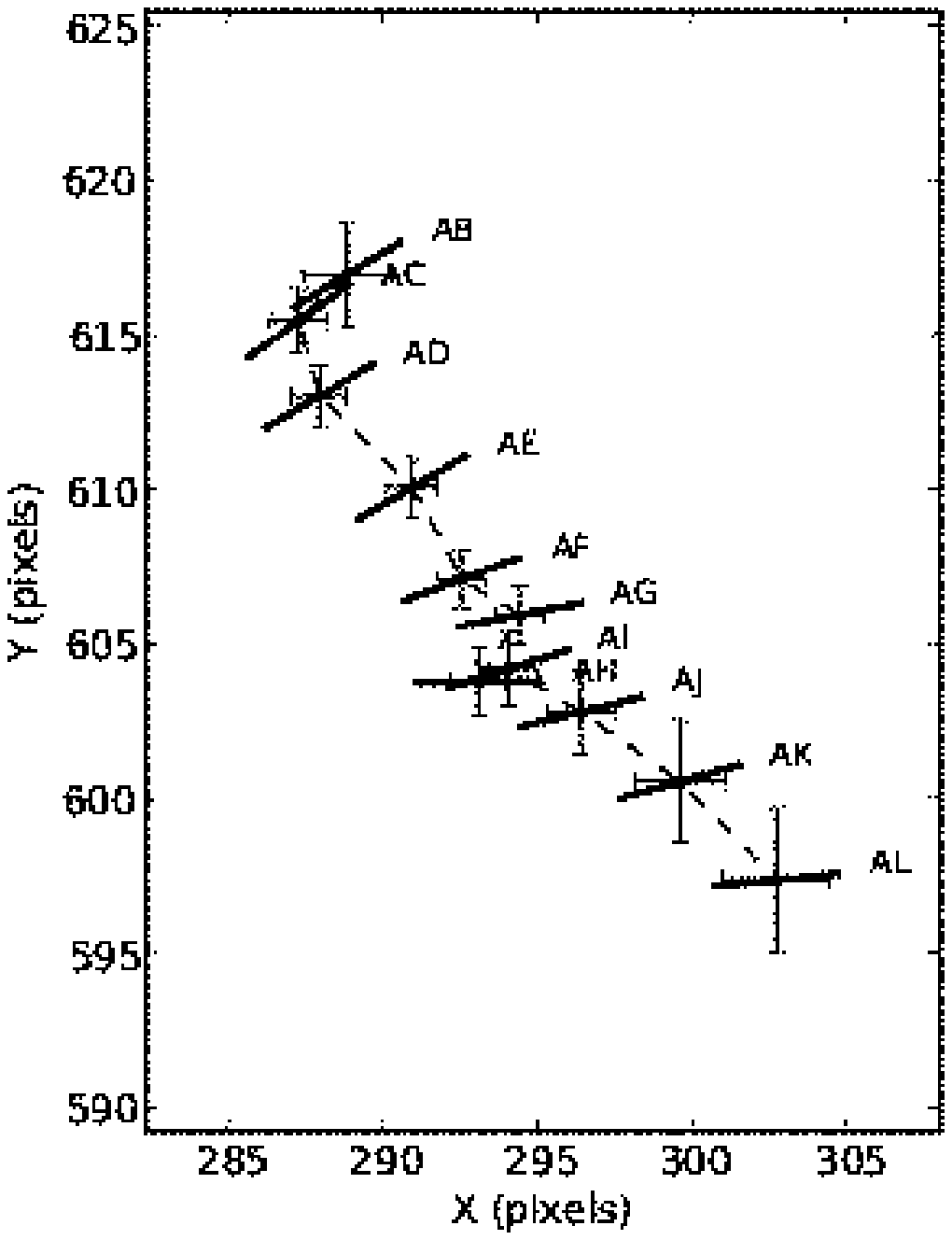}

\caption{A sequence of weighted-mean component positions for the eastern
component trajectory shown in Figure ~\ref{fig-pcntr-e}. Error bars in
x and y are shown for each weighted-mean position, and the EVPA is
drawn at each epoch as a bold vector, of uniform, constant length over epoch.}

\label{fig-traj-e} 
\end{figure}

A second isolated component is visible over epochs \{AA-AL\} to the
east of the mid-latitude of the projected shell, and we plot the
counterparts of Figures~\ref{fig-pcntr-n}-\ref{fig-traj-n} for this
eastern component as Figures~\ref{fig-pcntr-e}-\ref{fig-traj-e}. This
component is infalling over this epoch range and the EVPA is primarily
either parallel or perpendicular to the projected proper motion
vector.

This preliminary polarization proper motion analysis for the initial
components considered here, which were chosen for their location in
the outer shell where radiative anisotropic pumping effects are less
likely to predominate, show two key properties: i) the inferred
magnetic field orientation is likely to be parallel to the proper
motion vector, as the EVPA orientation is either parallel or
perpendicular to this proper motion; and ii) the motion is sometimes
non-radial relative to the central star, in contrast to what is
expected for ballistic motion in the stellar gravitational
field. These properties are consistent with component motion along
magnetic field lines in the extended atmosphere. An alternative
explanation is that a local magnetic field is being dragged and
elongated by the gas motion, and the component motion is not governed
dynamically by the magnetic field. This explanation is not excluded
however, but we believe it is not favored by the preponderance of the
current observational evidence. This issue is also discussed in
connection with jet-like feature extensions in the SiO images obtained
by \citep{cotton08}.

We also observe several 90-degree EVPA reversal at the inner-shell
boundary, as noted in the first TX Cam polarization imaging
\citep{kemball97}. These features are particularly prominent in the
south-west edge of the shell, from epochs \{AJ-AR\}. These EVPA
reversals can be explained either as transitions across the break
angle in the \citet{goldreich73} solutions or, in the case of
$m$-anisotropic pumping, a transition in the degree of anisotropy
\citep{ramos05}. These issues will be considered in more detail in a
future analysis.

\section{Conclusions}

We have presented a finely-sampled time series of 42 images of the
linearly-polarized $v=1,\ J=1-0$ SiO maser emission toward the Mira
variable, TX Cam over a pulsation phase range $\phi=0.68$ to
$\phi=1.82$. These data augment a total intensity animated sequence
previously published for this pulsation phase range by
\citet{diamond03}. From our initial analysis we conclude:

\begin{enumerate}

\item{The predominant linear polarization morphology is one in which
the EVPA of the linearly-polarized SiO emission is predominantly
tangential to the projected shell of SiO maser emission in total
intensity. This morphology is observed for all epochs presented here.}

\item{The linear polarization EVPA of individual SiO maser
components is relatively stable over their component lifetimes for
many individual components visible in the images.}

\item{The magnitude of the linear polarization appears brighter at the
inner region of the shell boundary, and decreases for larger radii.}

\item{An analysis is presented that argues that we are in the
theoretical regime $g\Omega \gg R \gg \Gamma$ or $g\Omega > R \gg
\Gamma$ for maser polarization propagation. There is, as yet,
insufficient observational data to distinguish the relative influence
of $m$-anisotropic pumping on the magnitude or orientation of the
linearly-polarized SiO $v=1,\ J=1-0$ emission}.

\item{An initial polarization proper motion study is presented for two
outer maser components that provides some supporting evidence to the
argument that magnetic fields are dynamically significant in the
extended atmospheres of late-type, evolved stars.}

\end{enumerate}

\acknowledgments

\acknowledgments

This material is based upon work supported by the National Science
Foundation under Grant No. AST-0507473. Any opinions, findings, and
conclusions or recommendations expressed in this material are those of
the authors and do not necessarily reflect the views of the National
Science Foundation. We acknowledge with thanks data from the AAVSO
International Database based on observations submitted to the AAVSO by
variable star observers worldwide. We thank Dr. W. Watson for reading
a draft of the manuscript and for helpful comments from an anonymous
referee that improved the paper.

{\it Facilities:} \facility{VLBA}

\clearpage
\begin{deluxetable}{llc}
\tablewidth{0pt} 
\tablecaption{Observing dates and epochs\label{tbl-vlba-epochs}}
\tablecolumns{3}
\tablehead{  
\colhead{Epoch code}   &  \colhead{Observing date} & 
\colhead{Optical phase} \tablenotemark{a}\\
 & & $(\phi)$ \\
}
\startdata
BD46A  & 1997 May 24       & 0.68 $\pm$ 0.01 \\
BD46B  & 1997 June 7       & 0.70 $\pm$ 0.01 \\
BD46C  & 1997 June 22      & 0.73 $\pm$ 0.01 \\
BD46D  & 1997 July 6       & 0.75 $\pm$ 0.01 \\
BD46E  & 1997 July 19      & 0.78 $\pm$ 0.01 \\
BD46F  & 1997 August 2     & 0.80 $\pm$ 0.01 \\
BD46G  & 1997 August 16    & 0.83 $\pm$ 0.01 \\
BD46H  & 1997 August 28    & 0.85 $\pm$ 0.01 \\
BD46I  & 1997 September 12 & 0.87 $\pm$ 0.01 \\
BD46J  & 1997 September 26 & 0.90 $\pm$ 0.01 \\
BD46K  & 1997 October 9    & 0.92 $\pm$ 0.01 \\
BD46L \tablenotemark{b}  & 1997 October 25   & 0.95 $\pm$ 0.01 \\
BD46M  & 1997 November 8   & 0.98 $\pm$ 0.01 \\
BD46N  & 1997 November 21  & 0.00 $\pm$ 0.01 \\
BD46O  & 1997 December 5   & 1.03 $\pm$ 0.01 \\
BD46P  & 1997 December 17  & 1.05 $\pm$ 0.01 \\
BD46Q  & 1997 December 30  & 1.07 $\pm$ 0.01 \\
BD46R  & 1998 January 13   & 1.10 $\pm$ 0.01 \\
BD46S  & 1998 January 25   & 1.12 $\pm$ 0.01 \\
BD46T  & 1998 February 5   & 1.14 $\pm$ 0.01 \\
BD46U  & 1998 February 22  & 1.17 $\pm$ 0.01 \\
BD46V  & 1998 March 5      & 1.19 $\pm$ 0.01 \\
\nodata \tablenotemark{c}     & 1998 March 21 & 1.22 $\pm$ 0.01 \\
BD46W  & 1998 April 07 & 1.25 $\pm$ 0.01 \\
BD46X  & 1998 April 19     & 1.27 $\pm$ 0.01 \\
BD46Z  & 1998 May 10       & 1.30 $\pm$ 0.01 \\
BD46AA & 1998 May 22       & 1.33 $\pm$ 0.01 \\
BD46AB & 1998 June 6       & 1.35 $\pm$ 0.01 \\
BD46AC & 1998 June 18      & 1.37 $\pm$ 0.01 \\
BD46AD & 1998 July 3       & 1.40 $\pm$ 0.01 \\
BD46AE & 1998 July 23      & 1.44 $\pm$ 0.01 \\
BD46AF & 1998 August 9     & 1.47 $\pm$ 0.01 \\
BD46AG & 1998 August 23    & 1.49 $\pm$ 0.01 \\
BD46AH & 1998 September 9  & 1.52 $\pm$ 0.01 \\
BD46AI & 1998 September 25 & 1.55 $\pm$ 0.01 \\
BD46AJ & 1998 October 14   & 1.59 $\pm$ 0.01 \\
BD46AK & 1998 October 29   & 1.61 $\pm$ 0.01 \\
BD46AL & 1998 November 17  & 1.65 $\pm$ 0.01 \\
BD46AM & 1998 December 6   & 1.68 $\pm$ 0.01 \\
BD46AN & 1998 December 23  & 1.71 $\pm$ 0.01 \\
BD46AO & 1999 January 5    & 1.74 $\pm$ 0.01 \\
BD46AP & 1999 January 23   & 1.77 $\pm$ 0.01 \\
BD46AQ & 1999 February 6   & 1.79 $\pm$ 0.01 \\
BD46AR & 1999 February 19  & 1.82 $\pm$ 0.01 \\
\enddata
\tablenotetext{a}{The optical phase is computed using the optical
maximum at MJD = 50773 cited by \citet{gray99}, and assuming their
quoted uncertainty $\triangle \phi \sim 0.01$. A mean period of 557.4
days is adopted \citep{kholopov85}.}

\tablenotetext{b}{Failed epoch, that could not be successfully
reduced in full.}
\tablenotetext{c}{This epoch was not scheduled.}
\end{deluxetable}

\clearpage
\begin{deluxetable}{lccccrrl}
\tabletypesize{\scriptsize}
\tablecaption{VLA observations of J0359+509\label{tbl-vla}}
\tablewidth{0pt}
\tablehead{
\colhead{} & \colhead{Source} & \colhead{Date} & \colhead{Start} &
\colhead{End} & \colhead{$\chi$} & \colhead{$\epsilon_\chi$} & 
\colhead{Config.\tablenotemark{c}} \\
\colhead{Epoch\tablenotemark{d}} & \colhead{} & \colhead{} & \colhead{(IAT)} &
\colhead{(IAT)} & \colhead{(deg)\tablenotemark{a}} & \colhead{(deg)\tablenotemark{b}} & 
\colhead{}
}
\startdata
 A &  J0359+509 & 1997 May 24 & 17:08:39 & 20:01:20 & 124.8 &  5.5 &  B \\
 C &  J0359+509 & 1997 Jun 23 & 17:37:39 & 19:00:20 & 112.1 &  3.6 &  CB \\
 E &  J0359+509 & 1997 Jul 24 & 12:35:00 & 14:01:49 & 130.2 &  2.6 &  C \\
 G &  J0359+509 & 1997 Aug 16 & 13:02:20 & 14:31:09 & 143.1 &  2.4 &  C \\
 J &  J0359+509 & 1997 Sep 30 & 10:38:20 & 11:34:10 & 157.1 &  0.6 &  CD \\
 K\tablenotemark{e} &  J0359+509 & 1997 Oct 13 & 07:48:49 & 09:03:19 & 161.8 &  3.1 &  CD \\
 L &  J0359+509 & 1997 Oct 31 & 07:04:50 & 08:01:59 & 144.0 &  0.8 &  D \\
 T &  J0359+509 & 1998 Feb 08 & 00:02:00 & 00:59:29 & 111.6 &  1.7 &  AD \\
 W\tablenotemark{f} &  J0359+509 & 1998 Apr 06 & 22:42:40 & 00:40:59 &  - & - & A \\
AA &  J0359+509 & 1998 May 28 & 16:54:00 & 18:17:00 & 128.5 &  8.4 &  A \\
AB &  J0359+509 & 1998 Jun 08 & 17:36:19 & 19:33:30 & 162.0 & 24.0 &  AB \\
AE &  J0359+509 & 1998 Jul 17 & 15:31:39 & 17:30:00 & 144.4 &  3.2 &  B \\
AF &  J0359+509 & 1998 Aug 15 & 15:37:10 & 17:05:50 & 149.3 &  2.5 &  B \\
AI &  J0359+509 & 1998 Sep 24 & 11:01:10 & 12:52:49 & 150.0 &  1.2 &  B \\
AJ &  J0359+509 & 1998 Oct 20 & 10:15:30 & 11:46:20 &   7.8 & 18.2 &  B \\
AL &  J0359+509 & 1998 Nov 15 & 09:06:10 & 11:03:49 &  24.5 &  8.6 &  BC \\
AN &  J0359+509 & 1998 Dec 22 & 07:38:59 & 09:38:19 &  75.6 &  1.5 &  C \\
AP &  J0359+509 & 1999 Jan 22 & 00:11:00 & 03:07:09 &  60.1 &  3.7 &  C \\
AQ &  J0359+509 & 1999 Feb 07 & 03:03:19 & 04:33:50 &  73.8 &  4.7 &  CD \\
AR &  J0359+509 & 1999 Mar 09 & 01:07:39 & 03:05:50 &  75.2 &  6.5 &  D \\
\enddata
\tablenotetext{a}{Absolute electric vector position angle (EVPA) of
the linearly-polarized emission from J0359+509, relative to an
assumed EVPA for 3C138 at 43 GHz of $\chi_{3C138}=-14$ deg \citep{perley03}.}

\tablenotetext{b}{Estimated standard error in $\chi$, derived from
independent anlyses of each 50 MHz VLA continuum spectral window, as
$\frac{1}{\sqrt{2}}|\chi_{1} - \chi_{2}|$.}

\tablenotetext{c}{VLA configuration code.}

\tablenotetext{d}{Nearest BD46 epoch in Table~\ref{tbl-vlba-epochs}.}

\tablenotetext{e}{EVPA measurement from VLBA project BK50 (Mitra
\etal, in preparation).}

\tablenotetext{f}{Data were not reducible at this VLA epoch.}

\end{deluxetable}

\clearpage

\clearpage

\clearpage

\clearpage

\clearpage

\clearpage

\clearpage

\clearpage

\clearpage

\clearpage

\clearpage

\clearpage

\clearpage

\clearpage

\clearpage

\clearpage
\end{document}